\documentclass[12pt]{article}

\usepackage[left=1.5cm,top=2.cm,right=1.5 cm,bottom=2.2cm]{geometry}
\usepackage{graphicx}
\usepackage{amsfonts}
\usepackage[T1]{fontenc}
\usepackage{lmodern}
\usepackage{amsmath}
\usepackage{color}
\usepackage{epsfig}
\usepackage{comment}
\usepackage{amssymb}

\numberwithin{equation}{section}

\def\gtwid{\mathrel{\raise.3ex\hbox{$>$\kern-.75em\lower1ex\hbox{$\sim$}}}}
\def\ltwid{\mathrel{\raise.3ex\hbox{$<$\kern-.75em\lower1ex\hbox{$\sim$}}}}
\def\square{\kern1pt\vbox{\hrule height 1.2pt\hbox{\vrule width 1.2pt\hskip 3pt
   \vbox{\vskip 6pt}\hskip 3pt\vrule width 0.6pt}\hrule height 0.6pt}\kern1pt}

\begin{document}

\begin{titlepage}

\begin{flushright}
Date: \today
\end{flushright}

\vskip .5cm

\begin{center}
{\bf\Large Scalar propagator for planar gravitational waves}
\end{center}

\vskip .5cm

\begin{center}
\bf  Rens van Haasteren$^{\star}$ and Tomislav Prokopec$^{\diamondsuit}$
\end{center}

\vskip .5cm

\begin{center}
{
Institute for Theoretical Physics, Spinoza Institute  \& EMME$\Phi$ \\
Utrecht University, Princetonplein 5,
3584 CC Utrecht, The Netherlands \\
}
\end{center}

\begin{center}
{\bf Abstract}
\end{center}

We construct the massive scalar propagator for planar gravitational wave backgrounds 
propagating on Minkowski space.
We represent the propagator in terms of the Bessel's function 
of suitably deformed nonlocal distance functions, the deformation being
caused by gravitational waves. 
 We calculate the propagator both for nonpolarized, for the {\it plus} ($+$) 
 and {\it cross} ($\times$) polarized plane waves,
 as well as for more general planar waves in $D$ spacetime dimensions. 
The propagator is useful for studying interactions of scalar fields on 
planar gravitational wave backgrounds in the context of interacting quantum fields,
in which renormalization plays a crucial role.
As simple applications of the propagator we calculate the one-loop effective action,
the scalar mass generated by the scalar self-interaction term and the one-loop energy-momentum
tensor, all renormalized by using dimensional regularization and renormalization. 
While the effective action and scalar mass remain unaffected by gravitational waves,  
the energy-momentum tensor exhibits flow of energy density in 
the direction of gravitational waves which grows quadratically with the scalar field mass,
gravitational wave amplitude and its frequency.

\begin{flushleft}
\end{flushleft}

\vskip .5cm

\begin{flushleft}
$^{\star}$ e-mail: renshaas@hotmail.com 
\\
$^{\diamondsuit}$ e-mail: T.Prokopec@uu.nl \\
\end{flushleft}

\end{titlepage}


\section{Introduction}
\label{Introduction}

 In this work we consider how a massive scalar field responds to planar gravitational waves. 
In order to facilitate such studies, we first construct the Wightman functions 
and then the Feynman propagator, which is a fundamental building block for any perturbative 
studies of an interacting scalar field theory. Anticipating dimensional regularization and 
renormalization generally required for perturbative studies, we construct the propagator 
in general (complex) $D$ spacetime dimensions.

This work builds on earlier studies~\cite{Garriga:1990dp,Jones:2016zqw, Jones:2017ejm,Siddhartha:2019yjm,Xu:2020pbj,Chen:2021bcw,Zhang:2017rno,Zhang:2018srn},
which address some aspects of the problem how planar gravitational waves affect scalar fields.
However, none of these works attempts to construct two-point functions and the corresponding 
Feynman propagator for massive scalar field in classical gravitational wave backgrounds, which is 
the main goal of this work. This propagator is an essential building block for understanding 
how quantum scalar fields respond to gravitational waves in various perturbative settings, 
either through self-interactions or through interactions with other matter fields.

Ref.~\cite{Garriga:1990dp} computes the classical cross section, and pays a particular
attention to the focusing of geodesics induced by gravitational waves, 
a very interesting phenomenon which, to our knowledge, has not been further investigated. 
In Ref.~\cite{Jones:2016zqw} the 2-gravitons-to-2-photons decay rate was computed. 
Ref.~\cite{Jones:2017ejm} obtained a non-vanishing complex scalar field condensate and
a scalar charge current, the former they interpret as a time-dependent, BEH-like mechanism, driven by the gravitational wave background.
In Ref.~\cite{Siddhartha:2019yjm} the imprint of gravitational waves 
on fermions was investigated for the first time, and in particular
how the thus-induced changes in the energy density of neutrinos could be observed in cosmological settings.
Ref.~\cite{Xu:2020pbj} has demonstrated that quantum entanglement in atomic systems can be affected by passing gravitational waves, thus paving a way for studies of decoherence induced by 
gravitational wave backgrounds.
The authors of  Ref.~\cite{Chen:2021bcw} investigate the response of freely-falling 
and accelerating Unruh-DeWitt detectors in the presence of gravitational waves.

\bigskip
{\bf The model.}
In this work we cosider a real, self-interacting scalar field  $\phi(x)$ whose 
action and lagrangian are,
\begin{equation}
 S[\phi]=\int d^Dx\sqrt{-g}\,{\cal L}_\phi
 \,,\qquad {\cal L}_\phi = -\frac12 (\partial_\mu\phi)(\partial_\nu\phi) g^{\mu\nu}
 -\frac{m^2}{2}\phi^2
         -\frac{\lambda}{4!}\phi^4
         \,,\quad
\label{scalar field action}
\end{equation}
where $g={\rm det}[g_{\mu\nu}]$, $g^{\mu\nu}$ is the inverse of the metric tensor $g_{\mu\nu}$,
$m$ is the field's mass and $\lambda$ is the self-interaction coupling strength.
We work in natural units in which $c=1$, but keep the dependence on $\hbar$ explicit.
This means that the dimension of the field $\phi$ and the mass $m$ is ${\rm m}^{-1}$,
and $\lambda$ is dimensionless. To restore the physical dimension of $m$, one ought to rescale it as, $m\rightarrow mc/\hbar$.

\subsection{Two representations for gravitational waves}
\label{Two representations for gravitational waves}

We are interested in understanding the effects of gravitational waves on scalar fields
beyond linear order in metric perturbations, so
it matters how one represents gravitational wave perturbations. There are two common
metric decompositions used in the literature, to which we shall refer to 
as {\it linear} and {\it exponential} representation. These representations are 
not in general equivalent, and that necessitates clarification.

{\bf Linear representation.} One way of representing 
 a general gravitational wave background is,
 \begin{equation}
g_{\mu\nu}(x)= \eta_{\mu\nu}+ h_{\mu\nu}(x)
\,,
 \label{metric perturbation}
 \end{equation}
where $h_{\mu\nu}(x)$ is a perturbation of the metric tensor
$g_{\mu\nu}$ around flat Minkowski space, which is characterised by
Minkowski metric $\eta_{\mu\nu}$, which is a $D\times D$ symmetric matrix of the form, 
$\eta_{\mu\nu}={\rm diag}(-1,1,1,\dots)$.
 
  In the traceless-transverse gauge (in which the gravitational field perturbation $h_{\mu\nu}$ 
  is gauge invariant to linear order in the gravitational field), 
  planar gravitational waves moving in the $x^{D-1}$ direction satisfy $h_{0\mu} = 0 $ and 
 \begin{equation}
 h_{ij}(x) = h_{ij}^{(0)}\cos\big[\omega_g t - k x^{D-1}+\psi_{ij}\big]
\,, \qquad \delta_{ij}  h_{ij}(x)= 0\,,\qquad  \partial_i h_{ij}(x) = 0\;\; (\forall j)
\,, 
 \label{gravitational wave D}
 \end{equation}
 where $\omega_g = \| \vec k\|\equiv k$ ($c=1$) denotes the gravitational wave frequency, $k$ is
 its momentum and $ h_{ij}^{(0)}$ and $\psi_{ij}\;(i,j,=1,2,\cdots D-1)$ are constant 
 amplitudes and phases, respectively. There are in total
 $D(D-3)/2$ polarizations. The phases $\psi_{ij}$ are free, but the amplitudes satisfy
 the conditions, $ \delta_{ij}h_{ij}^{(0)}=0$ and  
 $h_{i D-1}^{(0)}=h_{D-1 i}^{(0)}=0\; ( i=1,2,\cdots ,D-1)$.
 This means that there are $D-3$ linearly independent diagonal polarizations 
 and $(D-2)(D-3)/2$ purely off-diagonal gravitational wave polarizations.

In four spacetime dimensions ($D=4$), a general gravitational wave has two polarizations, 
known as the plus ($+$) and cross ($\times$) polarizations, and their amplitudes are denoted by 
$h_+$ and $h_\times$.
A general planar gravitational wave moving in the positive $z-$direction~\footnote{This can be always
achieved locally (in a small region of space) by a suitable choice of the coordinate system.} 
can be represented as, 
 \begin{equation}
 h_{ij}(x) = h_+\epsilon_{ij}^{+}\cos(\omega_gu+\psi_+)
 +h_\times\epsilon_{ij}^{\times}\cos(\omega_gu+\psi_\times)
\,, \qquad (u = t-z)
\,,
 \label{gravitational wave: planar D=4}
 \end{equation}
where $\psi_+$ and $\psi_\times$ denote constant phases
and $\epsilon_{ij}^{+}$ and $\epsilon_{ij}^{\times}$ are the two polarization tensors,
 \begin{equation}
 \epsilon_{ij}^{+}=\left(\begin{array}{ccc}
                                                 1 & 0 & 0 \cr
                                                 0 & -1 & 0 \cr
                                                  0 & 0 & 0 \cr
                                            \end{array}
                                            \right)
\,, \qquad \epsilon_{ij}^{\times}=\left(\begin{array}{ccc}
                                                 0 & 1 & 0 \cr
                                                 1 & 0 & 0 \cr
                                                  0 & 0 & 0 \cr
                                            \end{array}
                                            \right)
\,.
 \label{gravitational wave: polarization tensors}
 \end{equation}
These gravitational waves have phase velocity, $\vec v_{\rm ph} = \hat z$,
 and are often referred to as the positive frequency solutions. 
 In addition there are negative frequency gravitational waves, 
 with an opposite phase velocity ($\vec v_{\rm ph} = -\hat z$), which are obtained from 
(\ref{gravitational wave: planar D=4})
 by replacing $\omega_g(t-z)$ by $\omega_g(t+z)$. If the scalar field is located near 
 a gravitational wave source (such as a binary system), it is a very good approximation
 to assume that the negative frequency solution is absent, which is what we assume 
 in this paper.

\bigskip

{\bf Exponential representation.} 
The spatial metric in this representation for a general gravitational wave propagating 
in $D$ spacetime dimentions can be written as ({\it cf.} Eq.~(\ref{metric perturbation})),
\begin{eqnarray}
 g_{ij}(x) = 
   \left(\!\exp({\mathbf{\tilde h}})\right)_{ij}(x)
\,,\qquad
\label{metric exponential representation}
\end{eqnarray}
where $\mathbf{\tilde h}$ denotes a $(D-1)\times (D-1)$ symmetric matrix 
which is traceless and transverse, ${\rm Tr}\!\left[\tilde h_{ij}\right] = 0$, $\partial_i\tilde h_{ij} = 0$.
These two conditions reduce the number of independent components to $D(D-3)/2$.
If the gravitational wave is sourced by a binary system in the $xy$ plane, 
the wave is planar, has two polarizations and propagates in the $x^{D-1}$ direction, then
$\tilde h_{ij}$ is of the form,
\begin{eqnarray}
 \tilde h_{ij} \!\!&=&\!\! \left(\begin{array}{cc} 
                              \tilde h_+c_+(u)  &  \tilde h_\times c_\times(u)   \cr
                                \tilde h_\times c_\times(u)  &  - \tilde h_+c_+(u)   \cr
                             \end{array}\right)  
\,,\qquad (i,j = 1,2)\,;
\nonumber\\
 \tilde h_{ij} \!\!&=&\!\! \delta_{ij} \,,\qquad (i,j = 3,\cdots D\!-\!2)\,;
\nonumber\\
  \tilde h_{ij} \!\!&=&\!\!0 \,,\qquad 
  \big(i=D-1, j=1,\cdots, D\!-\!1 \,\wedge\, i=1,\cdots, D\!-\!1, j= D\!-\!1
\nonumber\\
 \!\!&&\!\!\hskip 1.5cm \,\wedge\, i=1,2, j= 3,\cdots,D\!-\!1
  \,\wedge\, i=3,\cdots,D\!-\!1, j=1,2\big)
  \,,
\label{metric exponential representation 2}
\end{eqnarray}
with $u=t-x^{D-1}$, where $x^{D-1}$ is the direction of propagation, 
\begin{equation}
c_+(u)  = \cos(\omega_g u + \psi_+)
\,,\qquad 
c_\times(u)  = \cos(\omega_g u + \psi_\times)
\,,\qquad 
\label{c+ and c x}
\end{equation}
$\psi_+$ and $\psi_\times$ are phases and 
\begin{equation}
\tilde h(u)=\sqrt{\tilde h_+^2c_+^2(u)+\tilde h_\times^2c_\times^2(u)}
\,.\qquad
\label{tilde h u}
\end{equation}
Eq.~(\ref{metric exponential representation}) can be brought to the form, 
\begin{eqnarray}
 g_{ij}(u) \!\!&=&\!\! \left(\begin{array}{cc} 
                              \cosh[\tilde h(u)]  +\frac{\sinh[\tilde h(u)]}{\tilde h(u)}\tilde h_+c_+(u) 
                                      &  \frac{\sinh[\tilde h(u)]}{\tilde h(u)}\tilde h_\times c_\times(u)   \cr
                              \frac{\sinh[\tilde h(u)]}{\tilde h(u)}  \tilde h_\times c_\times(u)  &
                            \cosh[\tilde h(u)]-\frac{\sinh[\tilde h(u)]}{\tilde h(u)}\tilde h_+c_+(u)   \cr
                             \end{array}\right)  
\,,\qquad (i,j = 1,2)\,;
\nonumber\\
 g_{ij} \!\!&=&\!\! \delta_{ij} \,,\qquad (i,j = 3,\cdots D\!-\!2)\,;
\nonumber\\
  g_{ij} \!\!&=&\!\!0 \,,\qquad 
  \big(i=D-1, j=1,\cdots, D\!-\!1 \,\wedge\, i=1,\cdots, D\!-\!1, j= D\!-\!1
\nonumber\\
 \!\!&&\!\!\hskip 1.5cm \,\wedge\, i=1,2, j= 3,\cdots,D\!-\!1
  \,\wedge\, i=3,\cdots,D\!-\!1, j=1,2\big)
  \,.
\label{metric exponential representation 3}
\end{eqnarray}
From this form one immediately sees that exponential representation 
is unimodular, ${\rm det}[g_{ij}]=1$, which makes this representation particularly 
convenient. Namely, linear representation~(\ref{metric perturbation}) is not unimodular, 
as ${\rm det}[g_{ij}]\equiv \gamma(u)
=1-h_+^2c_+(u)^2- h_{\times}^2c_\times(u)^2$, implying that, at higher orders 
in $h_{ij}$, this representation carries in it spatial, wave-like, scalar gravitational potentials. To see this
more clearly, in what follows we construct explicit transformations that relate
the two representations. For simplicity we shall do that for $D=4$ and two polarizations, 
and leave as an exercise to the 
reader to generalize it to the $D$ dimensional case. 
To make the mapping possible, we need to modify Eqs.~(\ref{metric perturbation}) and~(\ref{gravitational wave: planar D=4}) by adding two spatial scalar gravitational potentials,
namely $\psi_G(u)$ and $\zeta_G(u)$ of the form,
\begin{eqnarray}
g_{ij} = \delta_{ij}(1+2\psi_G)  + \left(\frac13 \nabla^2 \delta_{ij}-\partial_i\partial_j\right)\zeta_G
    +h_{ij}
\,,\quad  h_{ij}(u) \!\!&=&\!\! \left(\begin{array}{ccc} 
                              h_+c_+(u)  &  h_\times c_\times(u)    & 0  \cr
                                h_\times c_\times(u)  &  - h_+c_+(u) & 0  \cr
                              0      &  0     & 0  \cr
                             \end{array}\right) 
\,.\qquad
\label{modified linear representation}
\end{eqnarray}
Comparing with Eq.~(\ref{metric exponential representation 3}) we see that, 
\begin{eqnarray}
h_{+}c_+(u)  \!\!&=&\!\!  \frac{\sinh[\tilde h(u)]}{\tilde h(u)}\tilde h_+c_+(u)
\,,\qquad 
h_{\times}c_\times(u)  = \frac{\sinh[\tilde h(u)]}{\tilde h(u)}\tilde h_\times c_\times(u)\,,
\nonumber\\
\psi_G(u) \!\!&=&\!\! \frac13\partial_z^2\zeta_G(u) = \frac13\left[\cosh(\tilde h(u))-1\right]
  = \frac13\left[\sqrt{1+ h(u)^2}-1\right]
\,,
\label{mapping: exponential onto linear}
\end{eqnarray}
maps exponential representatation $\exp(\tilde{\mathbf h})$ onto linear representation,
where we have introduced, $h(u) \equiv \sqrt{h_+^2c_+(u)^2+h_\times^2 c_\times(u)^2}$.
Conversely we have,  
\begin{eqnarray}
\tilde  h_{+}c_+(u)  \!\!&=&\!\!  \frac{\ln\left[h(u)+\sqrt{1+h(u)^2}\,\right]}{h(u)}h_+c_+(u)
\,,\qquad 
\tilde h_{\times}c_\times(u)  
= \frac{\ln\left[h(u)+\sqrt{1+h(u)^2}\,\right]}{h(u)}h_\times c_\times(u)
\,,\qquad
\label{mapping: linear onto exponential}
\end{eqnarray}
where we have used, 
$h(u)=\sinh\!\left[\tilde h(u)\right]$ and $\tilde h(u)=\ln\left[h(u)+\sqrt{1+h(u)^2}\,\right]$.
Since there are no gravitational scalars in exponential representation,~\footnote{Exponential 
representation also deforms lenghts in the $xy-$plane, which 
could be represented by scalar gravitational potentials. However,
this type of spatial deformations is naturally induced by propagating gravitational waves, and thus they are physical.}
 we shall also refer to it as pure {\it tensorial} representation.

Now, from the $S$-matrix theory we know that any two metric representations yield equivalent 
results of physical measurements as long as we know how to map one onto the other,
 {\it i.e.} we know 
the maps~(\ref{mapping: exponential onto linear}--\ref{mapping: linear onto exponential}). 
In this paper we shall use both representations, however we point out that 
exponential representation is advantageous, as it is an unimodular, 
purely tensorial representation, 
meaning that it does not carry in it any spatial gravitational potential $\psi_G$,
which is responsible for local volume deformations, ${\rm det }[\delta_{ij}(1+2\psi_G)]=
  (1+2\psi_G)^3$. At a first glance one may surmise that such local volume deformations are not permissible by Birkhoff's theorem, which states that any spherically symmetric 
matter configuration cannot source time dependent (propagating) scalar metric perturbations such as 
$\psi_G$. However, planar binary systems which source gravitational waves
are not sperically symmetric matter configurations, so Birkhoff's theorem is not directly applicable.
On a deeper level, general relativity couples to matter so that the total gravitational and matter 
energy is conserved, and the same holds for the energy flow. The local expression 
of these conservation laws are the $00$ and $0i$ Einstein's equations, 
which are known to be non-dynamical. 
Since gravitational waves carry energy, passage of a gravitational wave
can induce a change of the spatial volume in the plane of the wave,
making such local volume deformations travelling with the speed of light physically justified.
The transversal nature of gravitational waves should prevent spatial deformations 
in the direction of propagation however, thus favouring exponential representation.
For the sake of comparison and for completeness, in this paper we shall 
consider both linear and exponential representation, but we will pay a special attention
when making statements concerning the physical effects on matter fields
induced at higher orders in metric perturbations by passing gravitational waves.

\bigskip

In what follows, we consider canonical quantization in Cartesian and lightcone coordinates.
Lightcone coordinates are convenient as the Klein-Gordon equation obeyed by the scalar field 
simplifies in these coordinates.
However, canonical quantization in lightcone coordinates is quite subtle, and for that 
reason we devote the whole section~\ref{Quantization in lightcone coordinates} to review it.
For readers familiar with  lightcone quantization, we suggest to go directly to 
section~\ref{Scalar propagator}.


\section{Quantization in lightcone coordinates}
\label{Quantization in lightcone coordinates}

To emphasise the differences and similarities between quantization in Cartesian and lightcone coordinates, in this section we describe both quantizations. 
Anticipating the use of dimensional regularization, we perform quantization 
in $D$ spacetime dimensions, and then discuss the reduction to 
$4$ spacetime dimensions. 

We begin by recalling coordinate independent generalities of canonical quantization. 
The scalar field action~(\ref{scalar field action}) implies canonical momentum, 
$\Pi(x) = \delta S/\delta \partial_0\phi(x) = -\sqrt{-g}g^{0\mu}\partial_\mu \phi(x)$, 
where the subscript $0$ refers to the time coordinate, which must be non-spacelike.
Canonical quantization then promotes $\phi$ and $\Pi$ into operators which satisfy 
a canonical quantization commutator,
\begin{equation}
 \big[\hat \phi(x^0,\vec x),\hat \Pi(x^0,\vec x^{\,\prime})\big] 
            = i\hbar \delta^{D-1}(\vec x -\vec x\,')
\,.
\label{canonical quantization}
\end{equation}
The Feynman propagator equation can be written on a general gravitational 
background as,  
\begin{equation}
\big(\Box - m^2\big) i\Delta_F (x;x')  = i\hbar \frac{\delta^D(x-x')}{\sqrt{-g}}
 \,,
\label{Feynman propagator: covariant}
\end{equation}
where the d'Alembertian operator $\Box$ acts on a scalar as,
\begin{equation}
\Box  = \frac{1}{\sqrt{-g}}\partial_\mu {\sqrt{-g}}g^{\mu\nu}\partial_\nu
 \,,
\label {Box operator}
\end{equation}
and the quantum  field $\hat\phi(x)$ satisfies a Klein-Gordon equation,
\begin{equation}
 (\Box -m^2)\hat{\phi}(x) = 0
\,. 
\label{Klein-Gordon equation}
\end{equation}
In equation~(\ref{Feynman propagator: covariant}) both the d'Alembertian operator 
and the Dirac delta function 
are (bi-)scalars under general coordinate transformations, implying  
that one can seek the propagator solution
to be a biscalar ({\it i.e.} scalar on both legs: $x$ and $x'$). 
One can assume the propagator solution to be of the form, 
\begin{equation}
i\Delta_F(x;x') =\Theta(\Delta x^0)i\Delta^{(+)}(x;x')
                                      + \Theta(-\Delta x^0)i\Delta^{(-)}(x;x')
 \,,
\label{Feynman propagator form}
\end{equation}
where $\Theta(x)$ denotes the Heaviside theta function, $\Delta x^0 = x^0 \!-\! x^{0\,\prime}$
and $i\Delta^{(+)}(x;x')$ and $i\Delta^{(-)}(x;x')$ are homogeneous solutions known as 
the positive and negative frequency Wightman functions, respectively, satisfying,
\begin{equation}
\big(\Box - m^2\big) i\Delta^{(\pm)}(x;x')  = 0
 \,.
\label{Wightman function EOM}
\end{equation}
For a given state $|\Omega\rangle$, the Wightman functions can be written as the following
two-point functions, 
\begin{equation}
 i\Delta^{(+)}(x;x')  = \langle\Omega|\hat\phi(x)\hat \phi(x')|\Omega\rangle
 \,,\qquad 
  i\Delta^{(+)}(x;x')  = \langle\Omega|\hat\phi(x')\hat \phi(x)|\Omega\rangle
 \,,
\label{Wightman function: state}
\end{equation}
which -- in the absence of interactions -- clearly satisfy~(\ref{Wightman function EOM}). 
It is important to notice that the propagator equation~(\ref{Feynman propagator: covariant})
 is consistent with canonical quantization~(\ref{canonical quantization}). Indeed, upon
 inserting~(\ref{Feynman propagator form})  into~(\ref{Feynman propagator: covariant}) one obtains
that one time derivative $\partial_0$ from $\Box$ commutes through 
Heaviside functions~\footnote{When the first time derivative hits Heaviside functions, it produces 
$\delta(\Delta x^0) \Delta_J(x;x')$ (where 
$\Delta_J(x;x')\equiv\langle\Omega|[\hat\phi(x),\hat \phi(x')]|\Omega\rangle$ denotes 
the Jordan two-point function), which 
vanishes by causality. Indeed, if $x^0={\rm const.}$ hypersurface is space-like, 
fluctuations of quantum fields must be independent on spatially separated points,
implying that the Jordan two-point function must vanish on any equal-time hypersurface.
Surprisingly, the Jordan two-point function need not (and does not) vanish on equal-time hypersurfaces
in lighcone quantization. This is so because the time coordinate $u$ of lightcone quantization is lightlike.
 } 
 in~(\ref{Feynman propagator form}) and hits the field $\hat{\phi}(x)$ creating a canonical momentum term, and the other produces Dirac's delta functions
 according to,
 $\partial_0\Theta(\Delta x^0) = \delta(\Delta x^0) = -  \partial_0\Theta(-\Delta x^0)$,
 resulting in,
\begin{equation}
\big(\Box - m^2\big) i\Delta_F (x;x') 
= - \frac{1}{\sqrt{-g}}\delta(\Delta x^0)\langle\Omega|\big[\hat\Pi(x),\hat\phi(x')\big]\Omega\rangle 
 =\frac{1}{\sqrt{-g}}i\hbar \delta(x^0 \!-\! x^{0\,\prime}) \delta^{D-1}(\vec x \!-\! \vec x^{\,\prime})
 \,,
\label{Feynman propagator check}
\end{equation}
which equals the right-hand-side of~(\ref{Feynman propagator: covariant}).
Since this is true in any coordinate system,~\footnote{There is a subtlety in lightcone coordinates 
which makes this analysis not entirely correct.} 
this procedure should 
allow one to fix canonical quantization in any coordinates.


\subsection{Quantization in Cartesian coordinates}
\label{Quantization in Cartesian coordinates}

We now proceed with canonical quantization in Cartesian coordinates in
 Minkowski space,~\footnote{Recall that 
both Cartesian and lightcone coordinates represent Minkowski space.} in which 
$\Box\rightarrow \partial^2 = \eta^{\mu\nu}\partial_\mu\partial_\nu$, 
by rewriting the operators $ \hat \phi(x) $ and $ \hat \Pi(x) $  in
terms of the momentum space mode operators as, 
\begin{equation}
 \hat \phi(x)  = \int\! \frac{{\rm d}^Dk}{(2\pi)^D}
                        {\rm e}^{i k\cdot x}\hat\phi(k^\alpha)
                        \,, \quad 
 \hat \Pi(x)  = \int\! \frac{{\rm d}^Dk}{(2\pi)^D}
                        {\rm e}^{i k\cdot x}\hat\Pi(k^\alpha)
\,,\quad \big(k\cdot x = \eta_{\mu\nu} k^\mu x^\nu = -k^0 x^0 + \vec k\cdot\vec x\big)
\,.\qquad
\label{momentum space representation: Minkowski}
\end{equation}
The mode operators obey algebraic equations, which follow
 from the Klein-Gordon equation~(\ref{Klein-Gordon equation}), 
\begin{equation}
\big[\! (k^0)^2 -\omega^2\big]\hat\phi(k^\alpha) = 0
,\quad
\big[\! (k^0)^2 -\omega^2\big]\hat\Pi(k^\alpha) = 0
\,,\quad  \omega^2 =  \|\vec k\|^2+m^2
\,,\qquad
\label{momentum space representation: mode equations}
\end{equation}
whose general solutions can be written as, 
\begin{equation}
\hat\phi(k^\alpha) = \hat \phi_+(\vec k)2\pi\delta(k^0-\omega) 
+ \hat \phi_-(\vec k)2\pi\delta(k^0+\omega)
\,,\quad
\hat\Pi(k^\alpha) = \hat \Pi_+(\vec k)2\pi\delta(k^0-\omega) 
+ \hat \Pi_-(\vec k)2\pi\delta(k^0+\omega)
\,,
\label{momentum space representation: Minkowski2}
\end{equation}
where $\omega$ denotes the quasiparticle frequency and 
 $\hat \phi_+(\vec k)$ and $ \hat \phi_-(\vec k)$ are the positive and negative frequency
	 mode operators, respectively. 
 Upon  inserting~(\ref{momentum space representation: Minkowski2})  
 into~(\ref{momentum space representation: Minkowski}) and integrating over $k^0$
one obtains, 
\begin{equation}
 \hat \phi(x)  = \int\! \frac{{\rm d}^{D-1}k}{(2\pi)^{D-1}} {\rm e}^{i \vec k\cdot \vec x}
 \big[{\rm e}^{-i\omega t}\hat \phi_+(\vec k)
                 + {\rm e}^{-i\omega t}\hat \phi_-(\vec k)\big]
                        \,, \quad 
 \hat \Pi(x)  =  \int\! \frac{{\rm d}^{D-1}k}{(2\pi)^{D-1}} {\rm e}^{i \vec k\cdot \vec x}
 \big[{\rm e}^{-i\omega t}\hat \Pi_+(\vec k)
                 + {\rm e}^{-i\omega t}\hat \Pi_-(\vec k)\big]
\,.\quad
\label{momentum space representation: Minkowski D-1}
\end{equation}
The hermiticity conditions,
 $\hat\phi^\dagger (x)= \hat\phi(x)$ and $\hat\Pi^\dagger (x)= \hat\Pi(x)$ 
 impose on the mode operators, 
 \begin{equation}
 \left[\hat \phi_+(\vec k)\right]^\dagger = \hat \phi_-(-\vec k)
 \,,\qquad 
 \left[\hat \Pi_+(\vec k)\right]^\dagger = \hat \Pi_-(-\vec k)
 \,,
 \label{hermiticity conditions}
 \end{equation}
 which ensure that the momentum transformation~(\ref{momentum space representation: Minkowski D-1}) preserves the number of (on-shell) degrees of freedom. This implies that, for 
 a real scalar field,  
the  positive and negative frequency poles do not fluctuate independently.
 The mode operators obey canonical quantization relations, 
 \begin{equation}
 \big[\hat \phi_+(\vec k),\hat \Pi_-(\vec k')\big] 
      =\frac{i\hbar}{2}(2\pi)^{D-1}\delta^{D-1}(\vec k + \vec k') 
 = \big[\hat \phi_-(\vec k),\hat \Pi_+(\vec k')\big] 
 \,,
\label{momentum space representation: Minkowski3}
\end{equation}
and the other two vanish, $\big[\hat \phi_+(\vec k),\hat \Pi_+(\vec k')\big] 
=0=\big[\hat \phi_-(\vec k),\hat \Pi_-(\vec k')\big] $. One can now easily check that
 inserting~(\ref{momentum space representation: Minkowski D-1})
 into $\big[\hat\phi(x^0,\vec x),\hat\Pi(x^0,\vec x^{\,\prime})\big]$
and making use of~(\ref{momentum space representation: Minkowski3}) reproduces the
canonical quantization relation~(\ref{canonical quantization}). 
As a consequence of the hermiticity condition in~(\ref{hermiticity conditions})
the two commutators in~(\ref{momentum space representation: Minkowski3}) must be 
equal, which physically means that the vacuum fluctations on the positive and negative frequency
poles are of an equal strength.

Next, taking account of $\hat \Pi(x) = \partial_t \hat \phi(x)$ implies,
$\hat\Pi_\pm(\vec k) = \mp i\omega \hat \phi_\pm(\vec k)$, from which the more common 
{\it annihilation} and {\it creation} representation of the mode operators follows, 
\begin{equation}
     \hat\phi_+(\vec k) = \sqrt{\frac{\hbar}{2\omega}} \hat a(\vec k)
     \,,\quad \hat\phi_-(\vec k) = \sqrt{\frac{\hbar}{2\omega}} \hat a^\dagger(-\vec k)
      \,,\qquad  \hat\Pi_+(\vec k) =  -i\sqrt{\frac{\hbar\omega}{2}} \hat a(\vec k)
        \,,\quad  \hat\Pi_-(\vec k) =  i\sqrt{\frac{\hbar\omega}{2}} \hat a^\dagger(-\vec k)
\,,
\label{definition of a and a+}
 \end{equation}
which obey,
 \begin{equation}
 \big[\hat a(\vec k),\hat a^\dagger(\vec k')\big] 
      =(2\pi)^{D-1}\delta^{D-1}(\vec k - \vec k')\,;\quad
 \big[\hat a(\vec k),\hat a(\vec k')\big]  = 0 = 
       \big[\hat a^\dagger(\vec k),\hat a^\dagger(\vec k')\big] 
 \,.
\label{comutation relation for a and a+}
\end{equation}
The operators $\hat a(\vec k)$ and $\hat a^\dagger(\vec k)$ are interpreted as 
the operators that annihilate the vacuum state $|\Omega\rangle$, 
\begin{equation}
\hat a(\vec k)|\Omega\rangle=0
\,, 
\label{destruction of vacuum}
\end{equation}
and create a particle of momentum $\vec k$,
$\hat a^\dagger(\vec k)|\Omega\rangle=|1_{\vec k}\rangle$, respectively. Notice that,
as a consequence of the hermiticity conditions~(\ref{hermiticity conditions}), the two 
commutators in~(\ref{momentum space representation: Minkowski3}) collapse into
one (nontrivial) commutator~(\ref{comutation relation for a and a+}). 
\begin{figure}[h!]
\vskip -.4cm
\centerline{\hspace{.in}
\epsfig{file=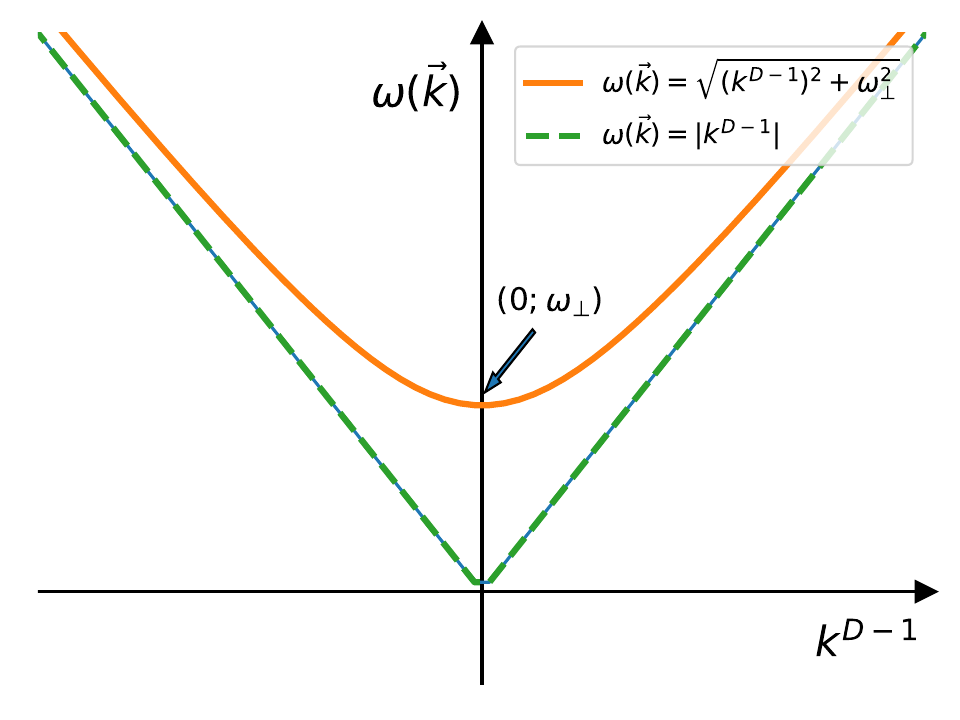, width=3.3in}
\hskip 1cm
\epsfig{file=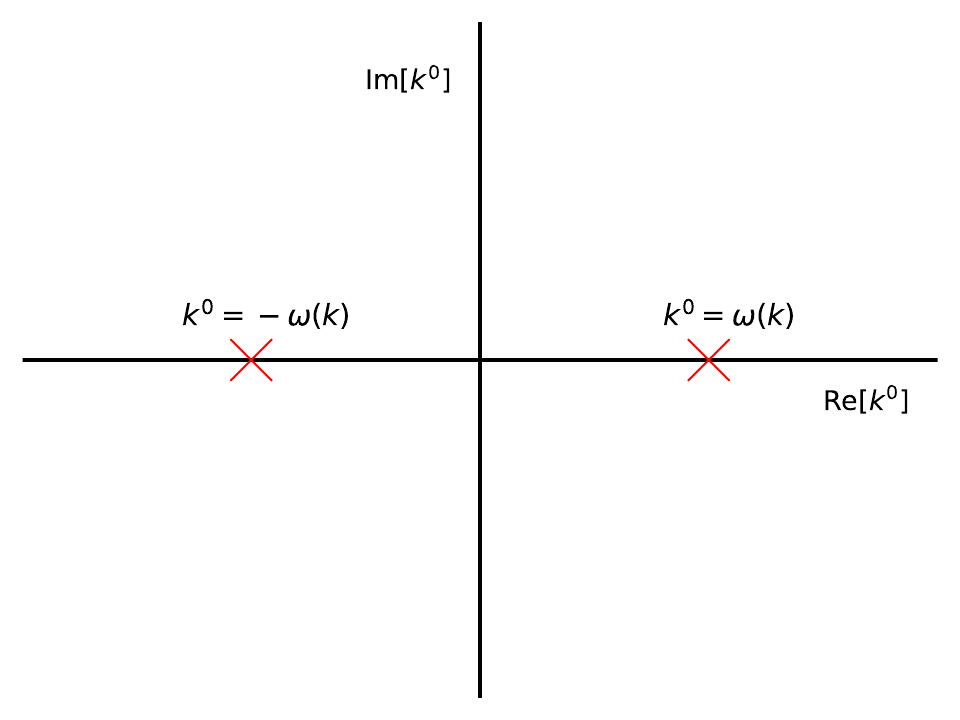, width=3.in}
}
\vskip -0.4cm
\caption{\small {\it Left panel:} Frequency versus $k^{D-1}$ for Cartesian coordinates.
{\it Right panel:} 
The quasiparticle poles $k^0=\pm\omega$ in the complex $k^0-$plane. The poles 
are symmetric with respect to ${\Re}[k^0]=0$.
}
\label{figure one}
\end{figure}

Clearly, if
$\hat a(\vec k)$ annihilate the vacuum, so do $\hat\phi_+(\vec k)$. One can construct
a more general (pure, Gaussian) state by relaxing that condition. In this case one can write,
\begin{equation}
 \hat\phi_+(\vec k) = \sqrt{\frac{\hbar}{2\omega}} \big[\alpha(\vec k)\hat a(\vec k)
                                    + \beta^*(-\vec k)\hat a^\dagger(-\vec k)\big]
     \,,\quad 
\hat\phi_-(\vec k) = \sqrt{\frac{\hbar}{2\omega}} 
   \big[\beta(\vec k)\hat a(\vec k) + \alpha^*(-\vec k) \hat a^\dagger(-\vec k)\big]
 \,,
\label{definition of a and a+ 2}
\end{equation}
where the form of the second relation is determined by 
the hermiticity condition~(\ref{hermiticity conditions}), 
and -- provided $|\alpha(\vec k)|^2-|\beta(\vec k)|^2=1$ --
canonical quantization~(\ref{canonical quantization}) remains satisfied.
The state~(\ref{definition of a and a+ 2}) is the most general pure Gaussian state,
and it corresponds to an excited state, in the sense that its energy per mode is enhanced by 
a factor $|\alpha(\vec k)|^2+|\beta(\vec k)|^2 >1$ when compared with 
the energy per mode of the vacuum state~(\ref{definition of a and a+}).

The final remark we make is that (as a consequence of the CPT theorem),
 the positive and negative frequency poles, $k^0=\pm\omega$,
 are symmetrically distributed around the origin in the complex $k^0-$plane, which is illustrated 
 in figure~\ref{figure one}. 
 In what follows, we show that quantization in lightcone coordinates yields 
a surprising result with regard to the distribution of the positive and negative frequency poles.


\subsection{Quantization in lightcone coordinates}
\label{Quantization in lightcone coordinates II}

We begin this section by an important observation
regarding canonical quantization in lightcone coordinates.
The Feynman propagator in lightcone coordinates is defined as, 
\begin{equation}
i\Delta_F(x;x') = \Theta(\Delta u) i\Delta^{(+)}(x;x')+\Theta(-\Delta u) i\Delta^{(-)}(x;x')
\,.
\label{Feynman propagator in LC}
\end{equation}
Inserting this into Eq.~(\ref{Feynman propagator: covariant}) one obtains
that canonical quantization in lightcone coordinates ought to change to,
\begin{equation}
  [\hat\phi(u,\vec x_\perp,v),\hat \Pi(u,\vec x_\perp^{\,\prime}, v')]=\frac{i}2\hbar 
\delta^{D-2}(\vec x_\perp-\vec x_\perp^{\,\prime})\delta(v-v')
\,
\label{canonical quantization LC}
\end{equation}
to make it consistent with the covariant propagator equation~(\ref{Feynman propagator: covariant}),
which should take preponderance~\footnote{This follows from the fact that the Feynman propagator 
is a fundamental building block of perturbation theory.} over canonical quantization 
as it is written in a coordinate independent way and therefore should hold in any coordinates. 
The question is what went wrong in the proof of consistency between
canonical quantization~(\ref{canonical quantization}) and the propagator equation~(\ref{Feynman propagator: covariant}). The principal mistake in the `proof' was the assumption 
that the d'Alembertian operator 
has two time derivatives, of which one commutes through the $\Theta$-functions on the account 
of a vanishing Jordan two-point function on spatial separations. Firstly, in the d'Alembertian operator 
in lightcone coordinates there is only {\it one} time coordinate $\partial_u$,
and therefore it {\it must} hit the $\Theta$-functions to produce the needed Dirac $\delta$-function,
$\delta(\Delta u)$. Secondly, equal time Jordan function needs not to vanish
(and does not vanish!)
at equal time ($u'=u$) hypersurfaces, because these hypersurfaces are lightlike, not spacelike.
Indeed,  the equal time Jordan function exhibits correlations of the type, 
\begin{equation}
 \left\langle \Omega\right|\left[\hat\phi(u,\vec x_\perp,v),\hat\phi(u,\vec x_\perp^{\,\prime}, v')\right]
 \left|\Omega\right\rangle
= \frac{i\hbar }{2}\delta^{D-2}(\vec x_\perp-\vec x_\perp^{\,\prime})\frac12{\rm sign}(v-v')
\,.
\label{equal u Jordan two point function}
\end{equation}
The factor $1/2$ in~(\ref{canonical quantization LC}) arises simply because the derivative $\partial_u$
appears twice in the d'Alembertian in lightcone coordinates,
\begin{equation}
 \Box = \frac{1}{\sqrt{-g}}\left[\partial_u \sqrt{-g}g^{uv}\partial_v 
                                       +\partial_v \sqrt{-g}g^{vu}\partial_u +\cdots\right]
\,,
\label{partial u occurs twice}
\end{equation}
thus generating two times the canonical commutator of $\hat\phi$ and $\hat\Pi$.

\bigskip

We are now ready to study vacuum fluctuations 
in lightcone coordinates~(\ref{light cone coordinates}). 
In what follows we repeat the analysis of 
section~\ref{Quantization in Cartesian coordinates}, thereby emphasising the particularities 
of lightcone quantization.

From the Minkowski metric in Cartesian coordinates, in 
which  $\eta_{\mu\nu}$ is a diagonal $D\times D$ matrix of the form, 
$\eta_{\mu\nu} = {\rm diag}(-1,1,1,\cdots,1)$ and coordinate transformations,
\begin{equation}
   u = t - x^{D-1}
   \,,\qquad  v = t + x^{D-1}
     \,,\qquad  \vec x_\perp  = \vec x_\perp
\,,
\label{light cone coordinates}
\end{equation}
one obtains the metric in lightcone coordinates and its inverse in the form, 
\begin{eqnarray}
  g_{uv} \!\!&=&\!\!  g_{vu} = - \frac12
   \,,\qquad  g_{ij}  = \delta_{ij} \;\; (i,j = 1,2,\cdots, D\!-\!2)
\,,
\label{metric in light cone coordinates}\\
  g^{uv} \!\!&=&\!\!  g^{vu} = - 2
   \,,\qquad  g^{ij}  = \delta_{ij} \;\; (i,j = 1,2,\cdots, D\!-\!2)
\,,
\label{inverse metric in light cone coordinates}
\end{eqnarray}
such that ${\rm det}[g_{\mu\nu}] = -1/4$.~\footnote{At a first sight the determinant $g$ differing from $-1$ is an inconvenience, and including a factor $1/\sqrt{2}$ in the definition of lightcone 
coordinates in~(\ref{light cone coordinates}) would fix that. We opt against that however, since 
that would result in unpleasant $\sqrt{2}$ factors in the definition of gravitational waves $h_{ij}(x)$ in 
Eq.~(\ref{gravitational wave: planar D=4}),
when expressed in terms of $u$.}

Performing the same type of Fourier transformation 
as in~(\ref{momentum space representation: Minkowski}), where now the invariant phase 
is~\footnote{The invariant phase is to be contrasted with,
$k\cdot x=-k^0 t + k^{D-1}x^{D-1} +  \vec k_\perp\cdot \vec x_\perp$,
in Cartesian coordinates.} 
\begin{equation}
k\cdot x =   \vec k_\perp\cdot \vec x_\perp -\frac12(k^uv+k^vu)
\,,
\label{invariant phase element LC}
\end{equation}
one obtains that the Klein-Gordon equation~(\ref{Klein-Gordon equation}) reduces to 
algebraic equations for the mode 
operators~(\ref{momentum space representation: mode equations}),
which in lightcone coordinates become,
\begin{equation}
\big(k^uk^v -\omega_\perp^2\big)\hat\phi(k^\alpha) = 0
,\quad
\big(k^uk^v -\omega_\perp^2\big)\hat\Pi(k^\alpha) = 0
\,,\qquad \omega_\perp^2 = \|\vec k_\perp\|^2 +m^2
\,.
\label{momentum space representation: mode equations LC}
\end{equation}
These equations tell us that, $k^u = \omega_\perp^2/k^v$,~\footnote{Alternatively, 
and completely equivalently, one can write, $k^v = \omega_\perp^2/k^u$. While in 
presence of gravitational waves the way one writes the poles matter, the two procedures 
in Minkowski space are equivalent and yield identical answers. This is 
a consequence of the fact that lightcone coordinates $(u,v)$ appear symmetrically 
in the action, and both can be used as time variables.
} and it seems that there is only 
one frequency pole. That is not true however, since $k^v$ can be both positive and negative,
and therefore we have two symmetric quasiparticle poles, $k^u= \omega_\perp^2/|k^v|$ 
(when $k^v>0$) and $k^u=-\omega_\perp^2/|k^v|$ (when $k^v<0$). This then implies 
the following general solution, 
\begin{eqnarray}
\hat\phi(k^\alpha) 
 \!\!&=&\!\!   \hat\phi_+^{\rm LC}(\vec k_\perp,k^v)
                         2\pi \delta\big(k^u-\omega_\perp^2/|k^v|\big)\Theta(k^v)
 +\hat\phi_-^{\rm LC}(\vec k_\perp,k^v)2\pi \delta\big(k^u+\omega_\perp^2/|k^v|\big)\Theta(-k^v)
\nonumber\\
\hat\Pi(k^\alpha) 
 \!\!&=&\!\!   \hat\Pi_+^{\rm LC}(\vec k_\perp,k^v)
                         2\pi \delta\big(k^u-\omega_\perp^2/|k^v|\big)\Theta(k^v)
 +\hat\Pi_-^{\rm LC}(\vec k_\perp,k^v)2\pi \delta\big(k^u+\omega_\perp^2/|k^v|\big)\Theta(-k^v)
\,,
\label{momentum space representation: mode equations LC}
\end{eqnarray}
where we introduced the superscript ${\rm LC}$ to distinguish the mode operators in
lightcone coordinates from those in Cartesian coordinates in 
Eq.~(\ref{momentum space representation: Minkowski D-1}).
When these are inserted into~(\ref{momentum space representation: Minkowski}) and integration
over $k^u$ is performed one obtains, 
\begin{eqnarray}
 \hat \phi(x)  \!\!&=&\!\!\! \int\! \frac{{\rm d}^{D-2}k_\perp {\rm d}k^v}{(2\pi)^{D-1}} 
 {\rm e}^{i \vec k_\perp\cdot \vec x_\perp}
 \big[{\rm e}^{-\frac{i}{2}( k^v u +v\omega_\perp^2/k^v)}
                   \hat \phi_+^{\rm LC}(\vec k_\perp,k^v)\Theta(k^v)
                 +{\rm e}^{-\frac{i}{2}( k^v u+v\omega_\perp^2/k^v)}
                  \hat \phi_-^{\rm LC}(\vec k_\perp,k^v)\Theta(-k^v)\big]
                        \,, \quad 
\nonumber\\
 \hat \Pi(x)  \!\!&=&\!\!\!  \int\! \frac{{\rm d}^{D-2}k_\perp {\rm d}k^v}{(2\pi)^{D-1}} 
            {\rm e}^{i \vec k_\perp\cdot \vec x_\perp}
 \big[{\rm e}^{-\frac{i}{2}( k^v u +v\omega_\perp^2/k^v)}
                  \hat \Pi_+^{\rm LC}(\vec k_\perp,k^v)\Theta(k^v)
                 +{\rm e}^{-\frac{i}{2}( k^v u+v\omega_\perp^2/k^v)}
                  \hat \Pi_-^{\rm LC}(\vec k_\perp,k^v)\Theta(-k^v)\big]
.
\nonumber\\
\label{momentum space representation: LC 1}
\end{eqnarray}
In analogy with canonical quantization in Cartesian coordinates, one can introduce 
the lightcone vacuum $|\Omega\rangle_{\rm LC}$ defined by,
\begin{equation}
  \hat \phi_+^{\rm LC}(\vec k_\perp,k^v)|\Omega\rangle_{\rm LC} = 0
\,.
  \label{lightcone vacuum: definition}
\end{equation}
A natural question that arises is: how are the two vacua $|\Omega\rangle_{\rm LC}$ 
and $|\Omega\rangle\equiv |\Omega\rangle_{\rm Cartesian}$
in Eq.~(\ref{destruction of vacuum}) related.

To answer this important question, we need to establish the relation between the mode 
operators $\hat \phi_\pm^{\rm LC}(\vec k_\perp,k^v)$,
 $\hat \Pi_\pm^{\rm LC}(\vec k_\perp,k^v)$ in lightcone coordinates in
 Eqs.~(\ref{momentum space representation: LC 1}) with those 
in Cartesian coordinates  $\hat \phi_\pm(\vec k_\perp,k^v)$,
 $\hat \Pi_\pm(\vec k_\perp,k^v)$ in Eqs.~(\ref{momentum space representation: Minkowski D-1}).
Firstly, observe that $k\cdot x $ is a scalar phase, and therefore invariant under coordinate 
transformations, which implies, 
\begin{equation}
\vec k_\perp\cdot \vec x_\perp -k^0 t + k^{D-1} x^{D-1} = 
\vec k_\perp\cdot \vec x_\perp -\frac12\big(k^u v + k^v u\big) 
\,,
\label{phase continuity}
\end{equation}
from where one infers, 
\begin{equation}
  k^{D-1} = \frac{k^v\!-\!k^u}{2} 
  \,,\quad   k^{0} = \frac{k^v\!+\!k^u}{2}
  \;\;\Longrightarrow \;\;  k^v = k^0 \!+\! k^{D-1}=-2k_u\,,\quad k^u = k^0 \!-\! k^{D-1}=-2k_v
  \,.\; 
  \label{scalar phase: Mink - LC}
\end{equation}
Next, by recalling that on the positive frequency shell, 
$k^u= \omega_\perp^2/k^v$ and $k^0=\omega$ ($k^v>0$),
and on the negative frequency shell, 
$k^u= \omega_\perp^2/k^v$ and $k^0=-\omega$ ($k^v<0$),
and upon dropping the trivial contribution $\vec k_\perp\cdot \vec x_\perp$
one obtains the following on-shell version of the phase relations~(\ref{phase continuity}),~\footnote{
Imposing the on-shell phase continuity requires solving simple equations for $k^v$, which 
for the positive frequency shell is given by, $k^vu+v\omega_\perp^2/k^v
 =2\omega t-2k^{D-1}x^{D-1}=(\omega+k^{D-1})u+(\omega-k^{D-1})v
 = \Omega_+u+\Omega_-v$, the unique solution of which is $k^v=\Omega_+$,
 $k^u=\omega_\perp^2/k^v=\Omega_-$. Similarly, for the negative frequency shell
 one obtains, $k^v=-\Omega_-$ and $k^u=\omega_\perp^2/k^v=-\Omega_+$.}
\begin{eqnarray}
{\rm Positive\; frequency \; shell:}\qquad  -\frac12\big(\Omega_- v + \Omega_+ u\big) 
\!\!&=&\!\! -\omega t + k^{D-1} x^{D-1}  
\nonumber\\
{\rm Negative\; frequency \; shell:}\qquad\quad\! \frac12\big(\Omega_+ v + \Omega_- u\big) 
\!\!&=&\!\! \omega t + k^{D-1} x^{D-1} 
\,,
\label{phase continuity: on-shell}
\end{eqnarray}
where we dropped the trivial parts $\vec k_\perp\cdot \vec x_\perp$, and we 
have introduced shifted frequencies,
\begin{eqnarray} 
\Omega_\pm(\vec k) = \omega \pm k^{D-1}
\,,
\label{Omega + and -}
\end{eqnarray}
such that $\Omega_\pm(-\vec k) =\Omega_\mp(\vec k)$. 
\begin{figure}[h!]
\vskip -.1cm
\centerline{\hspace{0.in}
\epsfig{file=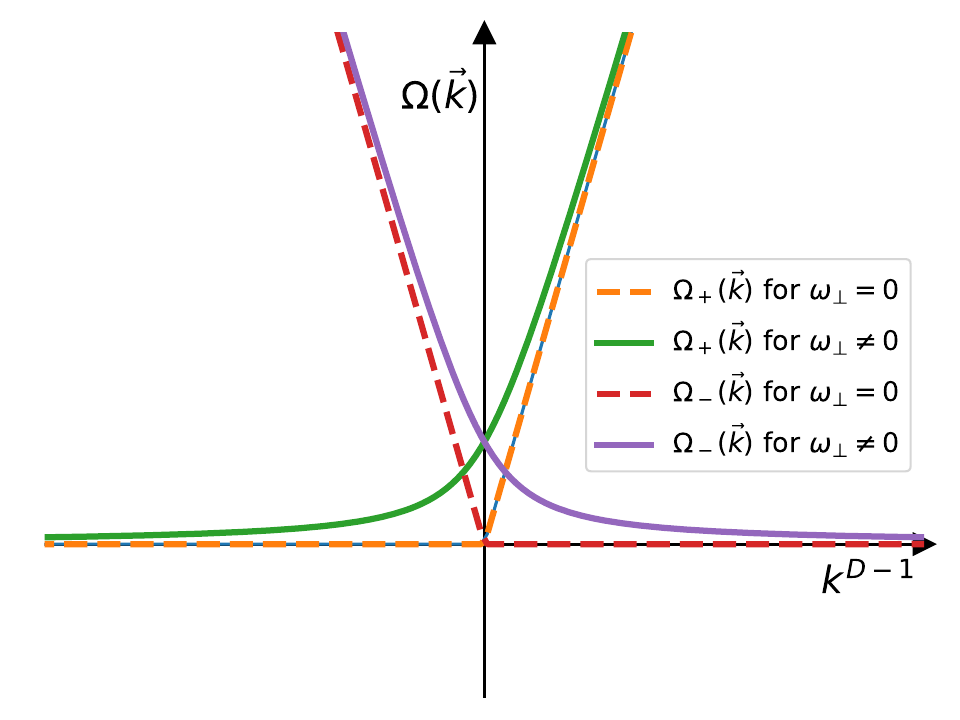, width=3.3in}
\hskip 1cm
\epsfig{file=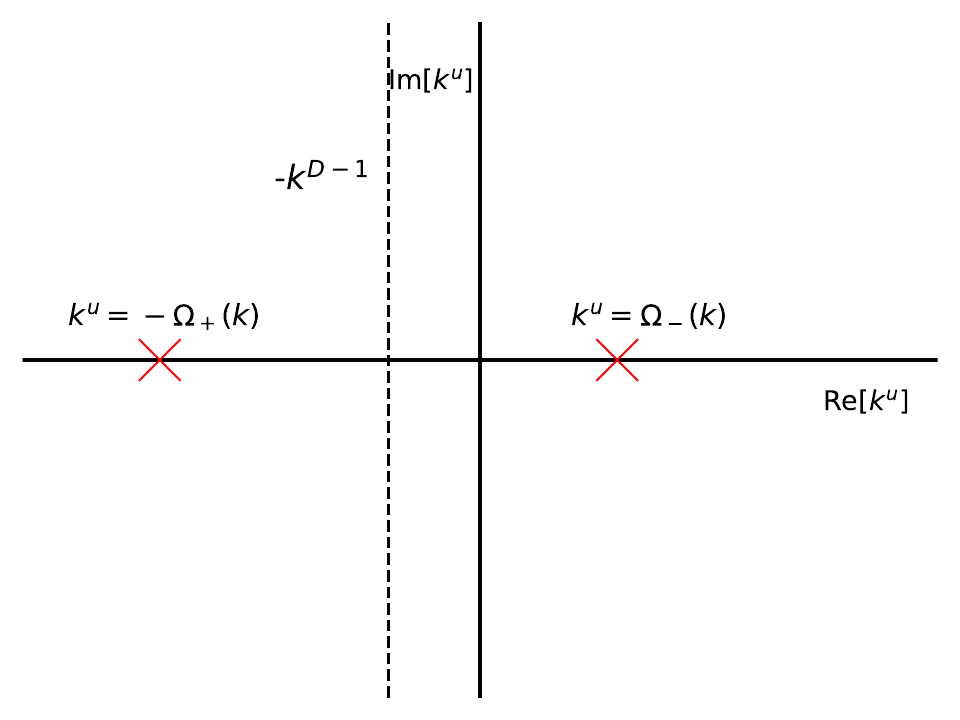, width=3.3in}
}
\vskip -0.1cm
\caption{\small {\it Left panel:} Quasiparticle frequencies $\Omega_\pm$ versus $k^{D-1}$ for lightcone coordinates for the cases when $\omega_\perp\neq 0$ (solid) and when $\omega_\perp = 0$ (dashed).
{\it Right panel:} Quasiparticle poles $k^u=-2k_v=\Omega_- =\omega-k^{D-1}$
and $k^u=-2k_v=-\Omega_+ =-(\omega+k^{D-1})$
in the complex $k^u-$plane.
The poles are {\it not} symmetric with respect to ${\Re}[k^u]=0$, 
but they are symmetric with respect
to ${\Re}[k^u]=-k^{D-1}$. 
}
\label{figure two}
\end{figure}
Therefore, when expressed 
in Cartesian coordinates, the positive and negative frequency poles 
of lightcone coordinates appear not to be symmetric, as can be seen in figure~\ref{figure two}.
The next step is to convert the integral measure 
in~(\ref{momentum space representation: LC 1})
into Cartesian coordinates to yield,
\begin{eqnarray}
 \hat \phi(x)  \!\!&=&\!\!\! \int\! \frac{{\rm d}^{D-2}k_\perp {\rm d}k^{D-1}}{(2\pi)^{D-1}} 
 {\rm e}^{i \vec k_\perp\cdot \vec x_\perp}
 \bigg[{\rm e}^{-\frac{i}{2}(\Omega_- v + \Omega_+ u)}
                   \frac{\Omega_+}{\omega}\hat \phi_+^{\rm LC}(\vec k_\perp,\Omega_+)
                 +{\rm e}^{\frac{i}{2}(\Omega_+ v + \Omega_- u)}
                  \frac{\Omega_-}{\omega}\hat \phi_-^{\rm LC}(\vec k_\perp,-\Omega_-)\bigg]
                        \,, \quad 
\nonumber\\
 \hat \Pi(x)  \!\!&=&\!\!\!  \int\! \frac{{\rm d}^{D-2}k_\perp {\rm d}k^{D-1}}{(2\pi)^{D-1}} 
            {\rm e}^{i \vec k_\perp\cdot \vec x_\perp}
 \bigg[{\rm e}^{-\frac{i}{2}(\Omega_- v + \Omega_+ u)}
                 \frac{\Omega_+}{\omega} \hat \Pi_+^{\rm LC}(\vec k_\perp,\Omega_+)
                 +{\rm e}^{\frac{i}{2}(\Omega_+ v + \Omega_- u)}
                 \frac{\Omega_-}{\omega} \hat \Pi_-^{\rm LC}(\vec k_\perp,-\Omega_-)\bigg]
\,,\qquad\;\;
\label{momentum space representation: LC 1BB}
\end{eqnarray}
where we made use of,
 ${\rm d}k^v = {\rm d}\Omega_+ = (\Omega_+/\omega){\rm d}k^{D-1}$,
for the positive frequency shell and of, 
${\rm d}k^v = {\rm d}(-\Omega_-) = (\Omega_-/\omega){\rm d}k^{D-1}$,
for the negative frequency shell. Finally, by
comparing~(\ref{momentum space representation: LC 1BB}) with 
Eq.~(\ref{momentum space representation: Minkowski D-1}), one obtains
the desired relations,
\begin{eqnarray}
\hat \phi_+(\vec k)
       \!\!&=&\!\!\frac{\Omega_+}{\omega}\hat \phi_+^{\rm LC}(\vec k_\perp,\Omega_+)
       =\sqrt{\frac{\hbar}{2\omega}}\hat a(\vec k\,)
\,,\qquad
\hat \phi_-(\vec k)
       =\frac{\Omega_-}{\omega}\hat \phi_-^{\rm LC}(\vec k_\perp,-\Omega_-)
       =\sqrt{\frac{\hbar}{2\omega}}\hat a^\dagger(-\vec k\,)
\,,\qquad
\label{mapping mode operators: LC to M}\\
\hat{\tilde \Pi}_\pm(\vec k)
       \!\!&=&\!\!\frac{\Omega_\pm}{\omega}\hat \Pi_\pm^{\rm LC}(\vec k_\perp,\pm\Omega_\pm)
\,.
\label{mapping mode operators: LC to M: momentum}
\end{eqnarray}
The mapping for the momentum operators
$\hat{\bar\Pi}_\pm(\vec k)$ in~(\ref{mapping mode operators: LC to M: momentum})
yields canonical momenta in Cartesian coordinates. However, these momenta correspond 
to the momenta defined by $\hat{\tilde \Pi}=\partial_t\hat \phi$, and therefore they are not the canonical 
momenta we need for quantization in lightcone coordinates, in which they are defined 
as, $\hat\Pi=\partial_v\hat\phi=\frac12(\partial_t+\partial_z)\hat\phi$. 
Incorporating this definition
 changes the relation between the field mode operators and the corresponding canonical momenta
in lightcone coordinates from $\hat {\tilde \Pi}_\pm=\mp i\omega\phi_\pm$ to, 
\begin{equation}
\hat \Pi_+(\vec k)= -\frac{i}{2} \Omega_-(\vec k) \hat \phi_+(\vec k)
\,,\quad
\hat \Pi_-(\vec k)= \frac{i}{2} \Omega_+(\vec k)\hat \phi_-(\vec k)
\,.\quad
\label{canonical momentum mode operators LC AA}
\end{equation}
Relations~(\ref{mapping mode operators: LC to M}) 
and~(\ref{canonical momentum mode operators LC AA})
completely define the mapping between the positive and negative 
frequency mode operators in lightcone coordinates to those in Cartesian coordinates.

Since the mapping does not mix the positive and negative frequency mode operators, 
the two vacua are identical ({\it cf.} Eqs.~(\ref{destruction of vacuum}) 
and~(\ref{lightcone vacuum: definition})),
\begin{equation}
|\Omega\rangle_{\rm LC} =|\Omega\rangle
\,,
\label{lightcone and Mink vacuum}
\end{equation}
which answers the principal question we have asked above. 
The trivial identification of the two vacua in Eq.~(\ref{lightcone and Mink vacuum}) is a consequence of 
the judicious split into the positive and negative frequency operators in 
Eq.~(\ref{momentum space representation: mode equations LC}--\ref{momentum space representation: LC 1}).
With these observations in mind, we can
write the following convenient on-shell decomposition of the field operators,
\begin{eqnarray}
 \hat \phi(x)  \!\!&=&\!\!\! \int\! \frac{{\rm d}^{D-1}k}{(2\pi)^{D-1}} 
 {\rm e}^{i \vec k_\perp\cdot \vec x_\perp}
 \Big[{\rm e}^{-\frac{i}{2}\Omega_- v}
                  \phi_+(u,\vec k\,)\hat a(\vec k\,)
                 +{\rm e}^{\frac{i}{2}\Omega_+ v}
                  \phi_-(u,\vec k\,)\hat a^\dagger(-\vec k\,)\Big]
                        \,, \quad 
\label{momentum space representation: LC 2BA}\\
 \hat \Pi(x)  \!\!&=&\!\!\!  \int\! \frac{{\rm d}^{D-1}k}{(2\pi)^{D-1}} 
            {\rm e}^{i \vec k_\perp\cdot \vec x_\perp}
 \Big[{\rm e}^{-\frac{i}{2}\Omega_- v}
                 \Pi_+(u,\vec k\,)\hat a(\vec k\,)
                 +{\rm e}^{\frac{i}{2}\Omega_+ v}
                  \Pi_-(u,\vec k\,)\hat a^\dagger(-\vec k\,)\Big]
\,,\qquad\;\;
\label{momentum space representation: LC 2BB}
\end{eqnarray}
where the vacuum mode functions are,
\begin{equation}
\phi_+(u,\vec k\,) = \sqrt{\frac{\hbar}{2\omega}}\;{\rm e}^{-\frac{i}{2}\Omega_+(\vec k) u}
\,,\quad
\phi_-(u,\vec k\,) = \sqrt{\frac{\hbar}{2\omega}}\;{\rm e}^{\frac{i}{2} \Omega_-(\vec k) u}
\,,\quad
\Pi_\pm(u,\vec k\,)  = \mp\frac{i}{2}\Omega_{\mp}(\vec k)\phi_\pm(u,\vec k\,) 
\,.\qquad
\label{positive and negative solutions in LC}
\end{equation}
This decomposition is convenient to study problems in lightcone coordinates in backgrounds 
which depend on $u=t-x^{D-1}$, in which case the mode functions 
in~(\ref{positive and negative solutions in LC}) become more general functions of $u$.
It is now not hard to show that 
Eqs.~(\ref{momentum space representation: LC 2BB}--\ref{positive and negative solutions in LC})
are consistent with the canonical quantization in 
lightcone coordinates in Eq.~(\ref{canonical quantization LC}).~\footnote{The only subtle point
in the integrations involves the realisation that the integral over $k^{D-1}$ ought to be performed as, 
\begin{equation}
\int_{-\infty}^\infty\frac{{\rm d}k^{D-1}}{2\pi}\frac{\hbar}{2\omega}
     \left[{\rm e}^{-\frac{i}{2}\Omega_-\Delta v}\frac{i}{2}\Omega_-
            + {\rm e}^{\frac{i}{2}\Omega_+\Delta v}\frac{i}{2}\Omega_+  \right]
   = \int_{-\infty}^0 \frac{{\rm d}z_-}{2\pi} \frac{i\hbar}{2}{\rm e}^{i z_-\Delta v}
     + \int_0^{\infty} \frac{{\rm d}z_+}{2\pi} \frac{i\hbar}{2}{\rm e}^{i z_+\Delta v}
     = \frac{i\hbar}{2}\delta(\Delta v)
\,,
\label{integration over  D-1 in LC}
\end{equation}
where $\Delta v = v-v'$, $z_- = -\Omega_-$ and $z_+=\Omega_+$.
}

Partial results on the quantization in lightcone coordinates have been known in literature
in this context. In particular in Refs.~\cite{Garriga:1990dp,Chen:2021bcw} the correct frequency poles 
were identified and Ref.~\cite{Chen:2021bcw}
obtained similar results for the Jordan function~(\ref{equal u Jordan two point function}).

To summarize, canonical quantization in lightcone coordinates yields to 
asymmetrically placed positive and negative frequency quasiparticle poles 
(when expressed in Cartesian coordinates $(\vec k_\perp,k^{D-1})$) with respect to 
the $\Re[k^u]=0$ vertical axis, as can be seen in figure~\ref{figure two}
(an analogous asymmetry of the quasiparticle poles with respect to $\Re[k^v]=0$ exists). 
The quasiparticle poles are symmetrically distributed
when viewed in the lighcone momenta $(k^u,k^v)$,
{\it cf.}  Eq.~(\ref{momentum space representation: mode equations LC}).
The origin of the asymmetry (when viewed in Cartesian coordinates) can be traced back 
to the fact that  $(k^u,k^v)$ coordinates are {\it linked}, 
in that when $k^u$ is projected on-shell, so is 
$k^v$, and {\it v.v.}, which can be seen in Eq.~(\ref{phase continuity: on-shell}). 
This pole coupling disappears in the limit when $\omega_\perp\rightarrow 0$,
which is the limit of the massless two-dimensional scalar field theory.~\footnote{The decoupling
between $k^u$ and $k^v$ also occurs in the massless higher dimensional theories,
in the $\vec k_\perp=0$ sector of the theory.} This is also the limit of enhanced symmetry,
namely the usual Poincar\'e symmetry gets enhanced to conformal symmetry.
Detailed ramifications of this symmetry enhancement will be discussed elsewhere. For now we 
just point out that, as can be seen in figure~\ref{figure two}, when $\omega_\perp=0$ 
the positive (negative) frequency
becomes a degenerate, non-invertible function of $k^{D-1}$ for negative (positive) $k^{D-1}$, 
resulting in a large degeneracy of the vacuum state, and 
a symmetric, $V$-shaped, dispersion relation $\Omega_\pm(\vec k)=|k^{D-1}|$ 
characterising massless particles.
Other noteworthly features of lightcone quantization include: 
the unexpected factor $1/2$ in the 
canonical quantization relation~(\ref{canonical quantization LC}), and 
the appearance of Heaviside functions 
in the mode decomposition~(\ref{momentum space representation: LC 1}), 
which is essential for the correct identification of the vacuum state in lightcone coordinates.

The next section generalizes of the results of this section,
by including planar gravitational wave backgrounds. In particular, we construct the two-point Wightman functions and the Feynman propagator.


\section{Scalar propagator}
\label{Scalar propagator} 
 
 In this section we derive the scalar field propagator in general $D$ spacetime dimensions,
which makes it suitable for dimensional regularization used in 
perturbative calculations. The extension to $D$ dimensions can be made 
 by assuming two gravitational wave polarizations propagating in the $x^{D-1}$ direction,
 or more generally $(D-2)(D-3)/2$ off-diagonal and $(D-3)$ diagonal  polarizations, all 
 propagating in the $x^{D-1}$ direction,  whereby most of our attention will be given 
to the former case. We shall not consider the case of stochastic gravitational waves,
which propagate in different directions.

Since binary systems typically generate nonpolarized gravitational waves, we shall 
start with the propagator for nonpolarized gravitational waves, and then discuss how to generalize 
it to polarized gravitational waves. For completeness, we shall separately consider two cases, linear 
representation~(\ref{gravitational wave: planar D=4}--\ref{gravitational wave: polarization tensors}), 
and exponential 
representation~(\ref{metric exponential representation}--\ref{metric exponential representation 3}).
For general planar gravitational waves the amplitudes $h_+$ and $h_\times$ and the corresponding phases $\psi_+$ and $\psi_\times$ are independent, while 
for nonpolarized gravitational waves the phases are related as, $\psi_\times =\psi_+-\pi/2$ 
($\psi_+$ can always be removed by a shift in time) and the amplitudes are equal, 
$h_+=h_\times\equiv h$. 
The amplitude of the two polarizations
differ when the binary
system has an excentric orbit, see {\it e.g.} Eq.~(62) of Ref.~\cite{Bourgoin:2022ilm,Amaro-Seoane:2022rxf};
however the phase difference between the two polarization remains $\pi/2$.
Since most of the binary systems have excentric orbits, 
polarized gravitational waves are ubiquitous.
In what follows, for simplicity, we first consider nonpolarized gravitational waves,
and then generalize to polarized gravitational waves.


 \subsection{Non-polarized gravitational waves}
\label{Non-polarized gravitational waves}

{\bf Linear representation.} 
Non-polarized gravitational waves moving in $z-$direction 
are characterized by $h_+=h_\times\equiv h$ and by
$\psi_+=0$ and $\psi_\times = -\pi/2$. 

The scalar operator field equation of motion~(\ref{Klein-Gordon equation})
in lightcone coordinates then becomes, 
\begin{eqnarray}
&&\hskip -1.1cm
\left(\Box - m^2\right)\hat \phi(x)
\label{EOM scalar in 4}\\
\!\!& =&\!\!\!
   \left[\!-4\partial_v\partial_u  
    \!+\!\frac{1}{1\!-\!h^2}\left(\partial_x^2\!+\!\partial_y^2\!-\!h_+\cos(\omega_gu)
        (\partial_x^2\!-\!\partial_y^2)
              \!-\!2 h_\times\sin(\omega_gu)\partial_x\partial_y\right) 
              \!+\!\sum_{i=3}^{D-2}\partial_i^2 \!-\! m^2\right]
       \hat\phi(u,\vec x_\perp,v) = 0
\,,\;
\nonumber
\end{eqnarray}
where we made use 
of~(\ref{gravitational wave: planar D=4}--\ref{gravitational wave: polarization tensors})
and $\vec x_\perp = (x,y,\cdots,x^{D-2})$. Note that the sum in
Eq.~(\ref{EOM scalar in 4}) is present only when $D>4$.
The simplicity of the equation ows to the fact that  
 $g= -\frac14\gamma(u)=-\frac14\left(1-h^2\right)={\rm constant}$ is spacetime independent
 and the inverse of $g_{ij} = \delta_{ij}+h_{ij}$ 
 is simply 
 $g^{ij}=(\delta_{ij}-h_{ij})/(1-h^2)\; (i,j=x,y)$, $g^{uv}=g^{vu}=-2$, 
 and $g^{uu}=g^{vv}=0$. The principal advantage of using lightcone coordinates
 is in that in these coordinates the Klein-Gordon equation~(\ref{EOM scalar in 4}) becomes
first order in the time coordinate $u$.

 Canonical quantization proceeds as in
Eqs.~(\ref{momentum space representation: LC 2BA}--\ref{momentum space representation: LC 2BB})
of section~\ref{Quantization in lightcone coordinates II}, 
 \begin{eqnarray}
  \hat\phi(u,\vec x_\perp,v) 
  \!\!&=&\!\! \int \frac{{\rm d}^{D-2}k_\perp {\rm d}k^{D-1}}{(2\pi)^{D-1}}
     {\rm e}^{i\vec k_\perp\cdot\vec x_\perp}
      \Big[{\rm e}^{-\frac{i}{2}\Omega_-(\vec k\,) v}\phi_+(u,\vec k\,)\hat a(\vec k\,)
       + {\rm e}^{\frac{i}{2}\Omega_+(\vec k\,) v}\phi_-(u,\vec k\,)\hat a^\dagger(-\vec k\,)\Big]
\,,
\label{field expansion in 4}
  \\
   \hat\Pi(u,\vec x_\perp,v) 
    \!\!&=&\!\! \int \frac{{\rm d}^{D-2}k_\perp {\rm d}k^{D-1}}{(2\pi)^{D-1}}
     {\rm e}^{i\vec k_\perp\cdot\vec x_\perp}
      \Big[{\rm e}^{-\frac{i}{2}\Omega_-(\vec k\,) v}\Pi_+(u,\vec k\,)\hat a(\vec k\,)
       + {\rm e}^{\frac{i}{2}\Omega_+(\vec k\,) v}\Pi_-(u,\vec k\,)\hat a^\dagger(-\vec k\,)\Big]
\,,
\label{canonical momentum expansion in 4}
\end{eqnarray}
but now 
with $\hat\Pi(x)= 2\sqrt{-g}\partial_v \hat\phi(x)$ 
\big($\Pi(x)=\delta S/\delta (\partial_u \phi(x))$\big) and
the vacuum mode functions~(\ref{positive and negative solutions in LC}) ought to be generalized
to solutions of the following mode equations, 
\begin{equation}
   \!\left(\!\partial_u
     \!\pm\! \frac{i}{2\Omega_\mp}\bigg[
        \frac{[1\!-\!h_+\cos(\omega_gu)]k_x^2\!+\![1\!+\!h_+\cos(\omega_gu)]k_y^2)
                           \!-\!2 h_\times\sin(\omega_gu) k_xk_y}{1\!-\!h^2}
                            \!-\!(k_x^2\!+\!k_y^2)\bigg]
      \!\pm\! \frac{i}{2}\Omega_\pm\right)\!
       \phi_\pm(u,\vec k) = 0
.\hskip 0.2cm
\label{EOM scalar in 4 mode functions}
\end{equation}
where we made use of, $\omega_\perp^2=\sum_{i=1}^{D-2}k_i^2 + m^2$ and
$\Omega_\pm=\omega_\perp^2/\Omega_\mp$.
Hermiticity of the field, $\hat\phi^\dagger(x) = \hat \phi(x)$, imposes the following condition on
the (complex) mode functions and their canonical momenta,
\begin{eqnarray}
                 \phi_-^*(u,-\vec k\,) = \phi_+(u,\vec k\,) 
                \,,\qquad 
                 \Pi_-^*(u,-\vec k\,) = \Pi_+(u,\vec k\,)
\,,
\qquad
\label{mode functions hermiticity}
\end{eqnarray}
where we used $\Omega_\pm(-\vec k\,) 
  =\Omega_\mp(\vec k\,)$. In fact, one can show that
the (hermiticity) conditions~(\ref{mode functions hermiticity})
hold for general gravitational waves. 
Eq.~(\ref{EOM scalar in 4 mode functions}) is easily solved, 
\begin{eqnarray}
 \phi_\pm(u,\vec k) \!\!&=&\!\! \frac{\sqrt{\hbar}}{\sqrt{2\omega}(1-h^2)^{1/4}}
 \exp\bigg\{\!\mp \frac{i}{2} \bigg[\Omega_\pm(\vec k\,)u 
              +\frac{\Psi(u)}{\Omega_\mp(\vec k\,)}\bigg]\bigg\}
 \,,\qquad
\label{mode function: general solution 4}
\\
 \!\!&&\!\!\hskip -2cm
 \Psi(u) =   \frac{1}{1\!-\!h^2}\int^u d\tilde u
 \big[h^2(k_x^2\!+\!k_y^2)
        \!-h_+\cos(\omega_g \tilde u)(k_x^2\!-\!k_y^2)
       -2h_\times\sin(\omega_g \tilde u)(k_xk_y)\big]
\nonumber\\
 \!\!&&\!\!\hskip -1cm
     =\,\frac{1}{\omega_g(1\!-\!h^2)}
 \big[h^2(k_x^2\!+\!k_y^2)\omega_gu\!-\!h_+\sin(\omega_g u)(k_x^2\!-\!k_y^2)
       \!+\!2h_\times\cos(\omega_gu)  
       (k_xk_y)\big]
\,,
\label{mode function: general solution phase 4}
\end{eqnarray}
where we dropped any momentum dependent, but time independent, integration constants,
as such constants are unphysical (recall that time independent phases can be absorbed in the definition 
of the wave function).
The $u$-dependence of the mode functions $\phi_\pm$ in~(\ref{mode function: general solution 4})
is written such that their phases are split into the part (${\exp}[\mp i\Omega_\pm u/2]$) 
which survives in the limit when $h_+$ and $h_\times$ vanish, 
and the part (${\exp}[\mp i\Psi(u)/(2\Omega_\mp)]$) 
which is suppressed by powers of  $h_+$ and/or $h_\times$. 
The mode function normalization in~(\ref{mode function: general solution 4}) 
comes from the Wronskian condition, 
$\phi_+(u,\vec{k})\Pi_-(u,-\vec{k}) - \phi_-(u,\vec{k})\Pi_+(u,-\vec{k}) = i\hbar /2$, 
whose individual contributions are,
\begin{eqnarray}
\phi_+(u,\vec k) \Pi_-(u,-\vec k) \!\!&=&\!\! 
               \phi_+(u,\vec k) \phi_-(u,-\vec k)
                     \frac{i}{2}\sqrt{1-h^2}\Omega_-(\vec k)
                            = \frac{i\hbar}{4\omega}\Omega_-(\vec k)
\nonumber\\
  \phi_-(u,\vec k)  \Pi_+(u,-\vec k)   \!\!&=&\!\!  
   \phi_-(u,\vec k) \phi_+(u,-\vec k)
     \left(-\frac{i}{2}\sqrt{1-h^2}\Omega_+(\vec k)\right)
                            = -\frac{i\hbar}{4\omega}\Omega_+(\vec k)
\,,
\label{Wronskian condition in 4}
\end{eqnarray}
 where we made use of $\Pi_\pm(u,\vec k)  = \mp \frac{i}{2}\Omega_{\mp}(\vec k\,)\sqrt{1-h^2}\phi_\pm(u,\vec k)$.

The result~(\ref{mode function: general solution 4}--\ref{mode function: general solution phase 4})
is exact to all orders in the gravitational field amplitude $h_{ij}$. The result is so simple due to the fact 
that nonpolarized gravitational waves affect the (spatial part) of the determinant 
of the metric tensor only by a constant,
${\rm det}[g_{ij}] =\gamma(u)= 1-h^2$.  In $D=4$ this can be interpreted as 
a linear combination of two spacetime independent gravitational potentials of a magnitude
({\it cf.} Eq.~(\ref{mapping: exponential onto linear})), 
$\psi_G=\frac13\partial_z^2\zeta_G = \frac13\big[\sqrt{1-h^2}-1\big]= -{h^2}/{6}+{\cal O}(h^4)$.
For eternal, nonpolarized, gravitational waves 
the effect is time independent, and therefore not measurable, {\it i.e.} it can be removed 
by a global gauge transformation.
However, 
for gravitational waves that turn on adiabatically in time, the effect becomes measurable,
and manifests itself as a contraction of the local spatial volume (between two instances in time), 
and it is a consequence of the fact  
that gravitational wave carries energy. As expected, the effect is quadratic in the field, just as is the  energy carried by gravitational waves. 
This is an example of 
the gravitational memory effect~\cite{Zeldovich:1974gvh,Braginsky:1985vlg,Christodoulou:1991cr}, which can be measured by 
gravitational wave detectors as a change (increase) 
in the (invariant spatial) distance between two points after passage of gravitational waves.
This effect may be different in the pure tensorial representation, implying that 
a more careful analysis is needed to understand the gravitational memory effects.

From Eq.~(\ref{mode function: general solution 4})
it follows that the scalar field responds to planar gravitational waves by an amplified average magnitude of vacuum fluctuations,
which is a response independent on the field momentum. As can be seen from 
Eq.~(\ref{mode function: general solution phase 4}), 
the scalar field oscillations (in the $u$ direction) exhibit a modulated frequency
whose amplitude is suppressed by the amplitude of the gravitational wave and 
delayed in phase by $\pi/2$ with repect to that of the gravitational wave.
Similarly as the vacuum amplitude, 
the amplitude of the phase modulations in~(\ref{mode function: general solution phase 4})
is enhanced at higher orders in $h_{ij}$ by a factor 
$1/(1-h^2)$, which is a consequence of the reduction in 
the spatial volume in which the scalar field fluctuates. Finally, 
Eq.~(\ref{mode function: general solution phase 4}) exhibits a potentially observable second order 
effect, the accumulated phase in~(\ref{mode function: general solution phase 4}),
$\delta \Psi(u) = \big[h^2/(1-h^2)\big]\big[k_\perp^2/\omega_g\big]u 
= [h^2k_\perp^2/(\omega_g)]u +{\cal O}(h^4)$, which is generated by 
the gravitational backreaction on the spatial volume, and speeds up
the scalar field oscillations with respect to those in Minkowski space.
Even though this effect is of the second order, 
it is cumulative, and therefore it can lead to observable 
effects (if observed over long time intervals), see figure~\ref{figure three}.~\footnote{
In more general situations, when $\omega_g$ and $h_{+,\times}$ are (adiabatic) functions of time, 
the accumulated phase generalizes to, $\delta \Psi(u) 
   \simeq\int^u_{u^0}  [(h_+^2(u')+h_\times^2(u')]k_\perp^2/(4\omega_g(u'))]{\rm d}u'
      +{\cal O}(h^4)$.
}

\bigskip

\begin{figure}[h!]
\vskip -.2cm
\centerline{\hspace{.in}
\epsfig{file=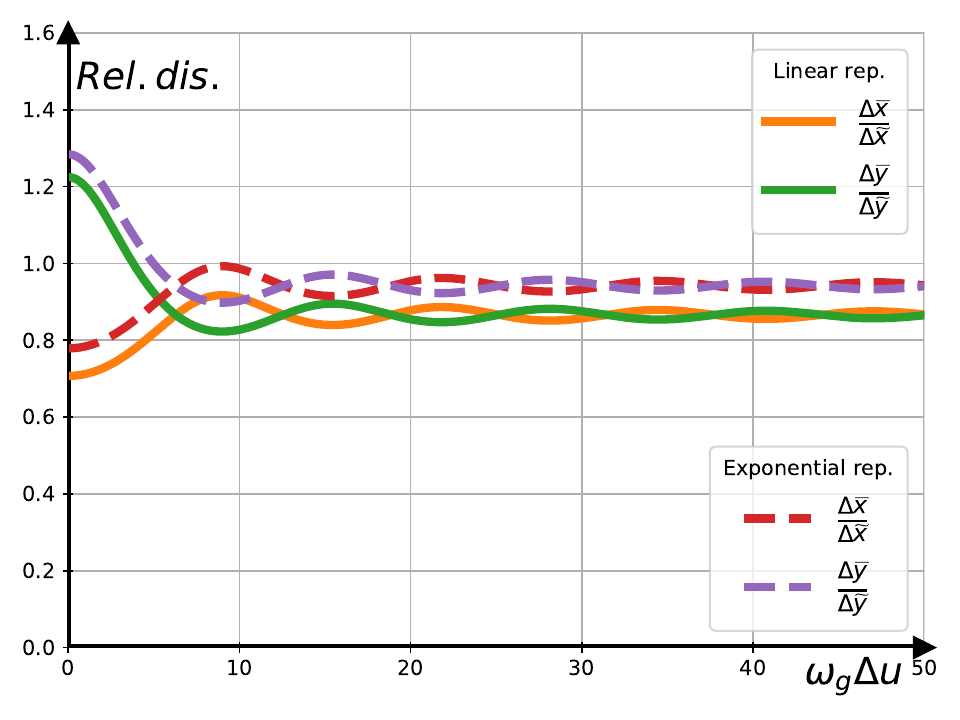, width=3.7in}
}
\vskip -0.4cm
\caption{\small Relative distances $\Delta {\bar x}/\Delta{\tilde x}$ 
and  $\Delta {\bar y}/\Delta{\tilde y}$ as function of $\Delta u$ in the diagonal basis 
$(\Delta\tilde x\;\;\Delta\tilde y)^T$.
}
\label{figure: relative distance}
\end{figure}

The next step is to construct the positive and negative frequency Wightman functions, which are 
defined as, 
\begin{eqnarray}
i\Delta^{(+)}(x;x') \!\!&=&\!\! \left\langle\Omega\right|\hat\phi(x)\hat\phi(x')\left|\Omega\right\rangle = 
 \int\! \frac{{\rm d}^{D-1}k}{(2\pi)^{D-1}}\,
{\rm e}^{i \vec k_\perp\cdot\,\Delta \vec x_\perp 
 -\frac{i}{2}\Omega_-(\vec k)\Delta v}
\phi_+(u,\vec k\,)\phi_-(u',-\vec k\,)
\,,
 \label{positive frequency Wightman function}\\
 i\Delta^{(-)}(x;x') \!\!&=&\!\! \left\langle\Omega\right|\hat\phi(x')\hat\phi(x)\left|\Omega\right\rangle 
                          =
\int\!  \frac{{\rm d}^{D-1}k}{(2\pi)^{D-1}}\,
{\rm e}^{i \vec k_\perp\cdot\,\Delta \vec x_\perp 
 +\frac{i}{2}\Omega_+(\vec k)\Delta v}
\phi_-(u,\vec k\,)\phi_+(u',-\vec k\,)
\nonumber\\
\!\!&& \hskip 3cm
=\, \int\!  \frac{{\rm d}^{D-1}k}{(2\pi)^{D-1}}\,
{\rm e}^{-i \vec k_\perp\cdot \,\Delta \vec x_\perp 
             +\frac{i}{2}\Omega_-(\vec k)\Delta v}
\phi_-(u,-\vec k\,)\phi_+(u',\vec k\,)
\,,\qquad\;\;
\label{negative frequency Wightman function}
\end{eqnarray}
where $\Delta \vec x_\perp = \vec x_\perp-\vec x_\perp^{\,\prime}$, $\Delta v = v\!-\!v'$, 
we made use of $\hat a(\vec k\,)|\Omega\rangle = 0$ and 
$\hat a(\vec k\,)\hat a^\dagger(\vec k')=(2\pi)^{D-1}\delta^{D-1}(\vec k\!-\!\vec k')
+\hat a^\dagger(\vec k')\hat a(\vec k\,)$. 
Note that the Wightman functions are related by transposition,  
$ i\Delta^{(-)}(x';x) = i\Delta^{(+)}(x;x')$, and by complex conjugation, 
$ i\Delta^{(-)}(x;x') = \left[ i\Delta^{(+)}(x;x')\right]^*$.
The Wightman functions 
in~(\ref{positive frequency Wightman function}--\ref{negative frequency Wightman function})
are already a half-baked result, based on which one can construct the propagator.
However, it is always better to perform the integrals 
in~(\ref{positive frequency Wightman function}--\ref{negative frequency Wightman function})
and get the explicit spacetime dependence. 
A principal obstacle to performing the integrals is the modulated 
phase of the mode function~(\ref{mode function: general solution phase 4}),
which breaks global Lorentz symmetry of the vacuum fluctuations, 
according to which the Wightman functions would 
depend on the invariant distance, $\Delta x^2(x;x') = -(t-t')^2 +\|\vec x-\vec x'\|^2$ (up to an infinitesimal 
$i\epsilon$ addition of imaginary time interval needed to regulate the integrals).
In Appendices~A and~B we perform these integrals in two steps.
In Appendix~A we evaluate the relevant integrals by expanding
the mode functions~(\ref{mode function: general solution 4}) 
in~(\ref{positive frequency Wightman function}--\ref{negative frequency Wightman function})
in powers of phases $\Psi(u)$. This representation is an expansion in powers of multipoles, 
and it is useful when one is interested in low powers 
in the gravitational wave amplitude. A better way is to perform the integrals
in~(\ref{positive frequency Wightman function}--\ref{negative frequency Wightman function}),
which is done in Appendix~B. The resulting Wightman functions are given in 
Eqs.~(\ref{propagator integral D: 4 Y}) and~(\ref{propagator integral D: 4 negative Y}),
\begin{equation}
i\Delta^{(\pm)}(x;x') = \frac{\hbar m^{D-2}}
               {(2\pi)^\frac{D}{2}[\gamma(u)\gamma(u')]^\frac14\sqrt{\Upsilon(u;u')}}
   \frac{K_{\frac{D-2}{2}}\big(m\sqrt{\Delta {\bar x}_{(\pm)}^2}\,\big)}
   {\big(m\sqrt{\Delta {\bar x}_{(\pm)}^2}\,\big)^\frac{D-2}{2}}
\,,
\label{Wightman functions: general Lor viol solution}
\end{equation}
where $K_{\nu}(z)$ denotes Bessel function of the second kind, and
$\Delta {\bar x}_{(\pm)}^2(x;x')$
are Lorentz-violating distance functions, which in lightcone coordinates can be recast as,
\begin{eqnarray}
\Delta {\bar x}_{(\pm)}^2(x;x') \!\!&=&\!\!
 -(\Delta u\!\mp\!i\epsilon)(\Delta v\!\mp\!i\epsilon)
  +\left(\Delta x\;\;\Delta y\right)\!\cdot\! {\mathbf\Upsilon}^{-1}
    \!\cdot\!\left(\begin{array}{c}
                         \Delta x\cr 
                        \Delta y\cr
                       \end{array}
                \right)
 + \sum_{i=3}^{D-2}\Delta x_i^2
\,,\qquad
\label{Lorentz breaking distance functions: LC coord}
\end{eqnarray}
and in Cartesian coordinates,
\begin{eqnarray}
\Delta {\bar x}_{(\pm)}^2(x;x')   \!\!&=&\!\! -(\Delta t \mp i\epsilon)^2
  +\left(\Delta x\;\;\Delta y\right)\!\cdot\! {\mathbf\Upsilon}^{-1}
    \!\cdot\!\left(\begin{array}{c}
                         \Delta x\cr 
                        \Delta y\cr
                       \end{array}
                \right)
 + \sum_{i=3}^{D-1}\Delta x_i^2
\,.\qquad
\label{Lorentz breaking distance functions: Cart coord}
\end{eqnarray}
This global Lorentz violation is mediated by the deformation 
matrix, 
\begin{eqnarray}
\mathbf{\cal G}\equiv{\mathbf\Upsilon}^{-1}(u;u') \!\!&=&\!\! \frac{1}{\Upsilon(u;u')}
               \left(\!\!\begin{array}{cc}
                         \Upsilon_y & - \Upsilon_{xy}\cr 
                      - \Upsilon_{xy} &  \Upsilon_x\cr
                       \end{array}
                \!\!\right)
\,,\qquad
\label{inverse rotation matrix: general}
\end{eqnarray}
which, in linear representation and for nonpolarized gravitational waves, is of the form,
\begin{eqnarray}
\mathbf{\cal G}(u;u')={\mathbf\Upsilon}^{-1}(u;u')
\!\!&=&\!\! \frac{1}{(1-h^2)\Upsilon(u;u')}\left(\!\!\begin{array}{cc}
                 1 \!+\! h \frac{\sin(\omega_g u)-\sin(\omega_g u')}{\omega_g\Delta u} &
                        -h\frac{\cos(\omega_g u)-\cos(\omega_g u')}{\omega_g\Delta u} \cr
                         -h\frac{\cos(\omega_g u)-\cos(\omega_g u')}{\omega_g\Delta u} &  
                  1 \!-\! h\frac{\sin(\omega_g u)-\sin(\omega_g u')}{\omega_g\Delta u}\cr
                  \end{array}\!\!\right)
\,,
\label{inverse rotation matrix}\\
\Upsilon(u;u') \!\!&\equiv &\!\!  {\rm det} \left[{\mathbf\Upsilon}\right]
   = \Upsilon_x\Upsilon_y-\Upsilon_{xy}^2=
 \frac{1}{(1-h^2)^2}\bigg[1\!-\!h^2j^2_0\Big(\frac{\omega_g\Delta u}{2}\Big)\bigg]
\,,\quad
\label{determinant upsilon: linear rep nonpol}
\end{eqnarray}
which deforms the distance
 function in position space~(\ref{Lorentz breaking distance functions: LC coord}--\ref{Lorentz breaking distance functions: Cart coord}), and it 
is the inverse of the corresponding momentum space deformation matrix ${\mathbf\Upsilon}$,
which is given in Eq.~(\ref{Appendix B2: phase matrix linear nonpolarized}),
and $j_0(z)=\sin(z)/z$ in~(\ref{determinant upsilon: linear rep nonpol})
denoting the spherical Bessel function.
At the time coincidence,  $\Upsilon(u;u)=1/(1-h^2)=1/{\rm det}[g_{ij}]>1$,
while at large time separations ($|\Delta u|\rightarrow \infty$) $\Upsilon(u;u)$ gets
further enhanced to,
$\Upsilon(u;u)\rightarrow 1/(1-h^2)^2$.
The precise dependence of $\Delta x$ and $\Delta y$ 
on $\Delta u$ (in the diagonal basis $(\Delta\tilde x\;\;\Delta\tilde y)^T$) \
is shown in figure~\ref{figure: relative distance}.
This means that the surface area of the plane in which the gravitational waves propagate
contracts by $1/\Upsilon(u;u')$. The origin of this area contraction can be understood 
by recalling that gravitational waves carry energy, which in turn induces a backreaction onto the spacetime by contracting the surface area of the plane in which the waves propagate.
%

%
%

\begin{figure}[h!]
\vskip -.4cm
\centerline{\hspace{.in}
\epsfig{file=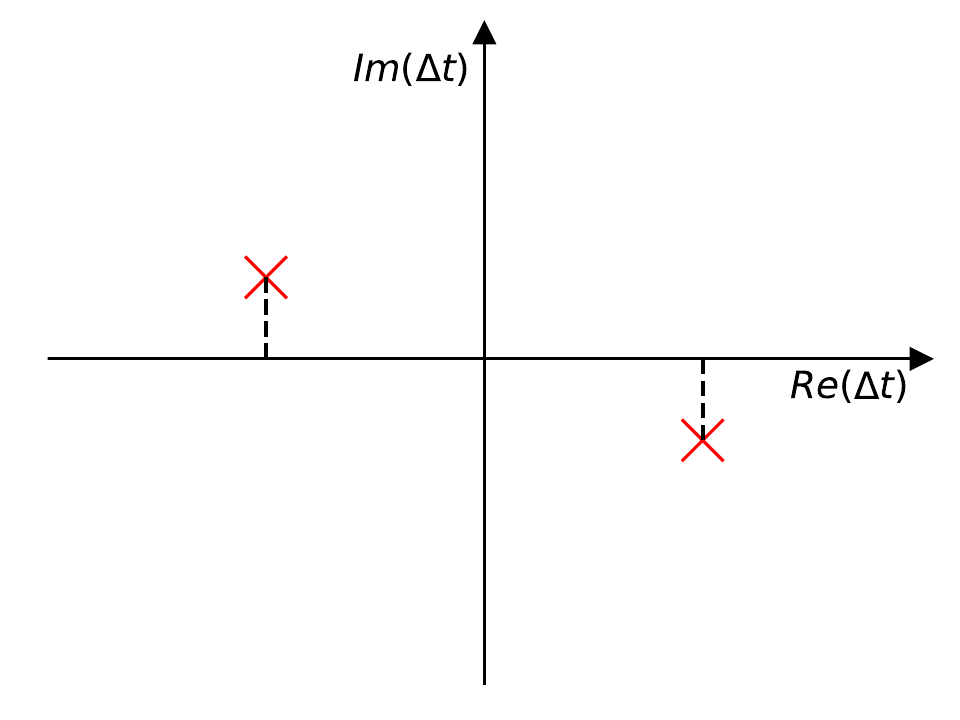, width=2.7in,height=2.2in}
\hskip 0.1cm
\epsfig{file=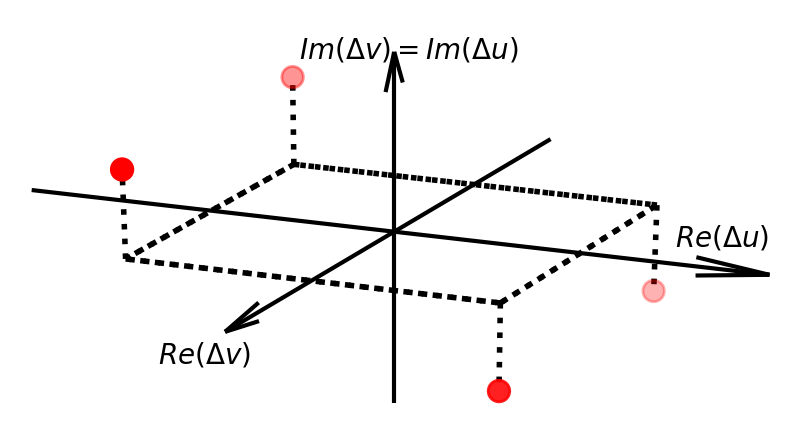, width=3.5in}\hskip -0.5cm
}
\vskip -0.4cm
\caption{\small {\it Left panel:} The standard $i\epsilon$ prescription for 
the Feynman propagator. 
{\it Right panel:} The $i\epsilon$ prescription for the Feynman propagator in lightcone coordinates.
}
\label{figure four}
\end{figure}

 Note the explicit $i\epsilon$ prescriptions 
 in~(\ref{Wightman functions: general Lor viol solution}), 
which give rise to the imaginary parts of 
the Wightman functions, and which can be traced back to the requirement 
that the integrals 
in~(\ref{positive frequency Wightman function 3}--\ref{negative frequency Wightman function 3}) 
converge.
From the Wightman functions~(\ref{Wightman functions: general Lor viol solution})
one can then construct the Feynman propagator in the standard way. In lightcone coordinates
we have, 
\begin{eqnarray}
i\Delta_{F,LC}(x;x') \!\!\!&\equiv&\!\!\! \Theta(\Delta u)i\Delta^{(+)}(x;x')
                                      \!+\! \Theta(-\Delta u)i\Delta^{(-)}(x;x')
\nonumber\\
\!\!&=&\!\! 
\frac{\hbar m^{D-2}}{(2\pi)^\frac{D}{2}[\gamma(u)\gamma(u')]^\frac14\sqrt{\Upsilon(u;u')}}
   \frac{K_{\frac{D-2}{2}}\big(m\sqrt{\Delta {\bar x}_{F,LC}^2}\,\big)}
   {\big(m\sqrt{\Delta {\bar x}_{F,LC}^2}\big)^\frac{D-2}{2}}
,\qquad\;\;
\label{Feynman propagator: lightcone coordinates}
\end{eqnarray}
and in Cartesian coordinates,
\begin{eqnarray}
i\Delta_F(x;x') \!\!&\equiv&\!\! \Theta(\Delta t)i\Delta^{(+)}(x;x')
                                      + \Theta(-\Delta t)i\Delta^{(-)}(x;x')
\nonumber\\
\!\!&=&\!\! 
\frac{\hbar m^{D-2}}{(2\pi)^\frac{D}{2}[\gamma(u)\gamma(u')]^\frac14\sqrt{\Upsilon(u;u')}}
   \frac{K_{\frac{D-2}{2}}\big(m\sqrt{\Delta {\bar x}_{F}^2}\,\big)}
   {\big(m\sqrt{\Delta {\bar x}_{F}^2}\big)^\frac{D-2}{2}}
\,.\qquad
 \label{Feynman propagator: Cartesian coordinates}
 \end{eqnarray}
Both propagators~(\ref{Feynman propagator: lightcone coordinates}--\ref{Feynman propagator: Cartesian coordinates})
are suitable for perturbative studies, the former for the initial
value problem defined on a $u={\rm constant}$ hypersurface,
the latter on a $t={\rm constant}$ hypersurface. However, 
the two $i\epsilon$ prescriptions differ,
\begin{eqnarray}
\Delta {\bar x}_{F,LC}^2(x;x')  
      \!\!&=&\!\! -(|\Delta u| - i\epsilon) (\pm\Delta v-i\epsilon)\!+\!\|\Delta\vec{\bar x}_\perp\|^2
\nonumber\\
\Delta {\bar x}_{F}^2(x;x')  
\!\!&=&\!\! -(|\Delta t| - i\epsilon)^2\!+\!\|\Delta\vec{\bar x}\|^2
\,,
 \label{Feynman propagator: i epsilon prescriptions}
 \end{eqnarray}
and they are illustrated in figure~\ref{figure four}.
That means that the imaginary parts of the propagators are different.
Both prescriptions are legitimate however, as they are designated to study inequivalent 
perturbative evolution problems.

From 
Eqs.~(\ref{Feynman propagator: lightcone coordinates}--\ref{Feynman propagator: Cartesian coordinates}) one easily obtains the corresponding Dyson propagators,
\begin{equation}
i\Delta_{D,LC}(x;x') = \big[ i\Delta_{F,LC}(x;x')\big]^*
\,,\qquad
i\Delta_{D}(x;x') = \big[ i\Delta_F(x;x')\big]^*
\,,\qquad
\label{Dyson propagator}
\end{equation}
which can be used in {\it in-out} problems to study evolution of {\it out} states 
backwards in time. Furthermore, 
the Dyson propagator is an essential ingredient for studying 
perturbative time evolution of Hermitian operators
in weakly interacting quantum field theories. 
 
\bigskip

{\bf Exponential representation.} 
In this representation the gravitational waves are 
written as in~(\ref{metric exponential representation}), or equivalently in 
Eq.~(\ref{metric exponential representation 3}), such that for {\it nonpolarized} gravitational
waves the Klein-Gordon equation~(\ref{EOM scalar in 4}) reduces to, 
\begin{eqnarray}
 &&\hskip -0.3cm
   \bigg\{\!-4\partial_v\partial_u  
    \!+\!\cosh(\tilde h)\left(\partial_x^2\!+\!\partial_y^2\right)
       \!-\!\frac{\sinh(\tilde h)}{\tilde h}\left[\tilde h_+\cos(\omega_gu)(\partial_x^2\!-\!\partial_y^2)
              \!+\!2 \tilde h_\times\sin(\omega_gu)\partial_x\partial_y\right]
 \nonumber\\
        && \hskip 5cm
              \!+\,\sum_{i=3}^{D-2}\partial_i^2 \!-\! m^2\bigg\}
       \hat\phi(u,\vec x_\perp,v) = 0
,\hskip -.3cm
\label{EOM scalar in 4: exp rep}
\end{eqnarray}
where we made use of ${\rm det}[g_{ij}]=1$.  The mode equations are then, 
\begin{eqnarray}
&&\hskip -1cm
\Bigg(\partial_u
     \!\pm\! \frac{i}{2\Omega_\mp}\Big[(\cosh(\tilde h)\!-\!1)(k_x^2\!+\!k_y^2)
             \!-\!\sinh(\tilde h)\Big(\!\cos(\omega_gu)(k_x^2\!-\!k_y^2)
             \!+\! 2\sin(\omega_gu)k_xk_y\Big)\Big]
            \!\pm\! \frac{i}{2}\Omega_\pm  \Bigg)\phi_\pm(u,\vec k) = 0
\,,\hskip .1cm
\nonumber\\
\label{mode equations: exp rep}
\end{eqnarray}
where we made use of, $\tilde h_+=\tilde h_\times =\tilde h$. The mode functions 
can be then written in the form~(\ref{mode function: general solution 4}), with 
\begin{eqnarray}
 \phi_\pm(u,\vec k) \!\!&=&\!\! \sqrt{\!\frac{\hbar}{2\omega}}
 \exp\bigg\{\!\mp \frac{i}{2} \bigg[\Omega_\pm(\vec k\,)u 
              +\frac{\Psi(u)}{\Omega_\mp(\vec k\,)}\bigg]\bigg\}
 \,,\qquad
\label{mode function: exp rep nonpol}\\
 \Psi(u) \!\!&=&\!\! (\cosh(\tilde h)\! -\!1)(k_x^2\!+\!k_y^2)u
  -\sinh(\tilde h)\frac{\sin(\omega_g u)}{\omega_g}(k_x^2\!-\!k_y^2)
       \!+\!2\sinh(\tilde h)\frac{\cos(\omega_g u)}{\omega_g}  (k_xk_y)
\,,\quad
\label{mode function: exp rep nonpol 2}
\end{eqnarray}
From this and Eq.~(\ref{Appendix B2: phase matrix exponential nonpolarized})
we can easily evaluate elements of the deformation matrix ${\mathbf{\Upsilon}}$,
\begin{eqnarray}
\Upsilon_{{}_x\atop {}^y}(u;u')\!\!&=&\!\! \cosh(\tilde h) 
      \mp\sinh(\tilde h)\frac{\sin(\omega_g u)-\sin(\omega_g u')}{\omega_g\Delta u}
\,,\qquad
\nonumber \\ 
\Upsilon_{xy}(u;u')\!\!&=&\!\!\sinh(\tilde h)
 \frac{\cos(\omega_g u)-\cos(\omega_g u')}{\omega_g\Delta u}
\,,
\label{Upsilon matrix exponential nonpolarized}
\end{eqnarray}
and its determinant evaluates to,
\begin{equation}
\Upsilon(u;u') = \Upsilon_x\Upsilon_{y}-\Upsilon_{xy}^2
=\cosh^2(\tilde h) - \sinh^2(\tilde h)\,j_0^2\Big(\frac{\omega_g\Delta u}{2}\Big)
\,.\qquad
\label{diagonalization of k matrix 6}
\end{equation}
At coincidence ($\Delta u = 0$) this evaluates to {\it one}, $\Upsilon(u;u)=1$, as was expected.
As $|\Delta u|$ increases, $\Upsilon(u;u')$ increases, plateauing at 
$\Upsilon\rightarrow \cosh^2(\tilde h)$ when $|\Delta u|\rightarrow \infty$.

\bigskip
  

 \subsection{Polarized gravitational waves}
\label{Polarized gravitational waves}
 
 \medskip
 
{\bf Linear representation.} 
General polarized gravitational waves are characterized 
by Eq.~(\ref{gravitational wave D}) (see also Eq.~(\ref{gravitational wave: planar D=4})),
but without any relation between
the amplitudes $h_+$, $h_\times$ and the corresponding phases $\psi_+$ and 
$\psi_\times$. The scalar field operator obeys (from Eq.~(\ref{EOM scalar in 4})), 
\begin{equation}
   \left\{\!-4\partial_v\partial_u 
    \!+\!\frac{(1\!-\! h_+c_+(u))\partial_x^2\!+\!(1\!+\!h_+c_+(u))\partial_y^2
                \!-\!2 h_\times c_\times(u)\partial_x\partial_y}
                {1\!-\! h_+^2c_+^2(u)\!-\!h_\times^2c_\times^2(u)}
                \!+\!\sum_{i=3}^{D-2}\partial_i^2 
  \!-\! m^2\right\}\!\big[\big(\gamma(u)\big)^{\frac14}
                                       \hat \phi(u,\vec x_\perp,v)\big]
         = 0
\,,\qquad
\label{field operator equation polarized}
\end{equation}
where $\gamma(u)\equiv 1-  h_+^2c_+^2(u)-h_\times^2c_\times^2(u)$
is the determinant of $h_{ij}(x)$
and we have introduced a shorthand notation,
\begin{equation}
  c_+(u) = \cos(\omega_g u + \psi_+)
 \,,\qquad 
  c_\times(u) = \cos(\omega_g u + \psi_\times)
\,.
 \label{notation: h+ u and h cross u}
 \end{equation}
One of the phases, for example $\psi_+$, can be absorbed by shifting $u$, leading to 
an equivalent representation,
\begin{equation}
 c_+(u) = \cos(\omega_g u- \psi/2)
 \,,\qquad 
  c_\times(u) = \cos(\omega_g u + \psi/2)
\,,\qquad (\psi\equiv \psi_\times-\psi_+)
\,.
 \label{notation: h+ u and h cross u 2}
 \end{equation}
Decomposing the field and its canonical momentum as
 in~(\ref{field expansion in 4}--\ref{canonical momentum expansion in 4})
yields the equation of motion for the mode functions 
({\it cf.} Eq.~(\ref{EOM scalar in 4 mode functions})), 
\begin{equation}
   \left(\!\partial_u
     \pm \frac{i}{2\Omega_\mp}
          \left[\frac{(1\!-\!h_+c_+(u))k_x^2\!+\!(1\!+\!h_+c_+(u))k_y^2
                           \!-\!2 h_\times c_\times(u) k_xk_y}
                           {1\!-\!h_+^2c_+^2(u)\!-\!h_\times^2c_\times^2(u)}
         \!-\!(k_x^2 \!+\! k_y^2)\right]\!
      \pm \frac{i}{2}\Omega_\pm\right)\!
      \big[\gamma(u)^{\frac14}
                                       \phi_\pm(u,\vec k)\big] = 0
\,.\quad
\label{EOM scalar in 4 mode functions: polarized}
\end{equation}
The solution can be formally written as 
({\it cf.} Eqs.~(\ref{mode function: general solution 4}--\ref{mode function: general solution phase 4})), 
\begin{eqnarray}
 \phi_\pm(u,\vec k) \!\!&=&\!\! \sqrt{\frac{\hbar}{2\omega}}
                            \frac{1}{[\gamma(u)]^{1/4}}
 \exp\Big[\mp \frac{i}{2} \Big(\Omega_\pm(\vec k\,)u 
              +\frac{\Psi(u)}{\Omega_\mp(\vec k\,)}\Big)\Big]
 \,,\qquad
\label{mode function: general solution 4 B} 
 \\
 \Psi(u) \!\!&=&\!\! \int^u{\rm  d}\tilde u
 \bigg[\frac{[1\!-\!h_+\cos(\omega_g \tilde u\!-\!\psi/2)]k_x^2
              \!+\![1\!+\!h_+\cos(\omega_g \tilde u\!-\!\psi/2)]k_y^2
                         \!-\!2h_\times\cos(\omega_g \tilde u\!+\!\psi/2)k_xk_y}
{1\!-\!h_+^2\cos^2(\omega_g\tilde u\!-\!\psi/2)-h_\times^2\cos^2(\omega_g\tilde u\!+\!\psi/2)}
\nonumber\\
              \!\!&&\!\!  \hskip 1.5cm \!-\,(k_x^2\!+\!k_y^2)\bigg]
\,,\qquad\;
\label{mode function: general solution phase 4 B}
\end{eqnarray}
such that, as in Eq.~(\ref{mode function: general solution phase 4}), $\Psi(u)\rightarrow 0$ 
as $h_+\rightarrow 0$ and $h_\times\rightarrow 0$.

\medskip

\begin{figure}[h!]
\vskip -.4cm
\centerline{\hspace{.in}
\epsfig{file=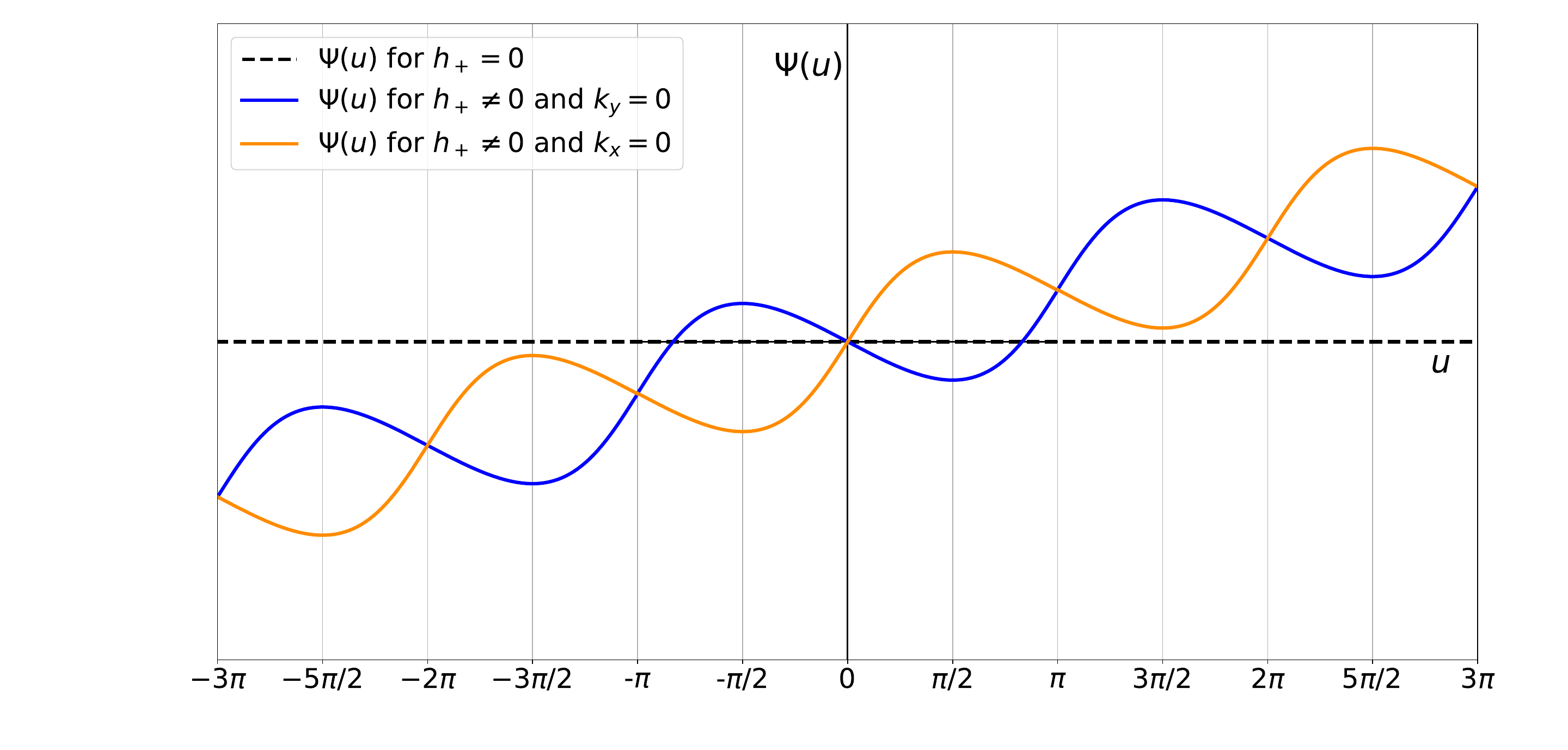, width=5.5in}
}
\vskip -0.4cm
\caption{\small   The analytically extended phase function 
$\Psi(u)$ defined in Eq.~(\ref{mode function:  phase 4 + pol 2B}). The choice 
of the parameters is $k^2_x/\omega_g^2=k_y^2/\omega_g^2 = 1$, and $h_+=1/2$ is chosen
(unrealistically) large in order to make its effect more visible.
}
\label{figure three}
\end{figure}

Let us first consider the simpler case of maximally polarized gravitational waves.
For the $(+)$ polarization ({\it i.e.} $h_\times=0$, $\psi=0$) 
the integral in~(\ref{mode function: general solution phase 4 B}) simplifies
to a standard integral performed in Eq.~(\ref{integral1}) of Appendix~C,
\begin{eqnarray}
 \Psi(u) \!\!&=&\!\! \int^u d\tilde u
 \bigg[\frac{k_x^2}
                 {1+h_+\cos(\omega_g\tilde u)}
                 +\frac{k_y^2}
                 {1-h_+\cos(\omega_g\tilde u)}\bigg]\!-\!(k_x^2\!+\!k_y^2)u
\nonumber\\
\!\!&&\!\!\hskip -1.6cm
=\frac{2}{\omega_g}\frac{k_x^2}{\sqrt{1\!-\!h_+^2}}
\text{Arctan}\bigg[\sqrt{\frac{1\!-\!h_+}{1\!+\!h_+} }\!\tan\Big( \frac{\omega_g u}{2} \Big)
                   \bigg]
 \!+\! \frac{2}{\omega_g}\frac{k_y^2}{\sqrt{1\!-\!h_+^2}}
\text{Arctan}\bigg[\sqrt{\frac{1\!+\!h_+}{1\!-\!h_+} }\!\tan\Big( \frac{\omega_g u}{2}\Big)\bigg]
\!-\!(k_x^2\!+\!k_y^2)u
.\qquad\;
\label{mode function:  phase 4 + pol}
\end{eqnarray}
This result is defined only in the interval $-\pi <\omega_g u<\pi$, but there is a unique
analytic extension of it to the real axis, $-\infty < u < \infty$ and to the complex $u-$plane.
Note first that in the limit when $h_+\rightarrow 0$, 
$\Psi(u)=0$ in the entire interval $-\pi <\omega_g u<\pi$, implying that its analytic extension 
to the complex $u-$plane is simply, $\Psi(u)=0$. When $h_+\neq 0$, note first 
that at the boundaries of the interval (when $\omega_g u\rightarrow \pm \pi$),
\begin{equation}
\Psi\Big(u\rightarrow \pm\frac{\pi}{\omega_g}\Big)  
= \pm \frac{\pi}{\omega_g}(k_x^2 + k_y^2)\left[\frac{1}{\sqrt{1-h_+^2}}-1\right]
 \,,\quad
\label{EOM scalar in 4 mode functions: polarized}
\end{equation}
such that $\Psi(u)$ increments by $\Psi(\pi/\omega_g)-\Psi(-\pi/\omega_g)$ 
accross the interval where it is defined. This can be utilized by extending the definition 
of ${\rm Arctan}(z)$ in Eq.~(\ref{mode function:  phase 4 + pol}) to ${\rm arctan}(z)$ defined
on the whole complex plane by,
\begin{eqnarray}
{\rm arctan}\bigg[\sqrt{\frac{1\pm h_+}{1\mp h_+}}\tan\Big(\frac{\omega_gu}{2}\Big)\bigg] 
\equiv \begin{cases}
&\!\!\!\!
{\rm Arctan}\bigg[\sqrt{\frac{1\pm h_+}{1\mp h_+}}
                     \tan\Big(\frac{\omega_gu-2n\pi}{2}\Big)
                                      \bigg]\!+\!\pi n ,
                           \quad \frac{(-1+2n)\pi}{\omega_g}<u<\frac{(1+2n)\pi}{\omega_g} \cr
 &\!\!\!\!
 \pi \left(n+\frac12\right)\,,\quad {\rm when} \quad u = \frac{(1+2n)\pi}{\omega_g }
 \,,\quad (n\in\mathbb{Z})                            \cr
 \end{cases}
 ,\quad\;
\label{EOM scalar in 4 mode functions:+  extended}
\end{eqnarray}
%
%
%
where $\mathbb{Z}$ denotes the set of integers. The analytically extended 
solution of~(\ref{mode function:  phase 4 + pol}) can then be written as, 
\begin{eqnarray}
 \Psi(u) \!\!&=&\!\!
\frac{2}{\omega_g}\frac{k_x^2}{\sqrt{1\!-\!h_+^2}}
\text{arctan}\bigg[\sqrt{\frac{1\!-\!h_+}{1\!+\!h_+} }\tan\Big( \frac{\omega_g u}{2} \Big)
                   \bigg]
 \!+\! \frac{2}{\omega_g}\frac{k_y^2}{\sqrt{1\!-\!h_+^2}}
\text{arctan}\bigg[\sqrt{\frac{1\!+\!h_+}{1\!-\!h_+} }\tan\Big( \frac{\omega_g u}{2}\Big)\bigg]
\quad
\nonumber\\
\!\!&-&\!\!(k_x^2\!+\!k_y^2)u
.
\label{mode function:  phase 4 + pol 2B}
\end{eqnarray}
and its form is illustrated in figure~\ref{figure three}.

 \bigskip
 
 The second case of interest is the $(\times)$-polarized wave (for which $h_+=0$).
 In this case, the phase modulation integral~(\ref{mode function: general solution phase 4 B})
 simplifies to,
\begin{eqnarray}
 \Psi(u) \!\!&=&\!\! \int^u d\tilde u
 \bigg[\frac{k_x^2\!+\!k_y^2}
                 {1-h_\times^2\cos^2(\omega_g\tilde u)}
                 \!-\!\frac{2h_\times\cos(\omega_g \tilde u)k_xk_y}
                 {1-h_\times^2\cos^2(\omega_g\tilde u)}
               \bigg]  \!-\!(k_x^2\!+\!k_y^2)u
\,,\quad
\nonumber\\
\!\!&=&\!\! \frac{k_x^2\!+\!k_y^2}{\omega_g\sqrt{1\!-\!h_\times^2}}\text{arctan}\bigg[\frac{1}{\sqrt{1\!-\!h_\times^2} }
                   \tan(\omega_g u)
                   \bigg]
\!-\!\frac{2k_xk_y}{\omega_g\sqrt{1\!-\!h_\times^2}}
                   \text{arctan}\bigg[\frac{h_\times}{\sqrt{1\!-\!h_\times^2} }
                   \sin(\omega_g u)
                   \bigg]\! \!-\!(k_x^2\!+\!k_y^2)u
, \qquad\;\;
\label{mode function: general solution phase 4 x}
\end{eqnarray}
where we set the (physically irrelevant) phase $\psi=0$, and
made use of the integrals~(\ref{integral2}--\ref{integral3}) in Appendix~C, whereby
we analytically extended ${\rm Arctan}(z)$ as in 
Eq.~(\ref{EOM scalar in 4 mode functions:+  extended}).

\bigskip

The integral in Eq. (\ref{mode function: general solution phase 4 B}) for the general case can be rewritten using the substitution $t= \tan(\omega_g u - \frac{\psi}{2})$ to obtain,
\begin{eqnarray}
	\Psi(u) =-(k_x^2 + k_y^2)u + \frac{1}{\omega_g} \int \frac{\text{d}t}{\sqrt{1+t^2}} \frac{(k_x^2 + k_y^2)\sqrt{1+t^2} + h_+(k_y^2 - k_x^2)-2h_\times k_xk_y\big[c_{\psi} - s_{\psi}t\big]}{1-h_+^2 - h_\times^2 c_{\psi}^2 + 2h_\times^2 c_{\psi} s_{\psi} t + (1-h_\times^2 s_{\psi}^2)t^2}
\,,\quad
\label{x}
\end{eqnarray}
where $c_{\psi}$ and $s_{\psi}$ are abbreviations of $\cos(\psi)$ and $\sin(\psi)$, respectively. Using integral~(\ref{integral6}) given in Appendix~C, 
$\Psi(u)$ evaluates to,
\begin{eqnarray}
\Psi(u) \!\!&=&\!\! 
 \frac{1}{\omega_g \sqrt{1-(h_+^2 + h_\times^2)+ h_+^2 h_\times^2s_{\psi}^2} }\Bigg\{\! \big(k_x^2 + k_y^2\big) \text{Arctan}\Bigg( \frac{h_\times^2 s_{\psi} c_{\psi} 
  + (1-h_\times^2 s_{\psi}^2)\tan(\omega_g u -\frac{\psi}{2}) }{\sqrt{1-(h_+^2 + h_\times^2)+ h_+^2 h_\times^2s_{\psi}^2}}\Bigg) 
\nonumber \\ 
&& \hskip -1.4cm
+\, \frac{i\big[\!\!-\!2h_\times k_x k_y  c_{\psi} 
\!+\! 2ih_\times k_x k_y s_{\psi}\sqrt{1\!-\!(h_+^2 \!+\! h_\times^2)
\!+\! h_+^2 h_\times^2 s_{\psi}^2} 
\!-\!(1\!-\!h_\times^2 s_{\psi}^2)
\big(k_x^2 \!-\! k_y^2\big)h_+ \big]
       \text{Arctanh}[g(u)]}
{2\sqrt{h_+^2 \!+\! h_\times^2 \!-\! h_\times^2 s_{\psi}^2 
    (2\!+\!h_+^2\!+\! h_\times^2)\!-\! 2ih_\times^2 c_{\psi} s_{\psi}
     \sqrt{1\!-\!(h_+^2 \!+\! h_\times^2)\!+\! h_+^2 h_\times^2 s_{\psi}^2}}}
\!+{\rm c.c.}\! \Bigg\} \;
\nonumber \\ 
&& \hskip -1.3cm
-\,\big(k_x^2 + k_y^2\big)u
\,,\quad
\label{general solution}
\end{eqnarray}
where ${\rm c.c.}$ denotes the complex conjugate of the term in the second line
and the function $g(u)$ is defined by, 
\begin{equation}
	g(u)\equiv\frac{(1\!-\!h_\times s_{\psi}^2)
	\cos(\omega_g \!-\! \psi/2) \!+\!\left[i\sqrt{1\!-\!(h_+^2 \!+\! h_\times^2)
	\!+\! h_+^2 h_\times^2 s_{\psi}^2}\!-\!h_\times^2 c_{\psi} s_{\psi}\right]
	\sin(\omega_g u \!-\! \psi/2)}
	{\sqrt{h_+^2 \!+\! h_\times^2 \!-\! h_\times^2s_{\psi}^2 
	(2\!+\!h_+^2\!+\! h_\times^2)
	\!-\!2ih_\times^2 c_{\psi} s_{\psi}
	\sqrt{1\!-\!(h_+^2 \!+\! h_\times^2)
	\!+\! h_+^2 h_\times^2s_{\psi}^2}}}
.\;
\label{general solution arctanh}
\end{equation} 
The corresponding ${\mathbf \Upsilon}$ matrix can be formally written from 
Eq.~(\ref{mode function: general solution phase 4 B}) as, 
\begin{eqnarray}
{\Upsilon}_{{}_x\atop{}^y}(u;u') \!\!&=&\!\! \frac{1}{\Delta u}\left[
\int^u{\rm  d}\tilde u
\frac{1\!\mp\!h_+\cos(\omega_g \tilde u\!-\!\psi/2)}
{1\!-\!h_+^2\cos^2(\omega_g\tilde u\!-\!\psi/2)-h_\times^2\cos^2(\omega_g\tilde u\!+\!\psi/2)}
          - (u\rightarrow u')\right]
\,, \;
\nonumber\\
{\Upsilon}_{xy}(u;u') \!\!&=&\!\! \frac{1}{\Delta u}\left[\int^u{\rm  d}\tilde u
\frac{\!-\!h_\times\cos(\omega_g \tilde u\!+\!\psi/2)}
{1\!-\!h_+^2\cos^2(\omega_g\tilde u\!-\!\psi/2)-h_\times^2\cos^2(\omega_g\tilde u\!+\!\psi/2)}
         - (u\rightarrow u')\right]
\,.\qquad\;
\label{general solution: Upsilon matrix lin rep: formal}
\end{eqnarray}
This can be integrated and the result follows from Eq.~(\ref{general solution}), 
\begin{eqnarray}
{\Upsilon}_{{}_x\atop{}^y} \!\!&=&\!\! 
 \frac{1}{\omega_g\Delta u \sqrt{1-(h_+^2 + h_\times^2)+ h_+^2 h_\times^2s_{\psi}^2} }
 \left\{\! \text{arctan}\Bigg( \frac{h_\times^2 s_{\psi} c_{\psi} 
  + (1-h_\times^2 s_{\psi}^2)\tan(\omega_g u -\frac{\psi}{2}) }{\sqrt{1-(h_+^2 + h_\times^2)+ h_+^2 h_\times^2s_{\psi}^2}}\Bigg) 
  \right.
\nonumber \\ 
&& \hskip -.2cm
\left.
\mp \, i\frac{(1\!-\!h_\times^2 s_{\psi}^2)h_+
       \text{arctanh}[g(u)]}
{2\sqrt{h_+^2 \!+\! h_\times^2 \!-\! h_\times^2 s_{\psi}^2 
    (2\!+\!h_+^2\!+\! h_\times^2)\!-\! 2ih_\times^2 c_{\psi} s_{\psi}
     \sqrt{1\!-\!(h_+^2 \!+\! h_\times^2)\!+\! h_+^2 h_\times^2 s_{\psi}^2}}}
\!+{\rm c.c.} - (u\rightarrow u')\! \right\}
, \;
\nonumber\\
{\Upsilon}_{xy} \!\!&=&\!\!
 \frac{1}{\omega_g\Delta u \sqrt{1-(h_+^2 + h_\times^2)+ h_+^2 h_\times^2s_{\psi}^2} }
 \nonumber\\
\!\!&\times&\!\!\!\!
 \left\{  -\frac{\big[ic_{\psi} 
   \!+\!   s_{\psi}\sqrt{1\!-\!(h_+^2 \!+\! h_\times^2)
\!+\! h_+^2 h_\times^2 s_{\psi}^2} \,
\big]h_\times
       \text{Arctanh}[g(u)]}
{\sqrt{h_+^2 \!+\! h_\times^2 \!-\! h_\times^2 s_{\psi}^2 
    (2\!+\!h_+^2\!+\! h_\times^2)\!-\! 2ih_\times^2 c_{\psi} s_{\psi}
     \sqrt{1\!-\!(h_+^2 \!+\! h_\times^2)\!+\! h_+^2 h_\times^2 s_{\psi}^2}}}
\!+{\rm c.c.}- (u\rightarrow u')\! \right\}
, \qquad\;
\label{general solution: Upsilon matrix lin rep}
\end{eqnarray}
and the special cases of singly-polarized gravitational waves can be obtained by taking 
the suitable limits of this general result. Alternatively, one can
extract the corresponding matrices  ${\mathbf \Upsilon}$ from 
Eq.~(\ref{mode function:  phase 4 + pol 2B})
\begin{equation}
{\mathbf \Upsilon}_+ = \frac{2}{\omega_g\Delta u}\frac{1}{\sqrt{1\!-\!h_+^2}}
\left\{\left(\!\!\begin{array}{cc}          
\text{arctan}\bigg[\sqrt{\frac{1\!-\!h_+}{1\!+\!h_+} }\tan\Big( \frac{\omega_g u}{2} \Big)
                   \bigg] \;\;& 0\cr
                      0  
\;\;& \text{arctan}\bigg[\sqrt{\frac{1\!+\!h_+}{1\!-\!h_+} }
                        \tan\Big( \frac{\omega_g u}{2}\Big)\bigg] \cr
                                      \end{array}\!\!\right)
\!-\! (u\rightarrow u')\right\}
,\;
\label{Upsilon matrix: plus lin}
\end{equation}
and Eq.~(\ref{mode function: general solution phase 4 x}),
\begin{equation}
{\mathbf \Upsilon}_\times = \frac{1}{\omega_g\Delta u}\frac{1}{\sqrt{1\!-\!h_\times^2}}
\left\{\left(\!\!\begin{array}{cc}          
\text{arctan}\bigg[\frac{1}{\sqrt{1\!-\!h_\times^2} }\tan(\omega_g u)\bigg]
                \;&   - \text{arctan}\bigg[\frac{h_\times}{\sqrt{1\!-\!h_\times^2} } \sin(\omega_g u)
                   \bigg] \cr
                    -\text{arctan}\bigg[\frac{h_\times}{\sqrt{1\!-\!h_\times^2} }\sin(\omega_g u)
                   \bigg]  
\;& \text{arctan}\bigg[\frac{1}{\sqrt{1\!-\!h_\times^2} }\tan(\omega_g u)\bigg] \cr
                                      \end{array}\!\!\right)
                                      - (u\rightarrow u')\right\}
,\;
\label{Upsilon matrix: plus lin}
\end{equation}
respectively.

 \bigskip
 
 {\bf Exponential representation.} Since in this purely tensorial 
 representation~(\ref{metric exponential representation}), 
 (\ref{metric exponential representation 3}),
 ${\rm det}[g_{ij}]=1$ is generally true, 
 the field operator $\hat \phi$ 
 obeys Eq.~(\ref{EOM scalar in 4: exp rep}), 
 but with $\tilde h=\tilde h(u)$ a function of $u$ given in Eq.~(\ref{tilde h u}),
 and the waves oscillate with general phases $\psi_\pm$ as in 
 Eq.~(\ref{notation: h+ u and h cross u 2}).
%
%

The mode function equations~(\ref{mode equations: exp rep}) generalizes to, 
\begin{eqnarray}
&&\hskip -1cm
\Bigg(\!\partial_u
     \!\pm\! \frac{i}{2\Omega_\mp}\bigg[
           \big(\!\cosh(\tilde h(u))\!-\!1\big)(k_x^2\!+\!k_y^2)
\label{mode equations: exp rep general}\\
&&\hskip 1.2cm
             \!-\,\frac{\sinh(\tilde h(u))}{\tilde h(u)}\Big(\!\tilde h_+ c_+(u)(k_x^2\!-\!k_y^2)
             \!+\! 2\tilde h_\times c_\times(u)k_xk_y\Big)\bigg]
            \!\!\pm\! \frac{i}{2}\Omega_\pm  \Bigg)\phi_\pm(u,\vec k) \!=\! 0
\,,\hskip 1.cm
\nonumber
\end{eqnarray}
which implies the following formal solution for $\Psi(u)$, 
\begin{eqnarray}
\Psi(u)
 \!\!&=&\!\!     
 \int^u {\rm d}\bar u\bigg[\big(\!\cosh(\tilde h(\bar u))\!-\!1\big)(k_x^2\!+\!k_y^2)
             \!-\!\frac{\sinh(\tilde h(\bar u))}{\tilde h(\bar u)}
              \Big(\!\tilde h_+ c_+(\bar u)(k_x^2\!-\!k_y^2)
             \!+\! 2\tilde h_\times c_\times(\bar u)k_xk_y\Big)\bigg]
\,,\hskip .5cm
\label{phase function: exp rep general}
\end{eqnarray}
from which one can extract elements of the deformation matrix
({\it cf.} Eqs.~(\ref{mode function: exp rep nonpol 2}--\ref{Upsilon matrix exponential nonpolarized})),
\begin{eqnarray}
\Upsilon_{{}_x\atop {}^y}(u;u')\!\!&=&\!\!
\frac{1}{\Delta u}
\bigg\{\left[ \int^u {\rm d}\bar u \cosh(\tilde h(\bar u)) -\int^{u'} {\rm d}\bar u \cosh(\tilde h(\bar u)) 
\right]
\nonumber\\
    \!\!&\mp&\!\!\tilde h_+ 
     \left[\int^u {\rm d}\bar u \frac{\sinh(\tilde h(\bar u))}{\tilde h(\bar u)}
      \cos\Big(\omega_g \bar u \!-\!\frac{\psi}{2}\Big)
          -  \int^{u'} {\rm d}\bar u \frac{\sinh(\tilde h(\bar u))}{\tilde h(\bar u)}
      \cos\Big(\omega_g \bar u \!-\!\frac{\psi}{2}\Big)\right]\bigg\}
\,,\qquad
\nonumber \\ 
\Upsilon_{xy}(u;u')\!\!&=&\!\!- \frac{\tilde h_\times}{\Delta u}
\left[\int^u {\rm d}\bar u \frac{\sinh(\tilde h(\bar u))}{\tilde h(\bar u)}
      \cos\Big(\omega_g \bar u \!+\!\frac{\psi}{2}\Big)
          -  \int^{u'} {\rm d}\bar u \frac{\sinh(\tilde h(\bar u))}{\tilde h(\bar u)}
      \cos\Big(\omega_g \bar u \!+\!\frac{\psi}{2}\Big)\right]
\,,\qquad\;
\label{Upsilon matrix: exp rep general}
\end{eqnarray}
in terms of which the determinant can be written as
in  Eq.~(\ref{diagonalization of k matrix 6}), 
$\Upsilon(u;u') =\Upsilon_{x}\Upsilon_{y}-\Upsilon_{xy}^2$.
The integrals~(\ref{Upsilon matrix: exp rep general}) cannot be evaluated 
for general gravitational waves
(see footnote~\ref{footnote 30}). However, one can expand the 
$\cosh(\tilde h(u))$ and $\sinh(\tilde h(u))$
in~(\ref{Upsilon matrix: exp rep general})
in powers of $\tilde h(u)$ and perform the integrals order by order. Keeping 
the terms up to the second order  in the gravitational wave amplitude, one obtains, 
\begin{eqnarray}
\Upsilon_{{}_x\atop {}^y}(u;u')\!\!&=&\!\!
\left(\!1+\frac{\tilde h_+^2+\tilde h_\times^2}{2}\right)
+\frac{\tilde h_+^2[\sin(2\omega_g u\!-\!\psi)\!-\!\sin(2\omega_g u'\!-\!\psi)]
               \!+\!\tilde h_\times^2[\sin(2\omega_g u\!+\!\psi)\!-\!\sin(2\omega_g u'\!+\!\psi)]}
         {4\omega_g \Delta u}
\,,\;\;
\nonumber\\
    \!\!&\pm&\!\! 
   \,\frac{\tilde h_+}{\omega_g\Delta u}
     \left[\sin\Big(\omega_g u \!-\!\frac{\psi}{2}\Big)
          -  \sin\Big(\omega_g u' \!-\!\frac{\psi}{2}\Big)\right]
+ {\cal O}(\tilde h_+^3,\tilde h_+\tilde h_\times^2)
\nonumber \\
\Upsilon_{xy}(u;u')\!\!&=&\!\! \frac{\tilde h_\times}{\omega_g\Delta u}
\left[\sin\Big(\omega_g u \!+\!\frac{\psi}{2}\Big)
          -  \sin\Big(\omega_g u' \!+\!\frac{\psi}{2}\Big)\right]
          + {\cal O}(\tilde h_\times^3,\tilde h_\times \tilde h_+^2)
\,,\qquad
\label{Upsilon matrix: exp rep general 2}
\end{eqnarray}
where we made use of, 
$\tilde h(u)^2 =\tilde h_+^2\cos^2(\omega_g u-\psi/2)
        +\tilde h_\times^2\cos^2(\omega_g u+\psi/2)$. 
From Eq.~(\ref{Upsilon matrix: exp rep general 2})
one sees that the general structure of ${\mathbf \Upsilon}$ is,
\begin{equation}
{\mathbf \Upsilon}(u;u') ={\mathbf A}_0+\sum_{n=1}^\infty\left[
            {\mathbf A}_n \frac{\cos(n\omega_g u)-\cos(n\omega_g u')}{\omega_g\Delta u}
           +{\mathbf B}_n  \frac{\sin(n\omega_g u) -\sin(n\omega_g u')}{\omega_g\Delta u} 
               \right]
\,,\qquad\;
\label{Upsilon matrix: exp rep general 3}
\end{equation}
where the matrix-valued coefficients are 
(up to the quadratic order in $ \tilde h_+$ and $\tilde h_\times$),
\begin{eqnarray}
{\mathbf A}_0  \!\!&=&\!\! \left(\!\!\begin{array}{cc}
                                1   &  0   \cr
                                 0   &   1\cr
                               \end{array}\!\!\right)
                                      \left(\! 1+\frac{\tilde h_+^2\!+\!\tilde h_\times^2}{2}\right)
\,,\qquad
{\mathbf A}_1 
=\left(\!\!\begin{array}{cc}
                                -\tilde h_+   & 
                                 \tilde h_\times   \cr
                                \tilde h_\times & 
                                   \tilde h_+ \cr
                               \end{array}\!\!\right)\sin(\psi/2) 
\,,\qquad
{\mathbf A}_2
=\left(\!\!\begin{array}{cc}
                                 1  &  0   \cr
                                 0   &   1\cr
                               \end{array}\!\!\right) 
                               \frac{-\tilde h_+^2\!+\!\tilde h_\times}{4}\sin(\psi) 
\quad
\nonumber\\
{\mathbf B}_1  \!\!&=&\!\! \left(\!\!\begin{array}{cc}
                                \tilde h_+   & 
                                 \tilde h_\times   \cr
                                \tilde h_\times & 
                                   -\tilde h_+ \cr
                               \end{array}\!\!\right)\cos(\psi/2)
  \,,\qquad
{\mathbf B}_2 
= \left(\!\!\begin{array}{cc}
                                 1  &  0   \cr
                                 0   &   1\cr
                               \end{array}\!\!\right) 
                               \frac{\tilde h_+^2+\tilde h_\times^2}{4}\cos(\psi) 
\,,\qquad
\label{Upsilon matrix: exp rep general 4}
\end{eqnarray}
which are all time independent. The coincident limit of~(\ref{Upsilon matrix: exp rep general 3})
is simply, 
\begin{equation}
{\mathbf \Upsilon}(u;u) ={\mathbf A}_0+\sum_{n=1}^\infty\big[n
             \big(\!-\!{\mathbf A}_n \sin(n\omega_g u)
           +{\mathbf B}_n  \cos(n\omega_g u)\big)
               \big]
\,,\qquad\;
\label{Upsilon matrix: exp rep general 5}
\end{equation}
from which one can also calculate the determinant in the usual way, 
$\Upsilon(u;u)=\Upsilon_x(u;u)\Upsilon_y(u;u)-\Upsilon_{xy}(u;u)^2 = 1$
({\it cf.} Eq.~(\ref{elements of Upsilon matrix: ij 4})).

The Wightman functions and the propagator for polarized gravitational waves
are then obtained by inserting the deformation matrix ${\mathbf \Upsilon}$ from this subsection into 
Eqs.~(\ref{Wightman functions: general Lor viol solution}) 
and~(\ref{Feynman propagator: lightcone coordinates}--\ref{Feynman propagator: Cartesian coordinates}), respectively.

In this section we considered planar gravitational waves generated by a single  binary system
(or several binary systems) orbiting in the $xy-$plane. 
While this situation is general in $D=4$, in $D$ dimensions more 
baroque situations are possible. Gravitational waves can be generated by multiple 
binary systems, and their gravitational waves can be decomposed into $(D-2)(D-3)/2$ orthogonal planes. 
Such systems can be analyzed, and general expressions can be obtained for the phase functions 
$\Psi(u)$ by expanding in powers of the gravitational wave amplitudes akin to 
Eq.~(\ref{Upsilon matrix: exp rep general 3}). Details of such an analysis are given in Appendix~D.


\section{One-loop effective action and scalar mass}
\label{One-loop effective action and scalar mass} 

In this section we perform two simple one-loop calculations: the one-loop effective action and 
the one-loop scalar mass. 
As we will now show, planar gravitational waves do not contribute to these two objects.

\subsection{One-loop effective action}
\label{One-loop effective action} 
 
\begin{figure}[t!]
\vskip -1.5cm
\centerline{\hspace{.in}
\epsfig{file=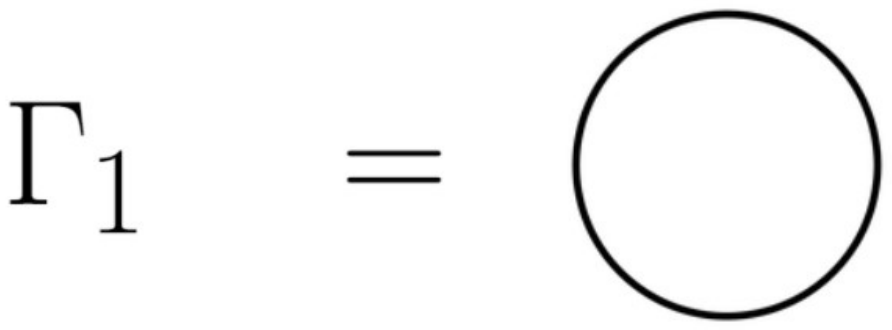, width=3.3in}
\hskip 1cm
\epsfig{file=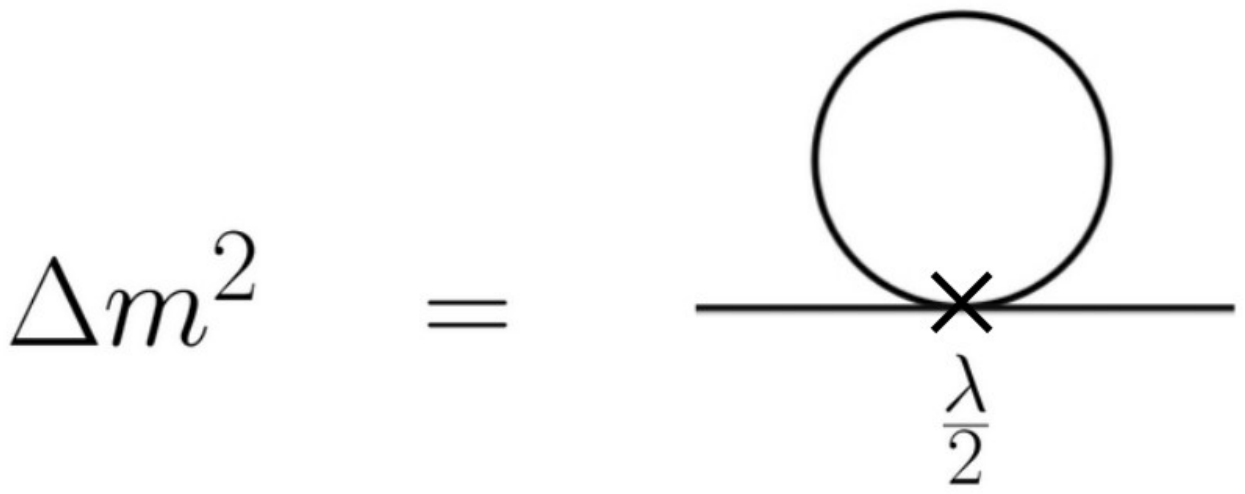, width=3.3in}
}
\vskip -2cm
\caption{\small {\it Left panel:} The Feynman diagram for the one-loop effective action.
{\it Right panel:} The Feynman diagram for the one-loop scalar self-mass.
}
\label{figure five}
\end{figure}

 It is of interest to investigate whether gravitational waves can have an effect on the one-loop
 effective action, whose Feynman diagram is shown in the left panel of figure~\ref{figure five}.
The one-loop effective action for a real scalar field is of the form,
\begin{equation}
\Gamma_1 = \frac{i}{2}{\rm Tr}\log\Bigg[\frac{\sqrt{-g}\big(\Box -m^2\big)}{\mu^2}\Bigg]
 = -\frac{1}{2}\int {\rm d}^Dx\sqrt{-g}\int ^{m^2}{\rm d}\bar m^2 \left[i\Delta_F(x;x)\right]
\,,
 \label{effective action}
\end{equation}
where $i\Delta_F(x;x')$ denotes the scalar propagator at spacetime coincidence, and $\mu$ is 
an arbitrary energy scale. 
The second equality in~(\ref{effective action}) follows from taking a derivative with respect to $m^2$ and integrating, and the two actions are identical up to an irrelevant field independent integration constant.

Note firstly that the coincident Wightman functions~(\ref{Wightman functions: general Lor viol solution}) 
are both real, implying that they are equal to the coincident 
Feynman propagators~(\ref{Feynman propagator: lightcone coordinates})
and~(\ref{Feynman propagator: Cartesian coordinates}),
\begin{equation}
i\Delta_F(x;x) = i\Delta^{(\pm)}(x;x)
= \frac{\hbar m^{D-2}}
               {(2\pi)^\frac{D}{2}\sqrt{\gamma(u)\Upsilon(u;u)}}
   \left[\frac{K_{\frac{D-2}{2}}\big(m\sqrt{\Delta {\bar x}_{(\pm)}^2}\,\big)}
   {\big(m\sqrt{\Delta {\bar x}_{(\pm)}^2}\,\big)^\frac{D-2}{2}}\right]_{x'\rightarrow x}
\,,
\label{Wightman functions: coincident}
\end{equation}
The scalar propagator at spacetime coincidence can be obtained
from the following power-series representation of Bessel function of the second kind,
\begin{equation}
\frac{K_\nu(z)}{z^\nu} = \frac{\Gamma(\nu)\Gamma(1-\nu)}{2^{1+\nu}}
                  \left[\sum_{n=0}^\infty\frac{(z/2)^{2n-2\nu}}{n!\Gamma(n+1-\nu)}
                  -\sum_{n=0}^\infty\frac{(z/2)^{2n}}{n!\Gamma(n+1+\nu)}\right]
\label{power series of Bessel function}
\end{equation}
where $z=m\sqrt{\Delta x_{\pm}^2}$ and $\nu=(D-2)/2$. 
Recalling that in dimensional regularization $D-$dependent powers of $z$ do not contribute
at coincidence, one arrives at,
\begin{equation}
 i\Delta_F(x;x) 
 =   \frac{m^{D-2}}{(4\pi)^{D/2}\sqrt{\gamma(u)\Upsilon(u;u)}}
                                  \Gamma\Big(1-\frac{D}{2}\Big)
\,,\qquad \Gamma\Big(1-\frac{D}{2}\Big) = \frac{2}{D\!-\!4} + \gamma_E -1 +{\cal O}(D-4)
\,.
\label{Wightman functions at coincidence II}
\end{equation}

The next step is to calculate the product $\gamma(u)\Upsilon(u;u)$.
In linear representation for nonpolarized gravitational waves, one 
has $\gamma(u) = (1-h^2)$ and from Eq.~(\ref{Appendix B: inverse rotation matrix}) one finds, 
$\Upsilon(u;u)=1/(1-h^2)$, such that,
\begin{equation}
 \gamma(u)\Upsilon(u;u) = 1
 \,.
\label{product gamma Upsilon: lin rep}
\end{equation}
In fact, this result holds for more general polarized gravitational waves. To show that, 
one can apply the l'Hospital rule to Eq.~(\ref{general solution: Upsilon matrix lin rep: formal}) to obtain,
\begin{eqnarray}
{\Upsilon}_{{}_x\atop{}^y}(u;u) \!\!&=&\!\! 
\frac{1\!\mp\!h_+\cos(\omega_g u\!-\!\psi/2)}
{1\!-\!h_+^2\cos^2(\omega_gu\!-\!\psi/2)-h_\times^2\cos^2(\omega_gu\!+\!\psi/2)}
\,, \;
\nonumber\\
{\Upsilon}_{xy}(u;u) \!\!&=&\!\! 
\frac{\!-\!h_\times\cos(\omega_gu\!+\!\psi/2)}
{1\!-\!h_+^2\cos^2(\omega_gu\!-\!\psi/2)-h_\times^2\cos^2(\omega_g u\!+\!\psi/2)}
\,,\qquad\;
\label{general solution: Upsilon matrix lin rep: coincident}
\end{eqnarray}
from which one easily obtains, 
\begin{eqnarray}
{\Upsilon}(u;u) =  \frac{1}
{1\!-\!h_+^2\cos^2(\omega_gu\!-\!\psi/2)-h_\times^2\cos^2(\omega_gu\!+\!\psi/2)}
\equiv \frac{1}{1-h(u)^2}
\,,\qquad 
\label{general solution: Upsilon matrix lin rep: coincident 2}
\end{eqnarray}
On the other hand, $\gamma(u) = {\rm det}[g_{ij}] = 1-h(u)^2$, such that,
\begin{eqnarray}
\gamma(u) {\Upsilon}(u;u) = 1
\,.\qquad 
\label{general solution: Upsilon matrix lin rep: coincident 3}
\end{eqnarray}
The calculation in exponential representation is even easier, as $\gamma(u)=1$ 
and $\Upsilon(u;u)=1$, which can be obtained by applying the l'Hospital rule to 
Eq.~(\ref{Upsilon matrix: exp rep general}). In fact, in Appendix~D we show that 
$\Upsilon(u;u)=1$ holds true 
for even more general gravitational waves in exponential representation.
The significant implication of this simple calculation
 is that planar gravitational waves do not influence 
scalar field coincident propagator~(\ref{Wightman functions at coincidence II}),
{\it i.e.} we have,
\begin{equation}
 i\Delta_F(x;x) 
 =   \frac{\hbar m^{D-2}}{(4\pi)^{D/2}}
                                  \Gamma\Big(1-\frac{D}{2}\Big)
\,,\qquad 
\Gamma\Big(1-\frac{D}{2}\Big) = \frac{2}{D\!-\!4} + \gamma_E -1 +{\cal O}(D-4)
\,.\qquad 
\label{Wightman functions at coincidence III}
\end{equation}
Inserting this into~(\ref{effective action}) gives,
\begin{eqnarray}
\Gamma_1 \!\!&=&\!\! \int {\rm d}^Dx\sqrt{-g}
                    \frac{\hbar m^{D}}{2(4\pi)^{D/2}}
                                  \Gamma\Big(\!-\frac{D}{2}\Big)
\nonumber\\
  \!\!&=&\!\! -\int {\rm d}^Dx\!\sqrt{-g}
             \frac{\hbar m^4}{64\pi^2}
                   \left[\frac{2\mu^{D-4}}{D\!-\!4}
                    +\ln\Big(\frac{m^2}{4\pi\mu^2}\Big)+\gamma_E-\frac32\right]
\,.
 \label{effective action II}
\end{eqnarray}
From this we see that the counterterm action is of the cosmological constant-type,
\begin{equation} 
 S^{\rm ct}_1 = \int {\rm d}^Dx\sqrt{-g} \Big(-\frac{\Lambda_0}{16\pi G_0}\Big)
 \,,\qquad
  \frac{\Lambda_0}{16\pi G_0}=-\frac{\hbar m^4}{32\pi^2}
                  \frac{\mu^{D-4}}{D\!-\!4}
\,,
\label{counterterm action: for Gamma1}
\end{equation}
where the counterterm is chosen in accordance with the minimal subtraction scheme. 
When~(\ref{counterterm action: for Gamma1})
is added to~(\ref{effective action II}) one obtains,  
\begin{eqnarray}
\Gamma_1^{\rm ren}= \int {\rm d}^Dx\sqrt{-g}\left\{\!-
             \frac{\hbar m^4}{64\pi^2}
                   \left[\ln\Big(\frac{m^2}{4\pi\mu^2}\Big)+\gamma_E-\frac32\right]\right\}
\,,
\label{effective action ren}
\end{eqnarray}
which is the standard vacuum result, interpreted as 
a positive contribution to the cosmological constant.

\subsection{One-loop scalar mass}
\label{One-loop scalar mass} 

The scalar self-interaction in Eq.~(\ref{scalar field action}) contributes at the one-loop order to 
the scalar mass term as,
\begin{eqnarray}
\Delta m^2   \!\!&=&\!\!  \frac{\lambda}{2}i\Delta_F(x;x) 
   = \frac{ \lambda\hbar m^{D-2}}{2(4\pi)^{D/2}}\Gamma\Big(1-\frac{D}{2}\Big)
\nonumber\\
     \!\!&=&\!\! \frac{\lambda\hbar m^{2}}{32\pi^{2}}
                       \left[\frac{2\mu^{D-4}}{D-4}+\ln\Big(\frac{m^2}{4\pi\mu^2}\Big)
                                    +\gamma_E-1\right]
\,,
\label{scalar one loop mass}
\end{eqnarray}
where we made use of Eq.~(\ref{Wightman functions at coincidence III}).
 The corresponding Feynman diagram is shown at the right panel
of figure~\ref{figure five}.
The divergence in~(\ref{scalar one loop mass}) can be removed by the counterterm action,
\begin{equation} 
 S^{\rm ct}_2 = \int {\rm d}^Dx\sqrt{-g} \left[-\frac{m_0^2}{2}\phi^2\right]
\,,\qquad m_0^2  = -\frac{\lambda \hbar m^{2}}{16\pi^{2}} \frac{\mu^{D-4}}{D-4}
\,,
\label{counterterm action II}
\end{equation}
resulting in the renormalized mass-squared contribution, 
\begin{eqnarray}
m^2_{\rm ren} = m^2 + m_0^2 +  \Delta m^2 
               =m^2 + \frac{\lambda\hbar m^{2}}{32\pi^2}
                       \left[\ln\Big(\frac{m^2}{4\pi\mu^2}\Big)+\gamma_E-1\right]
\,.
\label{scalar one loop mass: ren}
\end{eqnarray}
From this we conclude that planar gravitational waves do not affect 
in any observable way the one-loop mass. This also means that the 
Brout-Englert-Higgs (BEH) mechanism 
and the Coleman-Weinberg mechanism~\cite{Coleman:1973jx}
for mass generation are unaffected 
(at the one-loop level) by passage of gravitational waves.~\footnote{Our result holds at the one-loop
 level, and hence does not preclude non-trivial effects at two- and higher loops.}
The result~(\ref{scalar one loop mass: ren}) is in disagreement with 
Ref.~\cite{Jones:2017ejm}, where a nonvanishing effect on the scalar field condensate 
was found.


\section{One-loop energy-momentum tensor}
\label{One-loop energy-momentum tensor} 

\begin{figure}[t!]
\vskip -2.5cm
\centerline{\hspace{.in}
\epsfig{file=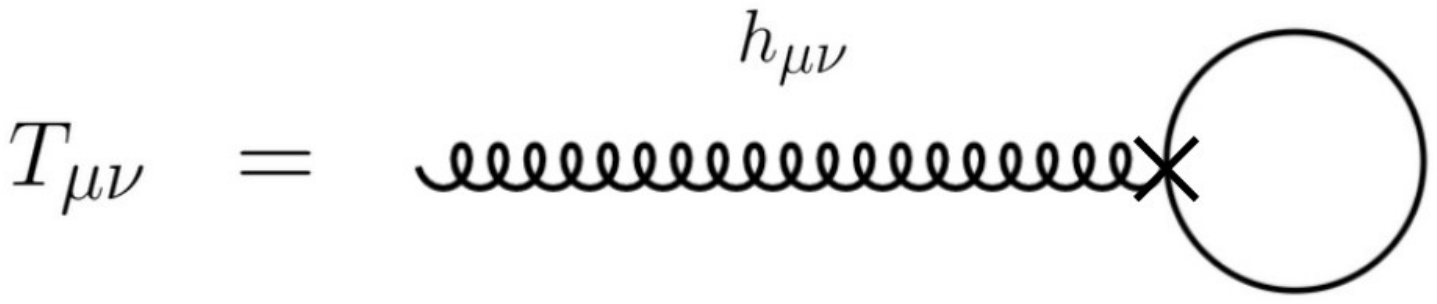, width=4.3in}
}
\vskip -3cm
\caption{\small  The Feynman diagram for the one-loop energy-momentum tensor.
}
\label{figure six}
\end{figure}

The one-loop energy-momentum tensor, $T_{\mu\nu}(x) = \textcolor{blue}{(}-2/\sqrt{-g})\delta S/\delta g^{\mu\nu}(x)$, 
of the scalar field whose action is given in~(\ref{scalar field action}) can be written as,
\begin{eqnarray}
\langle\Omega | T^*[ \hat T_{\mu\nu}(x)]|\Omega\rangle 
    \!\!&=&\!\! \big\langle\Omega \big| T^*\Big[(\partial_\mu\phi)(\partial_\nu\phi)
                  + g_{\mu\nu} \Big(\!-\frac12 g^{\alpha\beta}(\partial_\alpha\phi)(\partial_\beta\phi)
                                              -\frac{m^2}{2}\phi^2 \Big)\Big]\big| \Omega  \big\rangle
\nonumber\\
\!\!&=&\!\!  \Big(\delta_{(\mu}^\alpha \delta_{\nu)}^\beta -\frac12g_{\mu\nu}g^{\alpha\beta} \Big)
                     \Big[\partial_\alpha\partial_\beta'  \big\langle\Omega \big| T\left[\phi(x)\phi(x')\right]
                                     \big|\Omega\big\rangle\Big]_{x'\rightarrow x}
                  - \frac{m^2}{2}\big\langle\Omega \big|T\big[\phi(x)^2\big]\big|\Omega\big\rangle
\nonumber\\
\!\!&=&\!\! \Big(\delta_{(\mu}^\alpha \delta_{\nu)}^\beta -\frac12g_{\mu\nu}g^{\alpha\beta} \Big)
                     \big[\partial_\alpha\partial_\beta' i\Delta_F(x;x')\big]_{x'\rightarrow x}
                                      - \frac{m^2}{2}g_{\mu\nu}i\Delta_F(x;x)
\,,\qquad\qquad\qquad
\label{one-loop energy-momentum tensor}
\end{eqnarray}
where $A_{(\mu}B_{\nu)}=(1/2)(A_\mu B_\nu+A_\nu B_\mu)$ 
and we have dropped the self-interaction term which, 
in the absence of a scalar condensate, contributes
at the two- and higher-loops only. The symbol $T^*$ denotes the frequently-used $T$-star product
which 
-- after the vertex derivatives have been commuted outside Heaviside functions 
-- imposes the usual time ordering.
The corresponding Feynman diagram is shown in figure~\ref{figure six}, which is just
the one-loop contribution to the graviton one-point function.

{\bf Minkowski space contribution.}
Let us now evaluate the term
$\big[\partial_\alpha\partial_\beta' i\Delta_F(x;x')\big]_{x'\rightarrow x}$ in Eq.~(\ref{one-loop energy-momentum tensor}). To develop the intuition on what to expect,
we shall first consider the Minkowski space case in lightcone coordinates,
in which the metric tensor equals, 
\begin{equation}
g_{\mu\nu} = -\frac12\delta_{(\mu}^{\;u}\delta_{\nu)}^{\;v}(\delta_{uv}+\delta_{vu})
     +\delta_{(\mu}^{\;i}\delta_{\nu)}^{\;j}\delta_{ij}^\perp
\,,
\label{metric in LC coordinates}
\end{equation}
where $i,j=1,2,\cdots , D-2$. By making use of Eq.~(\ref{power series of Bessel function}) 
we see that the propagator can be written as two sums, the first having $D$ dependent powers,
the second integer powers. Upon taking two derivatives and then the coincident limit,
only the $n=1$ term from the second (integer) series survives, yielding,
\begin{equation}
 \big[\partial_\alpha\partial_\beta' i\Delta_F(x;x')\big]_{x'\rightarrow x}^{\rm Mink} 
 =- \frac{\hbar m^D}{D(4\pi)^{D/2}}\Gamma\left(1-\frac{D}{2}\right)g_{\alpha\beta}
 = - \frac{m^2}{D}g_{\alpha\beta}\times i\Delta_F(x;x)
 \,,\qquad
\label{energy momentum tensor: derivative term}
\end{equation}
where $g_{\mu\nu}$ is given in Eq.~(\ref{metric in LC coordinates}) and
the coincident propagator in~(\ref{Wightman functions at coincidence III})
(which follows from Eqs.~(\ref{Wightman functions at coincidence II})
and~(\ref{general solution: Upsilon matrix lin rep: coincident 3})).
Acting the 
tensor structure in Eq.~(\ref{one-loop energy-momentum tensor}) yields a factor $-(D-2)/2$,
\begin{equation}
 \Big(\delta_{(\mu}^\alpha \delta_{\nu)}^\beta -\frac12g_{\mu\nu}g^{\alpha\beta} \Big)
                     \big[\partial_\alpha\partial_\beta' i\Delta_F(x;x')\big]_{x'\rightarrow x}^{\rm Mink}
                  =    \frac{D-2}{2D}m^2g_{\mu\nu}\times i\Delta_F(x;x)
\,,\qquad\qquad
\label{one-loop energy-momentum tensor: derivs M}
\end{equation}
Adding to this the mass term in Eq.~(\ref{one-loop energy-momentum tensor}) one obtains, 
\begin{eqnarray}
\big[\langle\Omega | T^*[ \hat T_{\mu\nu}(x)]|\Omega\rangle\big] ^{\rm Mink}
    \!\!&=&\!\! -\frac{1}{D}m^2g_{\mu\nu}\times i\Delta_F(x;x)
     = -\left(\frac{1}{D}\right) \frac{\hbar m^{D}}{(4\pi)^{D/2}}
                                  \Gamma\Big(1-\frac{D}{2}\Big)g_{\mu\nu}
\nonumber\\
  \!\!&=&\!\! - \frac{\hbar m^4}{64\pi^2}
                                 \left[\frac{2\mu^{D-4}}{D-4}
                                   +\ln\left(\frac{m^2}{4\pi \mu^2}\right)+\gamma_E-\frac{3}{2}
                                   \right]g_{\mu\nu}
\,,\qquad\qquad\qquad
\label{one-loop energy-momentum tensor: Mink}
\end{eqnarray}
where to get the last equality we used~(\ref{Wightman functions at coincidence II})
(with $\gamma(u)=1$ and  $\Upsilon(u;u)=1$).
The divergence in~(\ref{one-loop energy-momentum tensor: Mink}) can be renormalized by 
the cosmological constant counterterm action~(\ref{counterterm action: for Gamma1}), 
which contributes to the energy-momentum tensor as, 
\begin{equation}
 T_{\mu\nu}^{(1)} =  -\frac{\Lambda_0}{16\pi G_0}g_{\mu\nu}
\,,\qquad
\label{counterterm action IIIb}
\end{equation}
from where we infer, 
\begin{equation}
\frac{\Lambda_0}{16\pi G_0}
                 = -\frac{\hbar m^{4}}{32\pi^{2}}\frac{\mu^{D-4}}{D-4}
\,,\qquad
\label{counterterm action IIIc}
\end{equation}
which agrees with~(\ref{counterterm action: for Gamma1}),
meaning that the counterterm action~(\ref{counterterm action: for Gamma1})
that renormalizes the one-loop effective action
also renormalizes this part of the one-loop energy momentum tensor, making the renormalization procedure 
internally consistent. Adding~(\ref{counterterm action IIIb}) to~(\ref{one-loop energy-momentum tensor: Mink})
yields the renormalized one-loop energy-momentum tensor in 
Minkowski space~\cite{Koksma:2011cq},
\begin{eqnarray}
\big[\langle\Omega | T^*[ \hat T_{\mu\nu}(x)]|\Omega\rangle\big] ^{\rm Mink, ren}
    \!\!&=&\!\! - \frac{\hbar m^4}{64\pi^2}
                                 \left[\ln\left(\frac{m^2}{4\pi \mu^2}\right)+\gamma_E-\frac{3}{2}
                                   \right]g_{\mu\nu}
\,,\qquad\qquad\qquad
\label{one-loop energy-momentum tensor: Mink 2}
\end{eqnarray}
 One can obtain this result also by varying Eq.~(\ref{effective action ren}) with respect 
 to $g^{\mu\nu}$, indicating a consistency of the framework. 

\bigskip

{\bf Nonpolarized gravitational waves.}

\noindent
{\bf Linear representation.}
Let us now study the gravitational wave contribution. We shall first consider nonpolarized gravitational waves.
From the structure of the Wightman functions~(\ref{Wightman functions: general Lor viol solution})
and~(\ref{Lorentz breaking distance functions: LC coord}) one sees that
there are two types of contributions that occur when two derivatives hit the Wightman functions:
two derivatives $(\partial_i\partial_j'$) hit the Lorentz violating distance 
function~(\ref{Lorentz breaking distance functions: LC coord}), and 
two derivatives ($\partial_u\partial_{u'}$) hit the prefactor $\sqrt{\gamma(u)\gamma(u')}\Upsilon(u;u')$
in~(\ref{Wightman functions: general Lor viol solution}); .

Let us first consider the spatial contributions.
Notice that acting $\partial_i\partial_j'$ on the propagator produces
({\it cf.} Eq.~(\ref{energy momentum tensor: derivative term})), 
 \begin{eqnarray}
  \big[ \partial_i\partial_{j}' i\Delta_F(x;x')\big]_{x'\rightarrow x}
  \!\!&=&\!\!  - \frac{m^2}{D} i\Delta_F(x;x)
    ({\mathbf \Upsilon}^{-1})_{ij}(u;u)
\nonumber\\
\!\!&=&\!\! - \frac{\hbar m^4}{64\pi^2}
          \left[\frac{2\mu^{D-4}}{D-4}+\ln\Big(\frac{m^2}{4\pi\mu^2}\Big)
                                    +\gamma_E-\frac{3}{2}\right] g_{ij}(u)
\,,\qquad
\label{Tmn: gravitational waves: lin rep nonpol 2}
\end{eqnarray}
where in the last step we made use of 
Eqs.~(\ref{Wightman functions at coincidence III}),
(\ref{inverse rotation matrix}--\ref{determinant upsilon: linear rep nonpol}) 
from which it follows that,
\begin{equation}
  ({\mathbf \Upsilon}^{-1})_{ij}(u;u) = \delta_{ij} + h_{ij}(u)
                                                         \equiv g_{ij}(u)
\,,\qquad
  h_{ij}(u) =  \left(\begin{array}{ccc}
                                                           h\cos(\omega_g u) &  h\sin(\omega_g u)  & 0 \cdots \cr
                                                           h\sin(\omega_g u) &  -h\cos(\omega_g u)  & 0 \cdots \cr
                                                                 0 &  0  & 0 \cdots        \cr
                                                              \vdots    &  \vdots  & \!\!\!\!\!\!\vdots \cr
                                                         \end{array}\right)
\,.\qquad
\label{Tmn: gravitational waves: lin rep nonpol 2B}
\end{equation}
Upon combining~(\ref{Tmn: gravitational waves: lin rep nonpol 2}) with the $uv$ and $vu$ 
contributions (which are identical as in Minkowski space~(\ref{one-loop energy-momentum tensor: Mink})), 
and inserting into 
Eq.~(\ref{one-loop energy-momentum tensor}) one obtains,  
\begin{eqnarray}
\big[\langle\Omega | T^*[ \hat T_{\mu\nu}(x)]|\Omega\rangle\big] ^{(1)}
    \!\!&=&\!\! -\frac{1}{D}m^2\times i\Delta_F(x;x)g_{\mu\nu}(u)
\nonumber\\
  \!\!&=&\!\! - \frac{\hbar m^4}{64\pi^2}
                                 \left[\frac{2\mu^{D-4}}{D-4}
                                   +\ln\left(\frac{m^2}{4\pi \mu^2}\right)+\gamma_E-\frac{3}{2}
                                   \right]g_{\mu\nu}(u)
\,,\qquad\qquad\qquad
\label{one-loop energy-momentum tensor: GW 1}
\end{eqnarray}
which is covariant, but not the same as the Minkowski result in
 Eq.~(\ref{one-loop energy-momentum tensor: Mink}),
since $g_{\mu\nu}(u)=\eta_{\mu\nu}+h_{\mu\nu}(u)$ depends on the gravitational wave strain,
$h_{ij}(u)$. As in the Minkowski vacuum case, Eq.~(\ref{one-loop energy-momentum tensor: GW 1})
can be renormalized by the cosmological constant
counterterm action~(\ref{counterterm action: for Gamma1}). Indeed, adding the corresponding 
energy momentum tensor~(\ref{counterterm action IIIb}) with 
the coupling constant~(\ref{counterterm action IIIc}) 
regularizes the energy momentum tensor in Eq.~(\ref{one-loop energy-momentum tensor: GW 1}),
and one obtains, 
\begin{eqnarray}
\big[\langle\Omega | T^*[ \hat T_{\mu\nu}(x)]|\Omega\rangle\big] ^{(1), \rm ren}
    \!\!&=&\!\! - \frac{\hbar m^4}{64\pi^2}
                                 \left[\ln\left(\frac{m^2}{4\pi \mu^2}\right)+\gamma_E-\frac{3}{2}
                                   \right]g_{\mu\nu}(u)
\,,\qquad\qquad\qquad
\label{one-loop energy-momentum tensor: GW 1B}
\end{eqnarray}
which is identical in form to Eq.~(\ref{one-loop energy-momentum tensor: Mink 2}).
 
 \bigskip 
 
\begin{figure}[t!]
\vskip -2.7cm
\centerline{\hspace{.in}
\epsfig{file=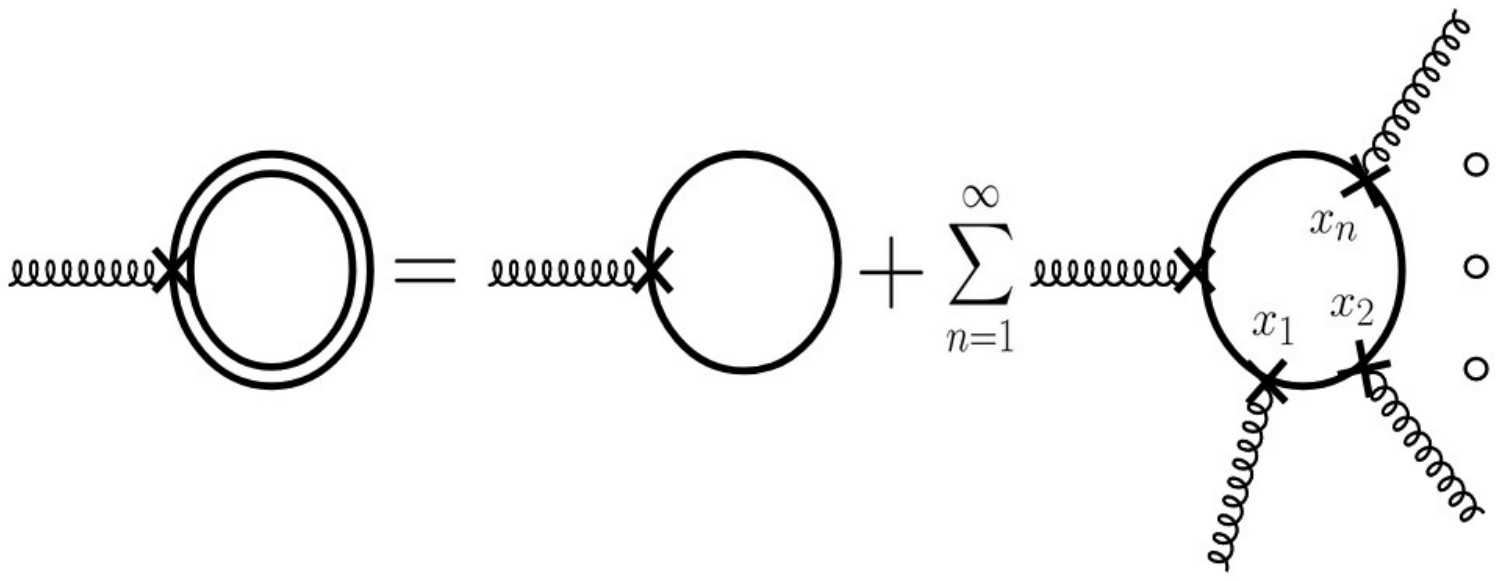, width=5.in}
}
\vskip -2.8cm
\caption{\small Diagramatic representation of the renormalized energy-momentum 
tensor in Eqs.~(\ref{Tmn: nonvanishing contribution with waves: renormalized B})
and~(\ref{Tmn: ren tensorial B}),
for which the one-loop correction to the Minkowski vacuum includes
resummation over (an even number of) the graviton insertions.
}
\label{figure seven}
\end{figure}

 Let us now evaluate the second contribution, which arises from acting the
 $\partial_u\partial_u'$ derivatives.
To facilitate the calculation, it is useful to expand
$\sqrt{\gamma(u)\gamma(u')}\Upsilon(u;u')$
in powers of $\Delta u$. From $\gamma(u)=1-h^2$ and 
Eq.~(\ref{determinant upsilon: linear rep nonpol}) we find,
\begin{eqnarray}
\sqrt{\gamma(u)\gamma(u')}\Upsilon(u;u') =
 \frac{1}{1-h^2}\bigg\{1\!-\!h^2\bigg[1 - \frac13\Big(\frac{\omega_g\Delta u}{2}\Big)^2\bigg]\bigg\}
        + {\cal O}(\Delta u^4)
\,,\quad
\label{determinant upsilon: linear rep nonpol 2B}
\end{eqnarray}
such that 
\begin{equation}
\left[\partial_u\partial_u'\Big(\sqrt{\gamma(u)\gamma(u')}\Upsilon(u;u')\Big)^{-\frac12}
       \right]_{u'\rightarrow u}
   = \frac1{12} \frac{\omega_g^2h^2}{1-h^2}
   \,,\qquad
\label{u derivatives: determinant upsilon: linear rep nonpol}
\end{equation}
which is the only contribution to the energy momentum tensor~(\ref{one-loop energy-momentum tensor}) 
from the $\partial_u\partial_{u'}$ derivatives.
Eq.~(\ref{u derivatives: determinant upsilon: linear rep nonpol}) implies,
\begin{eqnarray}
  \big[ \partial_u\partial_{u'} i\Delta_F(x;x')\big]_{x'\rightarrow x}
  \!\!&=&\!\! \big[ \partial_u\partial_{u'} i\Delta^{(\pm)}(x;x')\big]_{x'\rightarrow x}
=  \frac1{12} \frac{\omega_g^2h^2}{1-h^2} i\Delta_F(x;x)
\nonumber\\
\!\!&=&\!\!   \frac{\hbar m^{2}}{192\pi^{2}} \frac{\omega_g^2h^2}{1-h^2}
                       \left[\frac{2\mu^{D-4}}{D-4}+\ln\Big(\frac{m^2}{4\pi\mu^2}\Big)
                                    +\gamma_E-1\right]
\,,\qquad
\label{Tmn: gravitational waves: lin rep nonpol 2A}
\end{eqnarray}
where the first equality follows from the realization that when $\partial_u\partial_{u'}$
act on $\Theta(\pm\Delta u)$ in the definition of the Feynman propagator, 
they do not produce any contribution.~\footnote{To show that, note that the terms
produced by $\partial_u\partial_{u'}$ 
acting on the Heaviside functions in Eq.~(\ref{Feynman propagator: lightcone coordinates}) 
can be recast (after $\partial_u$ has acted) as,
\begin{equation}
\left\{\partial_{u'}\left[
 \delta(\Delta u)\langle\Omega|\big[\hat\phi(x),\phi(x')\big]  |\Omega\rangle
              \right]
       \right\}_{x'\rightarrow x}
= \left\{
 -\delta'(\Delta u)\frac{i\hbar}{2}\delta^{D-2}(\Delta\vec x_\perp)
          \frac12{\rm sign}(\Delta v)
       \right\}_{x'\rightarrow x}  = 0
\,,\qquad
\nonumber
\end{equation}
where, to obtain the first equality, we made use of Eq.~(\ref{canonical quantization LC}),
and to obtain the last equality we used, ${\rm sign}(0)=0$.
} 
Upon processing~(\ref{Tmn: gravitational waves: lin rep nonpol 2A})
through Eq.~(\ref{one-loop energy-momentum tensor}) one gets, 
\begin{eqnarray}
  \big[\langle\Omega | T^*[ \hat T_{uu}(x)]|\Omega\rangle\big] ^{(2)}
\!\!&=&\!\!   \frac{\hbar m^{2}}{192\pi^{2}} \frac{\omega_g^2h^2}{1-h^2}
                       \left[\frac{2\mu^{D-4}}{D-4}+\ln\Big(\frac{m^2}{4\pi\mu^2}\Big)
                                    +\gamma_E-1\right]
\,,\qquad
\label{Tmn: gravitational waves: lin rep nonpol 2B}
\end{eqnarray}
which can be also written as (in Cartesian coordinates),  
\begin{eqnarray}
  \big[\langle\Omega | T^*[ \hat T_{\mu\nu}(x)]|\Omega\rangle\big] ^{(2)}
\!\!&=&\!\!   \frac{m^{2}}{192\pi^{2}} \frac{\omega_g^2h^2}{1-h^2}
                       \left[\frac{2\mu^{D-4}}{D-4}+\ln\Big(\frac{m^2}{4\pi\mu^2}\Big)
                                    \!+\!\gamma_E\!-\!1\right]
\left(\begin{array}{ccccc}
             1 & 0 & 0 &\cdots & -1  \cr
             0 & 0 & 0 & \cdots & 0  \cr
             0 & 0 & 0 & \cdots & 0  \cr
              \vdots &   \vdots &   \vdots & \cdots &   \vdots  \cr
            -1 & 0 & 0 & \cdots & 1 \cr
            \end{array}\right)
,\qquad\;
\label{Tmn: gravitational waves: lin rep nonpol 2C}
\end{eqnarray}
The divergence in~(\ref{Tmn: gravitational waves: lin rep nonpol 2C})
can be renormalized by the Hilbert-Einstein counterterm action,
\begin{equation}
S_3^{\rm ct} = \int {\rm d}^Dx\sqrt{-g}\left[\frac{R}{16\pi G_0}\right]
\,,
\label{Hilbert-Einstein counterterm action}
\end{equation}
which contributes to the energy-momentum tensor as, 
\begin{equation}
T_{\mu\nu}^{(3)} = - \frac{1}{8\pi G_0}\;G_{\mu\nu}
\,,\qquad
\label{Hilbert-Einstein counterterm action: Tmn}
\end{equation}
where $G_{\mu\nu}=R_{\mu\nu}-\frac12 g_{\mu\nu}R$ 
denotes the Einstein tensor, which for the metric in consideration (nonpolarized gravitational waves
in linear representation) and in Cartesian
coordinates can be written as, 
\begin{equation}
G_{\mu\nu} = - \frac{\omega_g^2h^2}{2(1\!-\!h^2)}
\left(\begin{array}{ccccc}
             1 & 0 & 0 &\cdots & -1  \cr
             0 & 0 & 0 & \cdots & 0  \cr
             0 & 0 & 0 & \cdots & 0  \cr
              \vdots &   \vdots &   \vdots & \cdots &   \vdots  \cr
            -1 & 0 & 0 & \cdots & 1 \cr
            \end{array}\right)
\,.\qquad
\label{Hilbert-Einstein counterterm action: Gmn}
\end{equation}
 By comparing
Eq.~(\ref{Tmn: gravitational waves: lin rep nonpol 2C}) 
with Eqs.~(\ref{Hilbert-Einstein counterterm action: Tmn}) and~(\ref{Hilbert-Einstein counterterm action: Gmn}) we see that a suitable choice of the coupling constant is,
\begin{equation}
 \frac{1}{8\pi G_0} = -\frac{\hbar m^2}{48\pi^2}\times\frac{\mu^{D-4}}{D\!-\!4}
\,.
\label{Hilbert-Einstein counterterm action: Tmn 2}
\end{equation}
Upon adding~(\ref{Hilbert-Einstein counterterm action: Tmn})
 to~(\ref{Tmn: gravitational waves: lin rep nonpol 2C})
one obtains the renormalized energy-momentum tensor in Cartesian coordinates,  
\begin{eqnarray}
\langle\Omega | T^*[ \hat T^{\rm ren}_{\mu\nu}(x)]|\Omega\rangle
      \!\!&=&\!\! - \frac{\hbar m^4}{64\pi^2}
                                 \left[\ln\left(\frac{m^2}{4\pi \mu^2}\right)+\gamma_E-\frac{3}{2}
                                   \right]g_{\mu\nu}
 \label{Tmn: nonvanishing contribution with waves: renormalized A}\\
 \!\!&+&\!\!
 \frac{h^2}{1\!-\!h^2} 
            \frac{\omega_g^2\hbar m^2}{192\pi^2}\left[
            \log\left(\frac{m^2}{4\pi\mu^2}\right)
            +\gamma_E-1\right] 
            \left(\begin{array}{cccc}
            1 & 0 & 0 & -1  \cr
             0 & 0 & 0 & 0  \cr
             0 & 0 & 0 & 0  \cr
             -1 & 0 & 0 & 1 \cr
            \end{array}\right)
  \,,\qquad\;
 \label{Tmn: nonvanishing contribution with waves: renormalized B}
 \end{eqnarray}
 where we have also included the contribution from 
 Eq.~(\ref{one-loop energy-momentum tensor: GW 1}).
The result in Eq.~(\ref{Tmn: nonvanishing contribution with waves: renormalized B}) 
(and the corresponding result in 
Eq.~(\ref{Tmn: gravitational waves: exp rep nonpol 2}) for exponential representation)
is the renormalized one-loop energy-momentum tensor, and it includes the 
resummation over an arbitrary number of classical graviton insertions, the diagrammatic 
representation of which is shown in figure~\ref{figure seven}.

 {\bf Exponential representation.} From Eqs.~(\ref{Appendix B2: phase matrix exponential nonpolarized}),
  and~(\ref{Upsilon matrix exponential nonpolarized}--\ref{diagonalization of k matrix 6})
 we infer,
\begin{equation}
\Upsilon(u;u')  = \cosh^2(\tilde h) -\sinh^2(\tilde h)\left[1-\frac13\left(\frac{\omega_g\Delta u}{2}\right)^2\right]         + {\cal O}(\Delta u^4)
\,,\qquad \gamma(u) = 1
   \,,\qquad\;
 \label{det Upsilon: exp rep nonpol}
 \end{equation}
such that,
\begin{equation}
\left[\partial_u\partial_{u'}\Big(\sqrt{\gamma(u)\gamma(u')}\Upsilon(u;u')\Big)^{-1/2}
 \right]_{u'\rightarrow u}
  = \sinh^2(\tilde h) \frac{\omega_g^2}{12}
   \,,\qquad\;
 \label{det Upsilon: exp rep nonpol 2}
 \end{equation}
and 
\begin{equation}
{\mathbf\Upsilon}^{-1}(u;u) = \left(\!\!\begin{array}{ccc}
\cosh(\tilde h)+\sinh(\tilde h)\cos(\omega_g u) & \sinh(\tilde h)\sin(\omega_g u) & 0 \cdots\cr
\sinh(\tilde h)\sin(\omega_g u) & \cosh(\tilde h)-\sinh(\tilde h)\cos(\omega_g u)& 0 \cdots \cr
                                                                 0 &  0  & 1 \cdots        \cr
                                                              \vdots    &  \vdots  & \!\!\!\!\!\!\vdots \cr                                 \end{array}\!\!\right)
  \equiv \tilde g_{ij}(u)
\,,\qquad
\label{Upsilon inverse: exp rep nonpol 2}
\end{equation}
with $i,j=x,y$ and $\delta_{ij}$ for other values of $i,j$. These results imply 
that the primitive and renormalized spatial part of 
the energy-momentum tensor has the identical tensorial form as 
in~(\ref{one-loop energy-momentum tensor: GW 1})
and~(\ref{one-loop energy-momentum tensor: GW 1B}), respectively. 
The $uu$ component can be read off from Eq.~(\ref{det Upsilon: exp rep nonpol 2}),
\begin{eqnarray}
  \big[ \partial_u\partial_{u'} i\Delta_F(x;x')\big]_{x'\rightarrow x}
\!\!&=&\!\! \sinh^2(\tilde h) \frac{\omega_g^2}{12} i\Delta_F(x;x)
\nonumber\\
\!\!&=&\!\!  \sinh^2(\tilde h) \frac{\hbar m^{2}\omega_g^2 }{192\pi^{2}}
                       \left[\frac{2\mu^{D-4}}{D-4}+\ln\Big(\frac{m^2}{4\pi\mu^2}\Big)
                                    +\gamma_E-1\right]
\nonumber\\
\!\!&=&\!\!   \big[\langle\Omega | T^*[ \hat T_{uu}(x)]|\Omega\rangle\big] ^{(2)}
\,.\qquad
\label{Tmn: gravitational waves: exp rep nonpol 2}
\end{eqnarray}
Keeping in mind the field transformations in Eq.~(\ref{mapping: linear onto exponential}),
$\sinh^2(\tilde h) \rightarrow h^2$, we see that the two results differ by a factor 
$1/(1-h^2) = 1/\gamma(u)$, which is due the fact that, while exponential
representation is unimodular, linear representation is not. Nevertheless,
the result in Eq.~(\ref{Tmn: gravitational waves: exp rep nonpol 2}) is 
just as in linear representation, proportional to the Einstein tensor,
which in exponential representation and for nonpolarized gravitational waves reads
({\it cf.} Eq.~(\ref{Tmn: gravitational waves: exp rep nonpol 2})),
\begin{equation}
G_{\mu\nu} = - \frac{\omega_g^2\sinh^2(\tilde h)}{2}
\left(\begin{array}{ccccc}
             1 & 0 & 0 &\cdots & -1  \cr
             0 & 0 & 0 & \cdots & 0  \cr
             0 & 0 & 0 & \cdots & 0  \cr
              \vdots &   \vdots &   \vdots & \cdots &   \vdots  \cr
            -1 & 0 & 0 & \cdots & 1 \cr
            \end{array}\right)
\,.
\label{Hilbert-Einstein counterterm action: Gmn exp rep}
\end{equation}
Comparing this with Eqs.~(\ref{Tmn: gravitational waves: lin rep nonpol 2C})
and~(\ref{Tmn: nonvanishing contribution with waves: renormalized A}--\ref{Tmn: nonvanishing contribution with waves: renormalized B})
one sees that the primitive and renormalized energy momentum tensors
have the same tensorial form as in linear representation. 
Therefore, the renormalized result can be summarily written as, 
\begin{eqnarray}
\langle\Omega | T^*[ \hat T^{\rm ren}_{\mu\nu}(x)]|\Omega\rangle
      \!\!&=&\!\! - \frac{\hbar m^4}{64\pi^2}
                                 \left[\ln\left(\frac{m^2}{4\pi \mu^2}\right)+\gamma_E-\frac{3}{2}
                                   \right]g_{\mu\nu}
 \label{Tmn: ren tensorial A}\\
 \!\!&&\!\!
 - \frac{\hbar m^2}{96\pi^2}\left[
            \log\left(\frac{m^2}{4\pi\mu^2}\right)
            +\gamma_E-1\right] 
            G_{\mu\nu}
  \,.\qquad\;
 \label{Tmn: ren tensorial B}
 \end{eqnarray}
Remarkably, this result is both tensorial and representation independent.
The result~(\ref{Tmn: ren tensorial A}--\ref{Tmn: ren tensorial B}) disagrees 
with Ref.~\cite{Garriga:1990dp}, which used Pauli-Villars' regularization
and found that the (gravitational wave contribution to the) 
renormalized energy-momentum tensor vanishes in the vacuum state.

\bigskip
{\bf Polarized gravitational waves.}

\smallskip
{\bf Linear representation.} The calculation for this representation is complicated by the fact that 
this representation is not unimodular, $\gamma(u)={\rm det}[g_{ij}]=1-h^2(u)\neq 1$
\big($h^2(u)=h_+^2\cos^2(\omega_gu -\psi/2)+h_\times^2\cos^2(\omega_g u +\psi/2)$\big).
The deformation matrix is of the form,
\begin{eqnarray}
\mathbf\Upsilon(u;u')
 \!\!&&\!\!=\frac{1}{\Delta u}\left[ \left(\!\!\begin{array}{cc} 
\int^u {\rm d}\bar u\frac{1 - h_+\cos(\omega_g\bar u -\psi/2)}{\gamma(\bar u)}
    &
   \!-\!\int^u {\rm d}\bar u\frac{h_\times\cos(\omega_g\bar u +\psi/2)}{\gamma(\bar u)}
      \cr 
       \!-\!\int^u {\rm d}\bar u\frac{h_\times\cos(\omega_g\bar u +\psi/2)}{\gamma(\bar u)}
    &
     \int^u {\rm d}\bar u\frac{1 + h_+\cos(\omega_g\bar u -\psi/2)}{\gamma(\bar u)}
   \cr
                            \end{array}\!\!\right)
    \!\! -\!  (u\!\rightarrow\! u')   \right]
\!,\quad
 \label{Ypsilon matrix: lin rep pol}
\end{eqnarray}
and its determinant is denoted by $\Upsilon(u;u')$, with $\Upsilon(u;u)=1/\gamma(u)$. 
Expanding the deformation matrix around $U=(u+u')/2$
($u=U+\Delta u/2,\,u'=U-\Delta u/2$) to second order in $\Delta u = u-u'$, one obtains,
\begin{eqnarray}
\mathbf\Upsilon(u;u')
 \!\!&&\!\!= \left(\!\!\begin{array}{cc} 
\frac{g_y(U)}{\gamma(U)}
      +\frac{\Delta u^2}{24}\left(\frac{g_y(U)}{\gamma(U)}\right)^{\prime\prime}
    &
  -\frac{g_{xy}(U)}{\gamma(U)}
      -\frac{\Delta u^2}{24}\left(\frac{g_{xy}(U)}{\gamma(U)}\right)^{\prime\prime}
      \cr 
    -\frac{g_{xy}(U)}{\gamma(U)}
      -\frac{\Delta u^2}{24}\left(\frac{g_{xy}(U)}{\gamma(U)}\right)^{\prime\prime}
    &
    \frac{g_x(U)}{\gamma(U)}
      +\frac{\Delta u^2}{24}\left(\frac{g_x(U)}{\gamma(U)}\right)^{\prime\prime}
   \cr
                            \end{array}\!\!\right)
                            +{\cal O}(\Delta u^4)
\,,\quad
 \label{Ypsilon matrix: lin rep pol 2}
\end{eqnarray}
where $g_x(U)$,   $g_y(U)$, and $g_{xy}(U)$ are elements of the metric tensor
evaluated at $U$,
\begin{equation}
g^\perp_{ij}(U)= \left(\! \begin{array}{cc}
       g_x(U) & g_{xy}(U) \cr
      g_{xy}(U) & g_{y}(U) \cr
   \end{array}\!\right)
\,,\qquad g_{{}_x\atop{}^y}(U)  = 1\pm h_+\cos(\omega_g U -\psi/2)
\,,\quad  g_{xy}  = h_\times\cos(\omega_g U +\psi/2)
\,,
\label{metric tensor: U lin rep}
\end{equation}
such that $g_x(U)g_y(U)-g_{xy}(U)^2 = \gamma(U)$.

Note first that Eq.~(\ref{Ypsilon matrix: lin rep pol 2}) implies,
$\left[{\mathbf \Upsilon}^{-1}(u;u)\right]_{ij}= g_{ij}^\perp(u)$
({\it cf.} Eqs.~(\ref{Tmn: gravitational waves: lin rep nonpol 2B}) 
and~(\ref{Upsilon inverse: exp rep nonpol 2})),
from which we  immediately conclude that, after the renormalization is exacted, one obtains
the first part of the energy-momentum in the tensorial form as in Eq.~(\ref{Tmn: ren tensorial A}).

Next, upon inserting Eq.~(\ref{Ypsilon matrix: lin rep pol 2}) into 
$\Upsilon(u;u') =\Upsilon_x(u;u') \Upsilon_y(u;u') - \Upsilon_{xy}(u;u')^2$ yields,  
\begin{eqnarray}
\Upsilon(u;u') \!\!&=&\!\! \frac{1}{\gamma} 
      +\frac{\Delta u^2}{24}\left(
             \frac{g_xg_y^{\prime\prime}\!+\!g_yg_x^{\prime\prime}
                       \!-\!2g_{xy}g_{xy}^{\prime\prime}}{\gamma^2}
                \!-\! 2\frac{\gamma\gamma'' \!-\! \gamma'^2}{\gamma^3}
                \right)
\nonumber\\
                     \!\!&+&\!\! {\cal O}(\Delta u^4)
\,,\quad
 \label{Ypsilon matrix: lin rep pol 3}
\end{eqnarray}
where for notational brevity, in Eq.~(\ref{Ypsilon matrix: lin rep pol 3})
and in what follows,  we suppressed the dependence on $U$.
Multiplying this by,
\begin{equation}
 \sqrt{\gamma(u)\gamma(u')} = \gamma 
    + \frac{\Delta u^2}{8}\frac{\gamma'^2\!-\!\gamma\gamma''}{\gamma^3}
    +{\cal O}(\Delta u^4)
\,,
\label{Ypsilon matrix: lin rep pol 4}
\end{equation}
yields,
\begin{eqnarray}
 \sqrt{\gamma(u)\gamma(u')}\Upsilon(u;u') \!\!&=&\!\! 1 
      +\frac{\Delta u^2}{24}\left(
             \frac{g_xg_y^{\prime\prime}\!+\!g_yg_x^{\prime\prime}
                       \!-\!2g_{xy}g_{xy}^{\prime\prime}}{\gamma}
                \!+\!\frac{\gamma\gamma'' \!-\! \gamma'^2}{\gamma^2}
                \right)
\nonumber\\
                     \!\!&+&\!\! {\cal O}(\Delta u^4)
\,,\quad
 \label{Ypsilon matrix: lin rep pol 5}
\end{eqnarray}
from which it is easy to obtain,
%
%
\begin{eqnarray}
\left[\partial_u\partial_{u'}\Big(\sqrt{\gamma(u)\gamma(u')}\Upsilon(u;u')\Big)^{-1/2}\,
 \right]_{u'\rightarrow u}
  \!\!&=&\!\! \frac{1}{24}\left(
             \frac{g_xg_y^{\prime\prime}\!+\!g_yg_x^{\prime\prime}
                       \!-\!2g_{xy}g_{xy}^{\prime\prime}}{\gamma}
                \!+\!\frac{\gamma\gamma'' \!-\! \gamma'^2}{\gamma^2}
                \right)
\,,\qquad\;
 \label{det Upsilon: lin rep pol 2}
\end{eqnarray}
where the primes denote derivatives with respect to $u$
and we made use of, $\partial_u\partial_{u'} = \frac14 \partial_U^2-\partial^2_{\Delta u}$.
Inserting Eq.~(\ref{det Upsilon: lin rep pol 2}) into 
Eq.~(\ref{one-loop energy-momentum tensor}) results in,
%
\begin{eqnarray}
  \big[\langle\Omega | T^*[ \hat T_{uu}(x)]|\Omega\rangle\big] ^{(2)}
  \!\!\!&=&\!\!\! \big[ \partial_u\partial_{u'} i\Delta_F(x;x')\big]_{x'\rightarrow x}
=\frac{1}{24}\left(
             \frac{g_xg_y^{\prime\prime}\!+\!g_yg_x^{\prime\prime}
                       \!-\!2g_{xy}g_{xy}^{\prime\prime}}{\gamma}
                \!+\!\frac{\gamma\gamma'' \!-\! \gamma'^2}{\gamma^2}
                \right) i\Delta_F(x;x)
\,,\qquad\;\;\;
\label{Tmn: gravitational waves: lin rep pol 3}
\end{eqnarray}
from where one obtains the second contribution to the energy-momentum tensor,
which in Cartesian coordinates reads,
\begin{eqnarray}
  \big[\langle\Omega | T^*[ \hat T_{\mu\nu}(x)]|\Omega\rangle\big] ^{(2)}
  \!\!&=&\!\!
 \frac{\hbar m^{2}}{384\pi^{2}}\left(
             \frac{g_xg_y^{\prime\prime}\!+\!g_yg_x^{\prime\prime}
                       \!-\!2g_{xy}g_{xy}^{\prime\prime}}{\gamma}
                \!+\!\frac{\gamma\gamma'' \!-\! \gamma'^2}{\gamma^2}
                \right)
\nonumber\\     
\!\!&&\!\!\hskip 0cm
                    \times\;   \left[\frac{2\mu^{D-4}}{D-4}\!+\!\ln\Big(\frac{m^2}{4\pi\mu^2}\Big)
                                    \!+\!\gamma_E\!-\!1\right]
\left(\begin{array}{ccccc}
             1 & 0 & 0 &\cdots & -1  \cr
             0 & 0 & 0 & \cdots & 0  \cr
             0 & 0 & 0 & \cdots & 0  \cr
              \vdots &   \vdots &   \vdots & \cdots &   \vdots  \cr
            -1 & 0 & 0 & \cdots & 1 \cr
           \end{array}\right)
.\qquad\;\;\;
\label{Tmn: gravitational waves: lin rep pol 4}
\end{eqnarray}
The next step is to calculate the Einstein tensor, which one can show to be of the form,
\begin{equation}
G_{\mu\nu}(u) = - \frac{1}{4}\left(
             \frac{g_xg_y^{\prime\prime}\!+\!g_yg_x^{\prime\prime}
                       \!-\!2g_{xy}g_{xy}^{\prime\prime}}{\gamma}
                \!+\!\frac{\gamma\gamma'' \!-\! \gamma'^2}{\gamma^2}
                \right)
\left(\begin{array}{ccccc}
             1 & 0 & 0 &\cdots & -1  \cr
             0 & 0 & 0 & \cdots & 0  \cr
             0 & 0 & 0 & \cdots & 0  \cr
              \vdots &   \vdots &   \vdots & \cdots &   \vdots  \cr
            -1 & 0 & 0 & \cdots & 1 \cr
            \end{array}\right)
\,.
\label{Hilbert-Einstein counterterm action: Gmn lin rep: pol}
\end{equation}
This is proportional to Eq.~(\ref{Tmn: gravitational waves: lin rep pol 4}), 
and therefore the energy-momentum tensor in Eq.~(\ref{Tmn: gravitational waves: lin rep pol 4})
can be renormalized by the Hilbert-Einstein 
counterterm~(\ref{Hilbert-Einstein counterterm action}--\ref{Hilbert-Einstein counterterm action: Tmn}).
Upon choosing the bare Newton constant 
as in Eq.~(\ref{Hilbert-Einstein counterterm action: Tmn 2}) 
and adding the corresponding energy-momentum 
tensor~(\ref{Hilbert-Einstein counterterm action: Tmn})
to Eq.~(\ref{Tmn: gravitational waves: lin rep pol 4})
renormalizes the energy-momentum tensor, such 
that one again obtains the renormalized one-loop energy momentum tensor 
in the tensorial form given in Eqs.~(\ref{Tmn: ren tensorial A}--\ref{Tmn: ren tensorial B}).

\bigskip

{\bf Exponential representation.} 
 In this representation the deformation matrix is of the form,
\begin{eqnarray}
\hskip -0.6cm&&\hskip -.3cm
\mathbf\Upsilon(u;u')
 \label{Ypsilon matrix: exp rep pol}\\
 \!\!&&\!\!=\frac{1}{\Delta u}\left[ \left(\!\!\begin{array}{cc} 
\int^u {\rm d}\bar u\left(\cosh(\tilde h(\bar u))
     \!-\!\frac{\sinh(\tilde h(\bar u))}{\tilde h(\bar u)}h_+c_+(\bar u)\right)
    &
     \!-\!\int^u {\rm d}\bar u
     \frac{\sinh(\tilde h(\bar u))}{\tilde h(\bar u)}  h_\times c_\times(\bar u)
      \cr 
       \!-\!\int^u {\rm d}\bar u
     \frac{\sinh(\tilde h(\bar u))}{\tilde h(\bar u)}  h_\times c_\times(\bar u)
    &
      \int^u {\rm d}\bar u\left(\cosh(\tilde h(\bar u))
     \!+\!\frac{\sinh(\tilde h(\bar u))}{\tilde h(\bar u)}h_+c_+(\bar u)\right)
   \cr
                            \end{array}\!\!\right)
    \!\! -\!  (u\!\rightarrow\! u')   \right]
\!,\quad
\nonumber
\end{eqnarray}
where $\tilde h^2(u)= \tilde h_+^2c_+^2(u)+\tilde h_\times^2c_\times^2(u)$, 
see Eqs.~(\ref{c+ and c x}--\ref{tilde h u}),
and the determinant is denoted by $\Upsilon(u;u')$, with $\Upsilon(u;u)=1$. 

Upon denoting elements of the metric tensor in the $xy-$plane 
({\it cf.} Eq.~(\ref{metric exponential representation 2})
and~(\ref{metric exponential representation 3})) as, 
\begin{equation}
g_{ij}^\perp(u) \equiv \left(\!\!\begin{array}{cc} 
                            g_x(u) &  g_{xy}(u) \cr
                            g_{xy}(u) &  g_{y}(u) \cr
                                                 \end{array}\!\!\right)
\,, \qquad g_{{}_x\atop ^y} = \cosh(\tilde h) \mp \frac{\sinh(\tilde h)}{\tilde h}h_+c_+
\,,\quad g_{xy} = \frac{\sinh(\tilde h)}{\tilde h}h_\times c_\times 
\,,\quad
\label{g inverse matrix: def}
\end{equation}
one obtains that the structure of the deformation matrix 
$\mathbf\Upsilon$ is identical as in Eq.~(\ref{Ypsilon matrix: lin rep pol 3}), 
but with the simplification, $\gamma(u)=1$. With these remarks in mind
the calculation of the energy-momentum tensor, the Einstein tensor 
and renormalization proceeds as in the above subsection on linear representation, 
but with $\gamma\rightarrow 1$ and $\gamma'\rightarrow 0$, resulting again
in the renormalized energy-momentum tensor
in Eqs.~(\ref{Tmn: ren tensorial A}--\ref{Tmn: ren tensorial B}).

\bigskip

The energy-momentum tensor in Eq.~(\ref{Tmn: ren tensorial B}) is proportional to the (classical)
backreaction of gravitational waves,~\footnote{The quantum one-loop backreaction
from gravitational waves is known as the Lifshitz tensor, and it is defined by, 
$T_{\mu\nu}^{\rm Lifshitz}(x) \equiv 
- \langle \hat G^{(2)}_{\mu\nu}(x)\rangle/(8\pi G)
=\langle (\nabla_\mu \hat h^{\alpha\beta}(x))(\nabla_\nu \hat h_{\alpha\beta}(x))\rangle/(32\pi G)$,
where $G^{(2)}_{\mu\nu}(x)$ denotes the second order contribution
to the Einstein tensor from the gravitational waves.} 
which induces a backreaction, $T^{\rm gw}_{\mu\nu}(x) 
= -G_{\mu\nu}(x)/(8\pi G)$ (see 
Eq.~(\ref{Hilbert-Einstein counterterm action: Tmn})).
Comparing this with Eq.~(\ref{Tmn: ren tensorial B}) one obtains,~\footnote{Based on
Eqs.~(\ref{Tmn: ren tensorial B}) and~(\ref{scalar and gravitational wave backreaction})
one may be tempted to conclude that the effect studied here can be 
absorbed in a finite change of the Newton constant (changing thus the effective strength of gravity),
\begin{equation}
\frac{1}{8\pi G_{\rm eff}} = \frac{1}{8\pi G} 
+ \frac{\hbar m^2}{96\pi^2}\ln\left(\frac{m^2}{4\pi\bar\mu^2}\right)
\label{effective Newton constant}
\,.
\end{equation}
The situtation is more subtle however, since there are other contributions to 
the energy momentum tensor which 
affect the gravtational coupling strength differently. 
For example the quantum backreaction from gravitational waves, 
$\langle \hat G_{\mu\nu}^{(2)}\rangle$, contributes to the 
energy momentum tensor as, $-\langle \hat G_{\mu\nu}^{(2)}\rangle/(8\pi G)$, implying 
that one can resolve $G$ and $G_{\rm eff}$. Furthermore, 
Eq.~(\ref{effective Newton constant}) holds for the vacuum fluctuations 
of massive scalar fields, and there can be reagions of accummulated
scalar fields (for example, around stars),
where the effect in Eq.~(\ref{effective Newton constant}) would be enhanced.
This would make $G_{\rm eff}$ space and time dependent, and in this way observable.
}
\begin{equation}
 \frac{T^{\rm 1\,loop\; scalar \; gw}_{\mu\nu}}{T^{\rm gw}_{\mu\nu}} 
         = \frac{G \hbar m^2}{12\pi}\log\left(\frac{m^2}{4\pi\bar\mu^2}\right) 
\,,
\label{scalar and gravitational wave backreaction}
\end{equation}
where $\bar\mu^2 = \mu^2{\rm e}^{1-\gamma_E}$, which is suppressed 
as, $\sim m^2/m_{\rm P}^2$ ($m_{\rm P}^2=\hbar/G$).
This implies that: the larger is the scalar mass,
the  stronger is the gravitational backreaction. In particular, 
the backreaction vanishes in the massless scalar limit.
Such a matter backreaction can be used 
in controlled environments to detect very heavy scalar particles,
whose presence could not be detected by other means.

Even though the result in~(\ref{Tmn: ren tensorial B}) contains a directional 
flow of energy, we expect it to hold in more general situations,
such as multifrequency gravitational waves and stochastic gravitational backgrounds, for which 
the energy-momentum tensor induced by 
gravitational waves, $T_{\mu\nu}^{\rm gw}$, has a perfect 
relativistic fluid form, which is in the fluid rest frame diagonal,
$[T_\mu^{\;\nu}]^{\rm gw}
={\rm diag}(\rho_{\rm gw},{\cal P}_{\rm gw},{\cal P}_{\rm gw},{\cal P}_{\rm gw})$,
and characterized by the energy momentum tensor $ \rho_{\rm gw}$
and by the isotropic pressures, ${\cal P}_{\rm gw}$, of the form, 
${\cal P}_{\rm gw} = w_{\rm gw} \rho_{\rm gw}$, 
with the equation of state parameter, $w_{\rm gw} = 1/3$.

 \bigskip
 
 \section{Conclusions and outlook}
\label{Conclusions and outlook}

In this paper we construct the massive real scalar field propagator for 
classical, planar gravitational waves propagating on the Minkowski background.
Our principal results are in section~\ref{Scalar propagator}, where 
the Wightman functions~(\ref{Wightman functions: general Lor viol solution})
and the propagator~(\ref{Feynman propagator: lightcone coordinates}--\ref{Feynman propagator: Cartesian coordinates}) in general $D$ spacetime dimensions are constructed.
The solutions differ from the corresponding Minkowski vacuum two-point functions in two aspects: 
\begin{enumerate}
\item Gravitational waves change the strength of scalar field fluctuations, and its impact is 
reflected by the prefactor in Eq.~(\ref{Wightman functions: general Lor viol solution}); 
\item The Minkowski Poincar\'e invariant distance function is changed into the
deformed, non-local, Lorentz symmetry-breaking, distance function in 
Eqs.~(\ref{Lorentz breaking distance functions: LC coord}--\ref{Lorentz breaking distance functions: Cart coord}).
\end{enumerate}

In this work, we construct the two-point functions only for 
monochromatic gravitational waves moving in a definite direction, {\it i.e.} planar gravitational waves.
However, we consider both nonpolarized and polarized gravitational waves
(for which the amplitudes of the $(+)$ and $(\times)$ polarizations differ and/or 
the difference in phases of the two polarizations is not $\pi/2$). In Appendix~D we show how to generalize
to $D(D-3)/2$ polarization, which exist in $D$ spacetime dimensions. 
This work focuses on monochromatic gravitational waves. However, generalization to more general wave forms
(as observed in realistic systems~\cite{Bourgoin:2022ilm,Amaro-Seoane:2022rxf}) is feasible, 
and we leave that for future work.
Our two-point functions resum all powers of gravitational strain, and thus allow to study both linear and 
non-linear effects. To prevent interpretational ambiguities of the results that may occur at higher orders
in the gravitational strain, we use two inequivalent field representations for gravitational waves
-- which we dub linear representation~(\ref{metric perturbation}--\ref{gravitational wave D}) 
and exponential 
representation~(\ref{metric exponential representation}), (\ref{metric exponential representation 3}), 
which allows one to study how physical observables depend on the representation used. 
 
This has borne fruit in sections~\ref{One-loop effective action and scalar mass}
 and~\ref{One-loop energy-momentum tensor}, where we consider simple one-loop applications. 
 In section~\ref{One-loop effective action and scalar mass} we construct the renormalized 
 one-loop effective action~(\ref{effective action ren}), and show that it does not 
get affected by planar gravitational waves. 
The analogous result holds true for the one-loop mass~(\ref{scalar one loop mass: ren}) induced by 
the scalar self-interaction. 
In section~\ref{One-loop energy-momentum tensor} we construct the 
renormalized one-loop energy-momentum tensor induced by the scalar field fluctuations.
Our results~(\ref{Tmn: ren tensorial A}--\ref{Tmn: ren tensorial B}) show that the quantum scalar backreaction
induced by the planar gravitational waves harbors two distinct contributions:
\begin{enumerate}
\item The contribution~(\ref{Tmn: ren tensorial A}), which is proportional to the metric tensor $g_{\mu\nu}$
and thus contributes to the energy-momentum tensor as the cosmological constant. 
Notice that this contribution also depends on the gravitational wave strain through the metric tensor $g_{\mu\nu}$.
\item The contribution~(\ref{Tmn: ren tensorial B}), which is proportional to the Einstein tensor induced by the 
gravitational waves.  The leading order contribution is quadratic in the gravitational wave strain, however our 
calculation resums all powers of the gravitational wave strain.
\end{enumerate}
The energy-momentum tensor~(\ref{Tmn: ren tensorial A}--\ref{Tmn: ren tensorial B}) is tensorial, 
which gives us confidence that the results are correct.

This work assumes single frequency (monochromatic) waves moving in a definite direction.
In more realistic situations however, one expects that the gravitational waves carry 
higher frequency overtones,~\footnote{Gravitational waves with higher frequency overtones can be generated, 
for example, by spinning binary systems and by systems endowed with strong magnetic fields~\cite{Bourgoin:2022ilm}.}
 and -- if generated by multiple sources -- they 
will move in different directions, an important example being the stochastic gravitational wave 
background generated by cosmic inflation. Constructing the propagator for such more general 
gravitational wave backgrounds is a pursuit worth the effort.

While the initial results presented in this work represent simple applications, 
we emphasise that our propagator
constitutes an essential ingredient for understanding the important question,
namely how {\it matter fields respond to classical gravitational waves}
in the context of interacting quantum field theories. While this paper
discusses the massive real scalar field, we expect that generalization to 
matter fields with spin is feasible by making use of the techniques established 
in this work.

\section{Appendices}
\label{Appendices}

\section*{Appendix A: Propagator integral}
\label{Appendix A: Propagator integral} 

In this appendix we construct the Wightman functions and the propagator 
by writing them as a power-series in multipoles by expanding
the phase in the exponential induced by the gravitational waves in powers of spatial derivatives. 
The general form of the expansion 
for the Wightman 
functions~(\ref{positive frequency Wightman function}--\ref{negative frequency Wightman function}) is,
\begin{eqnarray}
i\Delta^{(+)}(x;x') \!\!&=&\!\! 
\sum_{n=0}^\infty \frac{i^n}{n!}
\left[\Delta\Psi_0(u;u')(\partial_x^2\!+\!\partial_y^2)
\!+\!\Delta\Psi_+(u;u')(\partial_x^2\!-\!\partial_y^2)
\!+\!\Delta\Psi_\times(u;u')\partial_x\partial_y\right]^n\!
   \frac{i\Delta_n^{(+)}(x;x')}{\left[\gamma(u)\gamma(u')\right]^\frac{1}{4}} 
,\qquad\;
 \label{positive frequency Wightman function 2}\\
 i\Delta^{(-)}(x;x') \!\!&=&\!\! \sum_{n=0}^\infty \frac{i^n}{n!}
 \left[\Delta\Psi_0(u;u')(\partial_x^2\!+\!\partial_y^2)\!+\!
 \Delta\Psi_+(u;u')(\partial_x^2\!-\!\partial_y^2)
 \!+\!2\Delta\Psi_\times(u;u')\partial_x\partial_y\right]^n\!
\frac{i\Delta_n^{(-)}(x;x') }{\left[\gamma(u)\gamma(u')\right]^\frac{1}{4}}
,\qquad\;
\label{negative frequency Wightman function 2}
\end{eqnarray}
where $\gamma(u)={\rm det}[g_{ij}]$ is the determinant of the spatial metric,
which differs from unity in linear representation
(for example, $\gamma(u)=1-h^2$ for nonpolarized waves), but in exponential representation, 
$\gamma(u)=1$, and can be omitted. 
Eqs.~(\ref{positive frequency Wightman function 2}--\ref{negative frequency Wightman function 2})
are obtained by expanding the mode functions~(\ref{mode function: general solution 4})
in Eqs.~(\ref{positive frequency Wightman function}--\ref{negative frequency Wightman function})
in powers of $\Psi(u)$. 

{\bf Linear representation.}
For nonpolarized gravitational waves in linear representation~(\ref{metric perturbation})
 from Eq.~(\ref{mode function: general solution 4}) one obtains, 
\begin{eqnarray}
\Delta\Psi_0(u;u') \!\!&=&\!\! \Psi_0(u)-\Psi_0(u')
\,,\qquad\;\;  \Psi_0(u) = \frac{h^2}{1-h^2}u
\nonumber\\ 
\Delta\Psi_+(u;u') \!\!&=&\!\! \Psi_+(u)-\Psi_+(u')
\,,\qquad  \Psi_+(u) = -\frac{h_+\sin(\omega_gu)}{\omega_g(1\!-\!h^2)}
\nonumber\\ 
\Delta\Psi_\times(u;u') \!\!&=&\!\!\Psi_\times(u)-\Psi_\times(u')
\,,\qquad  \Psi_\times(u) = \frac{h_\times\cos(\omega_gu)}{\omega_g(1\!-\!h^2)}
\,.
\label{wightman function integral 4}
\end{eqnarray}

For singly polarized gravitational waves, $h_+\neq 0$ and $h_\times=0$, 
one can easily read off from Eq.~(\ref{mode function:  phase 4 + pol 2B}) 
the phase functions, 
 %
 \begin{eqnarray}
\Psi_0(u)  \!\!&=&\!\! \frac{1}{\omega_g}\frac{1}{\sqrt{1\!-\!h_+^2}}
\bigg\{\text{arctan}\bigg[\sqrt{\frac{1\!-\!h_+}{1\!+\!h_+} }
                        \tan\Big( \frac{\omega_g u}{2} \Big) \bigg]
\!+\!\text{arctan}\bigg[\sqrt{\frac{1\!+\!h_+}{1\!-\!h_+} }
                  \tan\Big( \frac{\omega_g u}{2}\Big)\bigg]
\bigg\}  - u
\,,
 \nonumber\\
 \Psi_+(u)  \!\!&=&\!\! \frac{1}{\omega_g}\frac{1}{\sqrt{1\!-\!h_+^2}}
\bigg\{\text{arctan}\bigg[\sqrt{\frac{1\!-\!h_+}{1\!+\!h_+} }
                        \tan\Big( \frac{\omega_g u}{2} \Big) \bigg]
     \!-\!\text{arctan}\bigg[\sqrt{\frac{1\!+\!h_+}{1\!-\!h_+} }
                  \tan\Big( \frac{\omega_g u}{2}\Big)\bigg]
\bigg\} 
\,,
\nonumber\\
\Psi_\times(u)  \!\!&=&\!\!  0
\,.
\label{Psi functions for + polarization}
\end{eqnarray}
On the other hand, when $h_\times\neq 0$ and $h_+=0$, from 
Eq.~(\ref{mode function: general solution phase 4 x}) one infers, 
 %
 \begin{eqnarray}
  \Psi_0(u)  \!\!&=&\!\! \frac{1}{\omega_g\sqrt{1\!-\!h_\times^2}}\text{arctan}\bigg[\frac{1}{\sqrt{1\!-\!h_\times^2} }
                   \tan(\omega_g u)
                   \bigg]-u
\,,
 \nonumber\\
 \Psi_+(u)  \!\!&=&\!\! 0
 \,,
 \nonumber\\
 \Psi_\times(u)  \!\!&=&\!\!  -\,\frac{1}{\omega_g\sqrt{1\!-\!h_\times^2}}
                   \text{arctan}\bigg[\frac{h_\times}{\sqrt{1\!-\!h_\times^2} }
                   \sin(\omega_g u)
                   \bigg]
\,.
\label{Psi phases for x polarization}
\end{eqnarray}
For general polarized gravitational waves 
one can extract the phase functions 
from Eqs.~(\ref{general solution}--\ref{general solution arctanh}), 
 \begin{eqnarray}
  \Psi_0(u)  \!\!&=&\!\!  \frac{1}
  {\omega_g \sqrt{1-(h_+^2 + h_\times^2)+ h_+^2 h_\times^2s_{\psi}^2}}
  \text{Arctan}\Bigg( 
      \frac{h_\times^2 s_{\psi} c_{\psi} 
                   + [1-h_\times^2 s_{\psi}^2]\tan(\omega_g u -\frac{\psi}{2})}
            {\sqrt{1-(h_+^2 + h_\times^2)+ h_+^2 h_\times^2s_{\psi}^2}}\Bigg)  - u 
\,,
\label{Psi phases zero}\\
 \Psi_+(u)  \!\!&=&\!\! \frac{1}
  {\omega_g \sqrt{1-(h_+^2 + h_\times^2)+ h_+^2 h_\times^2s_{\psi}^2}}
\nonumber\\
\!\!&\times&\!\! \left[  \frac{ih_+(1-h_\times^2 s_{\psi}^2)\text{Arctanh}[g(u)]}{2\sqrt{h_+^2 + h_\times^2 - h_\times^2 s_{\psi}^2 (2+h_+^2+ h_\times^2)- 2ih_\times^2 c_{\psi} s_{\psi}\sqrt{1-(h_+^2 + h_\times^2)+ h_+^2 h_\times^2 s_{\psi}^2}}} + {\rm c.c.}\right]
 \,,
\label{Psi phases plus polarization}\\
 \Psi_\times(u)  \!\!&=&\!\!   \frac{1}
  {\omega_g \sqrt{1-(h_+^2 + h_\times^2)+ h_+^2 h_\times^2s_{\psi}^2}}
\nonumber\\
\!\!&\times&\!\! \left[ \frac{i\big[\!-h_\times^3 s_{\psi}^2 c_{\psi} 
 \!+\! i h_\times s_{\psi}
 \sqrt{1\!-\!(h_+^2 \!+\! h_\times^2)\!+\! h_+^2 h_\times^2 s_{\psi}^2} 
    \!-\!h_\times  c_{\psi} (1\!-\!h_\times^2 s_{\psi}^2)\big]\text{Arctanh}[g(u)]}
 {2\sqrt{h_+^2 + h_\times^2 - h_\times^2 s_{\psi}^2 
 (2\!+\!h_+^2\!+\! h_\times^2)\!-\!2ih_\times^2 c_{\psi} s_{\psi}
     \sqrt{1\!-\!(h_+^2 \!+\! h_\times^2)
 \!+\! h_+^2 h_\times^2 s_{\psi}^2}}}
 +{\rm c.c.}\right]
\,.
\qquad\;\;
\label{Psi phases cross polarization}
 \end{eqnarray}
This result is complicated, nevertheless its correctness can be checked 
by considering the simpler limits when either 
$h_\times = 0$ or $h_+=0$ vanishes. For example, 
in the $(+)-$polarized case Eqs.~(\ref{Psi phases zero}--\ref{Psi phases cross polarization}) 
imply,
\begin{eqnarray}
	\Psi(u)  \!\!&=&\!\! -(k_x^2 + k_y^2 ) u 
	+\frac{k_x^2 + k_y^2}{\omega_g \sqrt{1\!-\!h_+^2}} \text{Arctan} 
	\Bigg[ \frac{\tan(\omega_g u)}{\sqrt{1-h_+^2}}\Bigg] 
	+ 	\frac{k_y^2\!-\!k_x^2}{\omega_g \sqrt{1\!-\!h_+^2}} 
\label{checking the limit of plus polarization} \\
	\!\!&\times&\!\!\! \Bigg\{\!\! -\frac{i}{2} \text{Arctanh} 
	\Bigg[\!\frac{\cos(\omega_g u)+i\sqrt{1\!-\!h_+^2} \sin(\omega_g u)}{h_+}\!\Bigg] 
	\!+\!\frac{i}{2} \text{Arctanh} 
	\Bigg[\! \frac{\cos(\omega_g u)-i\sqrt{1\!-\!h_+^2} \sin(\omega_g u)}{h_+}\!\Bigg]
	\!\Bigg\}
\,.\;\;
\nonumber
\end{eqnarray}
Even though the form of~(\ref{checking the limit of plus polarization}) does 
not match~(\ref{mode function:  phase 4 + pol 2B}), one can show they are identical 
on the interval $-\frac{\pi}{2} < \omega_g u < \frac{\pi}{2}$. The equivalence 
on broader domains can be reestablished by analytically 
extending Eqs.~(\ref{Psi phases zero}--\ref{Psi phases cross polarization}) 
in analogous manner as it was 
done in the case of single polarizations in 
Eqs.~(\ref{mode function:  phase 4 + pol 2B}) 
and~(\ref{mode function: general solution phase 4 x}).

\bigskip

{\bf Exponential representation.} For nonpolarized gravitational waves in
exponential representation one obtains from Eq.~(\ref{mode function: exp rep nonpol 2}),
\begin{eqnarray}
\Delta\Psi_0(u;u') \!\!&=&\!\! \Psi_0(u)-\Psi_0(u')
\,,\qquad  \Psi_0(u) =  (\cosh(\tilde h)\! -\!1)u
\,,
\nonumber\\ 
\Delta\Psi_\times(u;u') \!\!&=&\!\! \Psi_+(u)-\Psi_+(u')
\,,\qquad  \Psi_+(u) = -\sinh(\tilde h)\frac{\sin(\omega_gu)}{\omega_g}
\,,
\nonumber\\ 
\Delta\Psi_\times(u;u') \!\!&=&\!\!\Psi_\times(u)-\Psi_\times(u')
\,,\qquad  \Psi_\times(u) = \sinh(\tilde h)\frac{\cos(\omega_gu)}{\omega_g}
\,.\qquad\;\;
\label{Appendix A: phase functions in exp rep: nonpol}
\end{eqnarray}
The integrals for polarized gravitational waves in exponential representation cannot 
be evaluated in terms of known functions, and for that reason we do not separately consider 
singly polarized cases. In the general case from Eq.~(\ref{phase function: exp rep general}) 
one can extract the phase functions in
Eqs.~(\ref{positive frequency Wightman function 2}--\ref{negative frequency Wightman function 2}),~\footnote{\label{footnote 30}
To see that these integrals are hard, let us convert the simplest looking 
one~(\ref{phase function: exp rep general: 0}) into an integral over $\tilde h(u)$,
\begin{eqnarray}
\Psi_0(u)
 \!\!&=&\!\!     
 \int^{\tilde h(u)}
  \frac{h{\rm d}h}{\sqrt{\bar h^4 -\left[h^2-(\tilde h_+^2+\tilde h_\times^2)/2\right]^2}}
 \big(\!\cosh(h)\!-\!1\big)
\,,\hskip .5cm
\label{phase function: exp rep general: 0 2}
\end{eqnarray}
where 
$\bar h^2 = \frac12\sqrt{\tilde h_+^4+\tilde h_\times^4 + 2\cos(2\psi)\tilde h_+^2\tilde h_\times^2}$. With $h_0^2 = (\tilde h_+^2+\tilde h_\times^2)/2$ and $h^2 = w$, the integral 
becomes, 
\begin{eqnarray}
\Psi_0(u)
 \!\!&=&\!\!     
 \frac12\int^{[\tilde h(u)]^2}
  \frac{{\rm d}w}{\sqrt{(\bar h^2+ h_0^2 -w)(\bar h^2-h_0^2+ w)}}
              \sum_{n=1}^\infty \frac{w^n}{(2n)!}
\,.\hskip .5cm
\label{phase function: exp rep general: 0 3}
\end{eqnarray}
The individual integrals in the sum can be expressed in terms of Appell functions
and elementary functions, illustrating how hard is 
the integral~(\ref{phase function: exp rep general: 0}). The other two
integrals~(\ref{phase function: exp rep general: +}) 
and~(\ref{phase function: exp rep general: x}) are not any easier.
}
\begin{eqnarray}
\Psi_0(u)
 \!\!&=&\!\!     
 \int^u {\rm d}\bar u\big(\!\cosh(\tilde h(\bar u))\!-\!1\big)
\,,\hskip .5cm
\label{phase function: exp rep general: 0}\\
\Psi_+(u)
 \!\!&=&\!\!     
  - \int^u {\rm d}\bar u\frac{\sinh(\tilde h(\bar u))}{\tilde h(\bar u)}\tilde h_+ c_+(\bar u)
\,,\hskip .5cm
\label{phase function: exp rep general: +}\\
\Psi_\times(u)
 \!\!&=&\!\!     
 - \int^u {\rm d}\bar u\frac{\sinh(\tilde h(\bar u))}{\tilde h(\bar u)}
           \tilde h_\times c_\times(\bar u)
\,,\hskip .5cm
\label{phase function: exp rep general: x}
\end{eqnarray}
with $\tilde h(\bar u)=\sqrt{\tilde h_+^2 c_+^2(\bar u)+\tilde h_\times^2 c_\times^2(\bar u)}$.
However, the integrals do simplify if one expands 
the original exponential representation~(\ref{metric exponential representation})
in powers of $\tilde h_+$ and $\tilde h_\times$, as it was done in 
Eqs.~(\ref{Upsilon matrix: exp rep general 2}--\ref{Upsilon matrix: exp rep general 5}),
\begin{eqnarray}
\Psi_0(u)
 \!\!&=&\!\!     
\frac{1}{4}\left(\tilde h_+^2+\tilde h_\times^2\right)u
+\frac{\tilde h_+^2}{8\omega_g}\sin\left(2\omega_gu-\psi\right)
+\frac{\tilde h_\times^2}{8\omega_g}\sin\left(2\omega_gu+\psi \right)
+{\cal O}\big(\tilde h_+^4,\tilde h_+^2\tilde h_\times^2,\tilde h_\times^4\big)
\,,\hskip .9cm
\label{phase function: exp rep general: 0}\\
\Psi_+(u)
 \!\!&=&\!\!     
    -\frac{\tilde h_+}{\omega_g}\sin\left(\omega_gu-\frac\psi 2\right)
    +{\cal O}\big(\tilde h_+^3,\tilde h_+\tilde h_\times^2\big)
\,,\hskip .5cm
\label{phase function: exp rep general: +}\\
\Psi_\times(u)
 \!\!&=&\!\!     
   -\frac{\tilde h_\times}{\omega_g}\sin\left(\omega_gu+\frac\psi 2\right)
   +{\cal O}\big(\tilde h_+^2\tilde h_\times,\tilde h_\times^3\big)
\,.\hskip .5cm
\label{phase function: exp rep general: x2}
\end{eqnarray}

\bigskip

{\bf Wightman functions and propagator.}
The biscalars  $i\Delta_n^{(\pm)}(x;x') $ in~(\ref{positive frequency Wightman function 2}) 
and~(\ref{negative frequency Wightman function 2}) denote the 
integrals in Eqs~(\ref{positive frequency Wightman function}--\ref{negative frequency Wightman function}),
\begin{eqnarray}
i\Delta_n^{(+)}(x;x') \!\!&\equiv&\!\! 
\int \frac{{\rm d}^{D-1}k}{(2\pi)^{D-1}}\frac{\hbar}{2\omega[2\Omega_-]^n}
\exp\left\{i \vec k_\perp\!\cdot \Delta\vec x_\perp
           -\frac{i}{2}\Big[\Omega_-(\vec k)\Delta v
      +\Omega_+(\vec k)\Delta u
              \Big]\right\}
\qquad
\label{positive frequency Wightman function 3}\\
i\Delta_n^{(-)}(x;x') \!\!&=&\!\!
 \int \frac{{\rm d}^{D-1}k}{(2\pi)^{D-1}}\frac{\hbar}{2\omega[2\Omega_-]^n}
\exp\left\{-i \vec k_\perp\!\cdot \Delta\vec x_\perp
           +\frac{i}{2}\Big[\Omega_-(\vec k)\Delta v
      +\Omega_+(\vec k)\Delta u
              \Big]\right\}
\label{negative frequency Wightman function 3}
\end{eqnarray}
where $n=0,1,2,\cdots$ is an integer,
$\Delta\vec x_\perp = \vec x_\perp\!-\!\vec x_\perp^{\,\prime}$,
$\Delta u = u\!-\!u'$,  $\Delta v = v\!-\!v'$.
The form of the integrals 
in~(\ref{positive frequency Wightman function 3}--\ref{negative frequency Wightman function 3})
follows from expanding the phases of the mode functions
in~(\ref{mode function: general solution 4}) in powers of $\Psi(u)$.
Recall that 
$i\Delta_n^{(+)}(x;x)=\big[i\Delta_n^{(-)}(x;x')\big]^*$
such that it suffices to evaluate $i\Delta_n^{(+)}(x;x)$.

Next, by making use of $\Omega_\pm(\vec k)=\omega_\perp^2/\Omega_\mp(\vec k)$ 
($\omega_\perp = \sqrt{k_\perp^2+m^2}$, $ k_\perp\equiv \|\vec k_\perp\|$)
 and 
 $\int_{-\infty}^\infty{\rm d}k^{D-1} =\int_0^\infty {\rm d}\Omega_+ (\omega/\Omega_+)$,
the integral~(\ref{positive frequency Wightman function 3}) becomes,
\begin{equation}
i\Delta_n^{(+)} = \frac{\hbar}{2(2\pi)^{D-1}}\int \frac{{\rm d}^{D-2} k_\perp}{\omega_\perp^{2n}}
   {\rm e}^{i\vec k_\perp\cdot  \Delta\vec x_\perp}
                   \int_0^\infty  {\rm d} \Omega_+ \frac{\Omega_+^{n-1}}{2^n}
 \exp\Big\{\!-\frac{i}{2} \Big[\Omega_+\Delta u + \frac{\omega_\perp^2}{\Omega_+}\Delta v\Big]
         \Big\}
  \,, \qquad (n=0,1,2,\cdots)
\,.
\label{propagator integral D: 1}
\end{equation}
The $\Omega_+$-integral can be performed with the help of Eq.~(3.471.9) 
in Ref.~\cite{Gradshteyn:2014} resulting in,
\begin{equation}
i\Delta_n^{(+)} = \frac{\hbar}{(2\pi)^{D-1}}\int \frac{{\rm d}^{D-2} k_\perp}{\omega_\perp^{2n}}
   \,{\rm e}^{i\vec k_\perp\cdot  \Delta\vec x_\perp}
\left[\frac{\omega_\perp}{2}\sqrt{\frac{\Delta v\!-\!i\epsilon}{\Delta u\!-\!i\epsilon}}\,\right]^n
                   K_n\big(\omega_\perp\sqrt{\!-(\Delta u\!-\!i\epsilon)(\Delta v\!-\!i\epsilon)}\,\big)
\,.\qquad
\label{propagator integral D: 1B}
\end{equation}
 The origin of the $i\epsilon$ prescriptions
 in~(\ref{propagator integral D: 1B}) is in
 the requirement on convergence of the integral~(\ref{propagator integral D: 1}).
Namely, convergence in the limit when $\Omega_+\rightarrow +\infty$ requires 
$\Delta u\rightarrow \Delta u-i\epsilon$, and when $\Omega_+\rightarrow 0$
it requires $\Delta v\rightarrow \Delta v-i\epsilon$.

 Next we rewrite the $\vec k_\perp-$integral in~(\ref{propagator integral D: 1B})
  in polar coordinates,
 \begin{eqnarray}
 \int {\rm d}^{D-2} k_\perp{\rm e}^{i\vec k_\perp\cdot  \Delta\vec x_\perp}
 \!\!&=&\!\!  \int_0^\infty {\rm d} k_\perp k_\perp^{D-3}\int {\rm d}\Omega_{D-4} 
  \int_0^{\pi} {\rm d}\theta \,
 {\rm e}^{i k_\perp \|\Delta\vec x_\perp\|\cos(\theta)}\big[\!\sin(\theta)\big]^{D-4}
 \nonumber\\
 \!\!&=&\!\! (2\pi)^{(D-2)/2}
 \int_0^\infty {\rm d} k_\perp k_\perp^{D-3} 
 \times 
 \frac{J_{\frac{D-4}{2}}\big(k_\perp \|\Delta\vec x_\perp\|\big)}
        {(k_\perp \|\Delta\vec x_\perp\|)^\frac{D-4}{2}}
 \,,
\label{propagator integral D: polar}
\end{eqnarray}
where $\int {\rm d}\Omega_{D-4} = 2\pi^{(D-3)/2}/\Gamma\big(\tfrac{D-3}{2}\big)$ 
is the volume of the $(D-4)-$dimensional sphere and we made use of Eq.~(8.411.7)
from Ref.~\cite{Gradshteyn:2014}. 
With this Eq.~(\ref{propagator integral D: 1B}) becomes, 
\begin{equation}
i\Delta_n^{(+)} = \frac{\hbar}{(2\pi)^\frac{D}{2}}
 \left[\frac{1}{2}\sqrt{\frac{\Delta v\!-\!i\epsilon}{\Delta u\!-\!i\epsilon}}\,\right]^n\!\!
\frac{1}{ \|\Delta\vec x_\perp\|^\frac{D-4}{2}}
\int_0^\infty\! {\rm d}k_\perp k_\perp^\frac{D-2}{2}
                   J_\frac{D-4}{2}\big(k_\perp \|\Delta\vec x_\perp\|\big)
                     \frac{K_n\big(\omega_\perp\sqrt{-(\Delta u\!-\!i\epsilon)(\Delta v\!-\!i\epsilon)}\,\big)}
                            {\omega_\perp^{n}}
\,,\quad
\label{propagator integral D: 2}
\end{equation}
with $\omega_\perp = \sqrt{k_\perp^2+m^2}$.
The last integral can be evaluated by using (6.596.7)
 in Ref.~\cite{Gradshteyn:2014}, which gives,
\begin{eqnarray}
&&\hskip -1cm\int_0^\infty\! {\rm d}k_\perp k_\perp^\frac{D-2}{2}
                   J_\frac{D-4}{2}\big(k_\perp \|\Delta\vec x_\perp\|\big)
                     \frac{K_n\big(\omega_\perp\sqrt{-(\Delta u\!-\!i\epsilon)(\Delta v\!-\!i\epsilon)}\,\big)}
                     {\omega_\perp^{n}}
\nonumber\\
&&\hskip 1cm = \frac{\|\Delta\vec x_\perp\|^\frac{D-4}{2}}
                       {\left[\sqrt{-(\Delta u\!-\!i\epsilon)(\Delta v\!-\!i\epsilon)}\right]^n}
 \left(\frac{\sqrt{\!-(\Delta u\!-\!i\epsilon)(\Delta v\!-\!i\epsilon)\!+\!\|\Delta \vec x_\perp\|^2}}{m}\,
    \right)^{n-\frac{D-2}{2}}
\nonumber\\
&&\hskip 6cm \times\,   K_{n-\frac{D-2}{2}}\big(m\sqrt{\!-(\Delta u\!-\!i\epsilon)(\Delta v\!-\!i\epsilon)
                      \!+\!\|\Delta\vec x_\perp\|^2}\,\big)
\,.\quad
\label{propagator integral D: 3}
\end{eqnarray}
Inserting this into~(\ref{propagator integral D: 2}) yields, 
\begin{eqnarray}
i\Delta_n^{(+)}(x;x')  \!\!\!&=&\!\! \frac{\hbar m^{D-2-n}}{(2\pi)^{D/2}}
 \left[\frac{-1}{2m(\Delta u\!-\!i\epsilon)}\,\right]^n\!
 \left(\!m\sqrt{\!-(\Delta u\!-\!i\epsilon)(\Delta v\!-\!i\epsilon)
                  \!+\!\|\Delta\vec x_\perp\|^2}\right)^{n-\frac{D-2}{2}}\!\!
\nonumber\\
&&\hskip 2.7cm \times\,  K_{n-\frac{D-2}{2}}\big(m\sqrt{\!-(\Delta u\!-\!i\epsilon)(\Delta v\!-\!i\epsilon)
                         \!+\!\|\Delta\vec x_\perp\|^2}\,\big)
 , \quad (n=0,1,2,\cdots)
.\qquad\;\;
\label{propagator integral D: 4}
\end{eqnarray}
The integration for the negative frequency integral in Eq.~(\ref{negative frequency Wightman function 3})
proceeds in analogous fashion, with the difference that the $i\epsilon$ prescription 
gets reversed \big(recall that 
$i\Delta_n^{(-)}(x;x')=\big[i\Delta_n^{(+)}(x;x')\big]^*$\big), resulting in 
\begin{eqnarray}
i\Delta_n^{(-)}(x;x')  \!\!\!&=&\!\! \frac{\hbar m^{D-2-n}}{(2\pi)^{D/2}}
 \left[\frac{-1}{2m(\Delta u\!+\!i\epsilon)}\,\right]^n\!
 \left(\!m\sqrt{\!-(\Delta u\!+\!i\epsilon)(\Delta v\!+\!i\epsilon)
                  \!+\!\|\Delta\vec x_\perp\|^2}\right)^{n-\frac{D-2}{2}}\!\!
\nonumber\\
&&\hskip 2.5cm \times\,   
         K_{n-\frac{D-2}{2}}\big(m\sqrt{\!-(\Delta u\!+\!i\epsilon)(\Delta v\!+\!i\epsilon)
                         \!+\!\|\Delta\vec x_\perp\|^2}\,\big)
 , \quad (n=0,1,2,\cdots)
.\qquad\;\;
\label{propagator integral D: 4 negative}
\end{eqnarray}
For $n=0$ these reduce to, 
\begin{equation}
i\Delta_0^{(\pm)}(x;x') = \frac{\hbar m^{D-2}}{(2\pi)^{D/2}}
   \frac{K_{\frac{D-2}{2}}\big(m\sqrt{\!-(\Delta u\!\mp\!i\epsilon)(\Delta v\!\mp\!i\epsilon)
                         \!+\!\|\Delta \vec x_\perp\|^2}\,\big)}
   {\big(m\sqrt{\!-(\Delta u\!\mp\!i\epsilon)(\Delta v\!\mp\!i\epsilon)
                  \!+\!\|\Delta \vec x_\perp\|^2}\,\big)^\frac{D-2}{2}}
\,,
\label{propagator integral D: 5}
\end{equation}
where we made use of $K_{-\nu}(z)=K_{\nu}(z)$
 (see Eq.~(8.486.16)
 in~\cite{Gradshteyn:2014}). 
Noting that $-(\Delta u\!\mp\!i\epsilon)(\Delta v\!\mp\!i\epsilon)+\|\Delta\vec x_\perp\|^2 \rightarrow
 -(\Delta t\mp i\epsilon)^2+\|\Delta\vec x\|^2\; (\Delta t = t-t')$, we see that the $n=0$ integral~(\ref{propagator integral D: 5}) 
 reduces to the positive and negative frequency Wightman functions for 
the massive scalar field in Minkowski vacuum.

To summarize, we have calculated the integral in 
Eq.~(\ref{positive frequency Wightman function 3}--\ref{negative frequency Wightman function 3}) 
and found that they evaluate to, 
\begin{equation}
i\Delta^{(\pm)}_n(x;x') = \frac{\hbar m^{D-2-n}}{(2\pi)^{D/2}}
 \left[\frac{-1}{2m(\Delta u\mp i\epsilon)}\,\right]^n\!
 \left(\!m\sqrt{\Delta x^2_{(\pm)}}\right)^{n-\frac{D-2}{2}}\!\!
   K_{n-\frac{D-2}{2}}\big(m\sqrt{\Delta x^2_{(\pm)}}\,\big)
 , \quad (n=0,1,2,\cdots)
\,,
\label{propagator integral D: 6}
\end{equation}
where 
\begin{eqnarray}
\Delta x^2_{(\pm)}(x;x') \!\!&=&\!\! -(\Delta t \mp i\epsilon)^2 + \|\Delta \vec x\,\|^2
                      = -(\Delta u \mp i\epsilon)(\Delta v\mp i\epsilon)^2 + \|\Delta \vec x_\perp\|^2
\,.\quad 
\label{invariant distance for Wightman}
\end{eqnarray}
In $D=4$ the expressions in~(\ref{propagator integral D: 6}) simplify to, 
\begin{equation}
i\Delta^{(\pm)}_n(x;x') = \frac{\hbar m^{2-n}}{(2\pi)^{2}}
 \left[\frac{-1}{2m(\Delta u\mp i\epsilon)}\,\right]^n\!
 \left(\!m\sqrt{\Delta x^2_{(\pm)}}\right)^{n-1}\!\!
   K_{n-1}\big(m\sqrt{\Delta x^2_{(\pm)}}\,\big)
 , \quad (n=0,1,2,\cdots)
\,.
\label{propagator integral 4: 6}
\end{equation}

The Wightman functions~(\ref{propagator integral D: 6}) can then 
be inserted into the propagator equation~(\ref{Feynman propagator in LC}) to yield the 
Feynman propagator in lighcone coordinates,
\begin{eqnarray}
i\Delta_F(x;x') \!\!&\equiv&\!\! \Theta(\Delta u)i\Delta^{(+)}(x;x')
                                      + \Theta(-\Delta u)i\Delta^{(-)}(x;x')
 \label{Feynman propagator}\\
 \!\!&=&\!\!
\sum_{n=0}^\infty \frac{i^n}{n!}
\left[\Delta\Psi_0(u;u')(\partial_x^2\!+\!\partial_y^2)
        \!+\!\Delta\Psi_+(u;u')(\partial_x^2\!-\!\partial_y^2)
            \!+\!2\Delta\Psi_\times(u;u')\partial_x\partial_y\right]^n
\frac{i\Delta^F_n(x;x')  }{[\gamma(u)\gamma(u')]^{1/4}}
,\;\;\qquad
\nonumber
 \end{eqnarray}
where 
\begin{eqnarray}
i\Delta_n^{F}(x;x') \!\!&=&\!\! \frac{\hbar m^{D-2-n}}{(2\pi)^\frac{D}{2}}
 \sum_{\pm}\left[\frac{\mp 1}{2m(|\Delta u|\!-\!i\epsilon)}\,\right]^n
 \left(m\sqrt{\!-(|\Delta u| \!-\! i\epsilon) (\pm\Delta v \!-\! i\epsilon)
      \!+\!\|\Delta\vec x_\perp\|^2}\right)^{n-\frac{D-2}{2}}
 \nonumber\\
 \!\!&&\times\,
   K_{n-\frac{D-2}{2}}\big(m\sqrt{\!-(|\Delta u| \!-\! i\epsilon) (\pm\Delta v \!-\! i\epsilon) 
      \!+\!\|\Delta\vec x_\perp\|^2}\big)
\,, \quad 
(n=0,1,2,\cdots)
\,,
\quad
\label{Feynman propagator integral}
\end{eqnarray}
where the projection property, $[\theta(\Delta u)]^n =\theta(\Delta u) \; (\forall  n\in {\mathbb N})$ 
was used.
These results represent a useful representation for the Wightman functions 
and the Feynman propagator, where the (global) Lorentz violation 
induced by passing gravitational waves 
is expressed as expansion in powers of the {\it multipoles}, the leading order one being the quadrupole. This representation is suitable for studying the leading order effects of gravitational 
waves. Remarkably, the Lorentz violation 
induced by gravitational waves can be incorporated into 
the Wightman functions and Feynman propagator (see Appendix~B),
and the corresponding exact results are presented in section~\ref{Scalar propagator}.

\bigskip

\section*{Appendix B: General propagator integral}
\label{Appendix B: General propagator integral} 

In this appendix we generalize the integrals from Appendix~A, but here we include 
all of the corrections from the gravitational waves.
The only approximation we make is that gravitational waves are generated by 
binary systems in a single plane, however the methods in this appendix can 
be straighforwardly generalized to the gravitational waves which propagator in general
$D\!-\!2$ spatial dimensions. Let us begin our analysis by 
inserting Eq.~(\ref{mode function: general solution 4 B}) into 
Eq.~(\ref{positive frequency Wightman function}). Upon converting 
${\int_{-\infty}^\infty \rm d}k^{D-1}$
into ${\int_{0}^\infty \rm d}\Omega_+ \omega/\Omega_+$
({\it cf.} Eq.~(\ref{propagator integral D: 1})) the integral becomes, 
\begin{eqnarray}
i\Delta^{(+)}(x;x') \!\!&=&\!\!\frac{\hbar}{[\gamma(u)\gamma(u')]^{1/4}} 
          \int \!\frac{{\rm d}^{D-2} k_\perp}{2(2\pi)^{D-1}}
    {\rm e}^{i\vec k_\perp\cdot  \Delta\vec x_\perp}
\nonumber\\
\!\!&\times&\!\!
                  \! \int_0^\infty \! \frac{{\rm d} \Omega_+}{\Omega_+}
 \exp\bigg\{\!-\frac{i}{2} 
 \bigg[\Omega_+\bigg(\!1\!+\!\frac{(\Upsilon_x\!-\!1) k_x^2\!+\!(\Upsilon_y\!-\!1) k_y^2
                         \!+\!2\Upsilon_{xy} k_xk_y}{\omega_\perp^2}\bigg)
                \Delta u 
                              \!+\! \frac{\omega_\perp^2}{\Omega_+}\Delta v\bigg]
         \bigg\}
,\quad
\label{propagator integral D: 1 Y}
\end{eqnarray}
where $\gamma(u)={\rm det}[g_{ij}(u)]$, we set $n=0$ and 
$\Upsilon_i=\Upsilon_i(u;u)\; (i=x,y,xy)$ 
denote the phase corrections (divided by $\Delta u$) 
generated by the gravitational waves, such that for the gravitational waves in 
linear representation~(\ref{gravitational wave: planar D=4}) we have from
Eqs.~(\ref{mode function: general solution 4}--\ref{mode function: general solution phase 4}) 
for nonpolarized gravitational waves, 
\begin{eqnarray}
\Upsilon_x(u;u')\!\!&=&\!\!\frac{\Psi_x(u)-\Psi_x(u')}{\Delta u}  
\,,\qquad
\Psi_x(u) = \frac{1}{1-h^2}u - \frac{h}{1-h^2}\frac{\sin(\omega_g u)}{\omega_g}
\nonumber \\ 
\Upsilon_y(u;u')\!\!&=&\!\!\frac{\Psi_y(u)-\Psi_y(u')}{\Delta u}  
\,,\qquad
\Psi_{y}(u) =  \frac{1}{1-h^2}u + \frac{h}{1-h^2}\frac{\sin(\omega_g u)}{\omega_g}
\nonumber \\ 
\Upsilon_{xy}(u;u')\!\!&=&\!\!\frac{\Psi_{xy}(u)-\Psi_{xy}(u')}{\Delta u}  
\,,\qquad
\Psi_{xy}(u) =  \frac{h}{1-h^2}\frac{\cos(\omega_g u)}{\omega_g}
\,,
\label{Appendix B2: phase matrix linear nonpolarized}
\end{eqnarray}
and from Eq.~(\ref{general solution})
for general polarized gravitational waves we have,
\begin{eqnarray}
\Psi_{{}_x\atop{^y}}(u)\!\!&=&\!\!
\frac{1}{\omega_g \sqrt{1-(h_+^2 + h_\times^2)+ h_+^2 h_\times^2s_{\psi}^2} }
\left[
 \text{arctan}\Bigg(\frac{h_\times^2 s_{\psi}c_{\psi}+[1-h_\times^2 s_{\psi}^2]
      \tan(\omega_g u \!-\!\frac{\psi}{2}) }
      {\sqrt{1-(h_+^2 + h_\times^2)+ h_+^2 h_\times^2s_{\psi}^2}}\Bigg)  
\right.
\nonumber \\ 
&& \hskip -0.1cm
\left.
\mp\, \frac{i(1\!-\!h_\times^2 s_{\psi}^2)h_+\,\text{arctanh}[g(u)]}
{2\sqrt{h_+^2 \!+\! h_\times^2 \!-\! h_\times^2 s_{\psi}^2 
    (2\!+\!h_+^2\!+\! h_\times^2)\!-\! 2ih_\times^2 c_{\psi} s_{\psi}
     \sqrt{1\!-\!(h_+^2 \!+\! h_\times^2)\!+\! h_+^2 h_\times^2 s_{\psi}^2}}}
     +\,{\rm c.c.} \right]
\nonumber \\ 
\Psi_{xy}(u)\!\!&=&\!\!
\frac{1}{\omega_g \sqrt{1-(h_+^2 + h_\times^2)+ h_+^2 h_\times^2s_{\psi}^2} }
\nonumber\\
\!\!&\times&\!\!
 \left[ 
 \frac{i h_\times \big[i s_{\psi}\sqrt{1\!-\!(h_+^2 \!+\! h_\times^2)
\!+\! h_+^2 h_\times^2 s_{\psi}^2} \!-\!c_{\psi}\big]
     \text{arctanh}[g(u)]}
{2\sqrt{h_+^2 \!+\! h_\times^2 \!-\! h_\times^2 s_{\psi}^2 
    (2\!+\!h_+^2\!+\! h_\times^2)\!-\! 2ih_\times^2 c_{\psi} s_{\psi}
     \sqrt{1\!-\!(h_+^2 \!+\! h_\times^2)\!+\! h_+^2 h_\times^2 s_{\psi}^2}}}
          +\,{\rm c.c.}\right]
\,,\qquad
\label{Appendix B2: phase matrix general linear polarization}
\end{eqnarray}
where
\begin{equation}
g(u) = \frac{(1\!-\!h_\times s_{\psi}^2)\cos(\omega_g \!-\! \psi/2) 
  \!+\!\left[i\sqrt{1\!-\!(h_+^2 \!+\! h_\times^2)
	\!+\! h_+^2 h_\times^2 s_{\psi}^2}\!-\!h_\times^2 c_{\psi} s_{\psi}\right]
	\sin(\omega_g u \!-\! \psi/2)}
	{\sqrt{h_+^2 \!+\! h_\times^2 \!-\! h_\times^2s_{\psi}^2 
	(2\!+\!h_+^2\!+\! h_\times^2)
	\!-\!2ih_\times^2 c_{\psi} s_{\psi}
	\sqrt{1\!-\!(h_+^2 \!+\! h_\times^2)
	\!+\! h_+^2 h_\times^2s_{\psi}^2}}}
\,.\qquad
\end{equation}

Next we consider nonpolarized gravitational waves in exponential 
representation~(\ref{metric exponential representation 2}). 
From Eq.~(\ref{mode function: exp rep nonpol 2})  we infer, 
\begin{eqnarray}
\Upsilon_x(u;u')\!\!&=&\!\!\frac{\Psi_x(u)-\Psi_x(u')}{\Delta u}  
\,,\qquad
\Psi_x(u) = \cosh(\tilde h)u - \sinh(\tilde h)\frac{\sin(\omega_g u)}{\omega_g}
\nonumber \\ 
\Upsilon_y(u;u')\!\!&=&\!\!\frac{\Psi_y(u)-\Psi_y(u')}{\Delta u}  
\,,\qquad
\Psi_{y}(u) =  \cosh(\tilde h)u +\sinh(\tilde h)\frac{\sin(\omega_g u)}{\omega_g}
\nonumber \\ 
\Upsilon_{xy}(u;u')\!\!&=&\!\!\frac{\Psi_{xy}(u)-\Psi_{xy}(u')}{\Delta u}  
\,,\qquad
\Psi_{xy}(u) =  \sinh(\tilde h)\frac{\cos(\omega_g u)}{\omega_g}
\,.
\label{Appendix B2: phase matrix exponential nonpolarized}
\end{eqnarray}

In the general polarized case the integrals in the $\Upsilon(u;u')$ matrix cannot 
be evaluated in a closed form,
\begin{eqnarray}
\Psi_{{}_x\atop {}^y}(u) \!\!&=&\!\! \int^u{\rm d}\bar u
      \left[\cosh\big(\tilde h(\bar u)\big) 
                   \mp\frac{\sinh(\tilde h(\bar u))}{\tilde h(\bar u)}
                         \tilde h_+ \cos\Big(\omega_g\bar u-\frac{\psi}{2}\Big)
     \right]
\nonumber \\  
\Psi_{xy}(u) \!\!&=&\!\! - \int^u{\rm d}\bar u
      \left[\frac{\sinh(\tilde h(\bar u))}{\tilde h(\bar u)}
                         \tilde h_\times \cos\Big(\omega_g\bar u+\frac{\psi}{2}\Big)
     \right]
\,,
\label{Appendix B2: phase matrix exponential polarized}
\end{eqnarray}
and in the main text some simple cases are discussed.  

Making use of Eq.~(3.471.9) in Ref.~\cite{Gradshteyn:2014} one can
evaluate the $\Omega_+$-integral in 
Eq.~(\ref{propagator integral D: 1 Y})
 ({\it cf.} Eq.~(\ref{propagator integral D: 1B})),
\begin{equation}
i\Delta^{(+)}(x;x') = \int\!\frac{{\rm d}^{D-2} k_\perp}{(2\pi)^{D-1}}
   {\rm e}^{i\vec k_\perp\cdot  \Delta\vec x_\perp}\!
                     K_0\Bigg(\omega_\perp\sqrt{-\Delta u_\epsilon\Delta v_\epsilon
                     \bigg(\!1\!+\!\frac{\Upsilon_x k_x^2\!+\!\Upsilon_y k_y^2
                         \!+\!2\Upsilon_{xy} k_xk_y}{\omega_\perp^2}\bigg)}\,\Bigg)
,\quad
\label{propagator integral D: 1B Y}
\end{equation}
where we have introduced a shorthand notation for 
the $i\epsilon$ prescriptions, $\Delta u_\epsilon\equiv \Delta u-i\epsilon$ 
and $\Delta v_\epsilon\equiv \Delta v-i\epsilon$. 
By noting that the argument of $K_0$ can be recast as, 
\begin{equation}
\omega_\perp^2 \bigg(\!1\!+\!\frac{(\Upsilon_x\!-\!1) k_x^2\!+\!(\Upsilon_y\!-\!1) k_y^2
                         \!+\!2\Upsilon_{xy} k_xk_y}{\omega_\perp^2}\bigg)
    = k_x^2\Upsilon_x+k_y^2\Upsilon_y+2\Upsilon_{xy} k_xk_y
       +\sum_{i=3}^{D-2}k_i^2 + m^2 \equiv \bar \omega_\perp^2 
    \,,\qquad
\label{Appendix B: rescaling perp frequency Y}
\end{equation}
 we see that the phase functions $\Upsilon_i\; (i=x,y,xy)$ deform the momentum
 quadratic polynomial
 in the plane of the gravitational waves, but leave the mass unaffected.
 The sum in~(\ref{Appendix B: rescaling perp frequency Y}) contributes only when $D>4$.
 
 To progress notice that the quadratic form $Q(k_x,k_y)$ in the $xy$-plane can be 
 diagonalized by an $(u;u')$-dependent rotation matrix 
 $R_\theta(u;u')$ (which is an element of $SO(2)$),
 \begin{eqnarray}
Q(k_x,k_y)\equiv (k_x\;\; k_y)\!\cdot\!\left(\begin{array}{cc}
                            \Upsilon_x &  \Upsilon_{xy} \cr
                             \Upsilon_{xy} & \Upsilon_y \cr
                           \end{array}\right)\!\cdot\!
                           \left(\begin{array}{c}
                             k_x \cr
                             k_y \cr
                           \end{array}\right)
  = (\tilde k_x\;\; \tilde k_y)\!\cdot\!\left(\begin{array}{cc}
                            A_+ & 0 \cr
                             0 & A_- \cr
                           \end{array}\right)\!\cdot\!
                           \left(\begin{array}{c}
                             \tilde k_x \cr
                             \tilde k_y \cr
                           \end{array}\right)
                           \equiv  (\bar k_x\;\; \bar k_y)\!\cdot\!\left(\begin{array}{c}
                             \bar k_x \cr
                             \bar k_y \cr
                           \end{array}\right)
,\quad\,
\label{Appendix B2: diagonalization of k matrix}
\end{eqnarray}
where $ A_\pm=A_\pm(u;u')$ are the eigenvalues of the matrix ${\mathbf\Upsilon}$ and 
\begin{equation}
  \left(\begin{array}{c}
                             \tilde k_x \cr
                             \tilde k_y \cr
                           \end{array}\right)
                           =
\left(\begin{array}{cc}
                            \cos(\theta) &  \sin(\theta) \cr
                             -\sin(\theta) & \cos(\theta) \cr
                           \end{array}\right)\!\cdot\!
                           \left(\begin{array}{c}
                             k_x \cr
                             k_y \cr
                           \end{array}\right) 
\,,\qquad   \left(\begin{array}{c}
                             \bar k_x \cr
                             \bar k_y \cr
                           \end{array}\right)
                           =  \left(\begin{array}{c}
                             \sqrt{A_+}\,\tilde k_x \cr
                             \sqrt{A_-}\,\tilde k_y \cr
                           \end{array}\right)
\,.
\label{Appendix B2: diagonalization of k matrix 2}
\end{equation} 
It is not hard to show that, 
\begin{equation}
\tan(2\theta) = \frac{2\Upsilon_{xy}}{\Upsilon_x-\Upsilon_y}
\,,\qquad
 A_\pm = \frac{\Upsilon_x+\Upsilon_y}{2}
         \pm\sqrt{\frac{(\Upsilon_x-\Upsilon_y)^2}{4}+\Upsilon_{xy}^2}
\,,
\label{Appendix B2: diagonalization of k matrix 3}
\end{equation}
such that the determinant of the matrix ${\mathbf\Upsilon}$,
\begin{equation}
\Upsilon(u;u')\equiv  {\rm det} \left[{\mathbf\Upsilon}\right]
    = {\rm det} \left[\left(\begin{array}{cc}
                            \Upsilon_x &  \Upsilon_{xy} \cr
                             \Upsilon_{xy} & \Upsilon_y \cr
                           \end{array}\right)\right]
                     = A_+ A_-      = \Upsilon_x\Upsilon_y-\Upsilon_{xy}^2
\,.
\label{Appendix B2: diagonalization of k matrix 4}
\end{equation}
These considerations suggest to name the matrix ${\mathbf\Upsilon}$ as
the {\it momentum space deformation matrix}.

For example, for nonpolarized gravitational waves in linear 
representation~(\ref{Appendix B2: phase matrix linear nonpolarized})
the determinant evaluates to,
\begin{equation}
\Upsilon(u;u')=\frac{1}{(1\!-\!h^2)^2}
      \left[1-h^2\left(\frac{2\sin\Big(\frac{\omega_g\Delta u}{2}\Big)}{\omega_g\Delta u}\right)^{\!2}\,\right]
\,.
\label{Appendix B2: diagonalization of k matrix 5}
\end{equation}
which at the coincident limit yields a finite $h$-dependent value, 
$\Upsilon(u;u)=1/(1-h^2)=1/{\rm det}[g_{ij}]$. On the other hand, for 
nonpolarized gravitational waves in exponential 
represenation~(\ref{Appendix B2: phase matrix exponential nonpolarized}) we have,
\begin{equation}
\Upsilon(u;u')=\cosh^2(\tilde h) - \sinh^2(\tilde h)
  \left(\frac{2\sin\big(\frac{\omega_g\Delta u}{2}\big)}{\omega_g\Delta u}\right)^{\!2}
\,,
\label{Appendix B2: diagonalization of k matrix 6}
\end{equation}
which at coincidence evaluates to one, $\Upsilon(u;u)=1$, as was expected.
If $|\Delta u|$ increases, $\Upsilon(u;u')$ increases, reaching a maximum
when $|\Delta u|\rightarrow \infty$, which for linear representation saturates at,
$\Upsilon\rightarrow 1/(1-h^2)^2$, and for exponential representation at,
$\Upsilon\rightarrow \cosh^2(\tilde h)$.

We are now ready to evaluate the angular part of the $\vec k_\perp-$integral 
in~(\ref{propagator integral D: 1B Y}) 
({\it cf.} Eq.~(\ref{propagator integral D: polar})). Notice first that, 
\begin{equation}
\vec k_\perp\cdot \Delta \vec x_\perp = \tilde k_x \Delta \tilde x + \tilde k_y \Delta \tilde y
            + \sum_{i=3}^{D-2} k_i \Delta x_i
            =\bar k_x \frac{\Delta \tilde x}{\sqrt{A_+}}
                       + \bar k_y \frac{\Delta \tilde y}{\sqrt{A_-}}
            + \sum_{i=3}^{D-2} k_i \Delta x_i
            \equiv \vec {\bar k}_\perp\cdot \Delta \vec{\bar x}
\,,
\label{Appendix B2: diagonalization of k matrix 7}
\end{equation}
where vector $(\Delta \tilde x ,\;\Delta \tilde y)^T$ is rotated in the opposite 
sense from $(\tilde k_x ,\; \tilde k_y)^T$.
With this in mind, the angular part of the $\vec k_\perp-$integral 
in~(\ref{propagator integral D: 1B Y})  can be evalutated
({\it cf.} Eq.~(\ref{propagator integral D: polar})),
 \begin{eqnarray}
 \int {\rm d}^{D-2} k_\perp{\rm e}^{i\vec k_\perp\cdot  \Delta\vec x_\perp}
 \!\!&=&\!\! 
\frac{1}{\sqrt{\Upsilon(u;u')}}  \int {\rm d}^{D-2} \bar k_\perp
 {\rm e}^{i\vec{\bar k}_\perp\cdot  \Delta\vec{\bar x}_\perp}
 \nonumber\\
 \!\!&=&\!\! 
 \frac{(2\pi)^\frac{D-2}{2}}{\|\Delta\vec{\bar x}_\perp\|^\frac{D-4}{2}\sqrt{\Upsilon(u;u')}}
 \int_0^\infty {\rm d} {\bar k}_\perp {\bar k}_\perp^{\frac{D-2}{2}} 
 \, J_{\frac{D-4}{2}}\big(\bar k_\perp \|\Delta\vec{\bar x}_\perp\|\big)
 \,,
\label{propagator integral D: polar Y}
\end{eqnarray}
such that Eq.~(\ref{propagator integral D: 1B Y}) becomes, 
\begin{equation}
i\Delta^{(+)}(x;x') =\frac{\hbar}{[\gamma(u)\gamma(u')]^{1/4}}
\frac{1}{ \|\Delta\vec{\bar x}_\perp\|^\frac{D-4}{2}\sqrt{\Upsilon(u;u')}}
\int_0^\infty\!\! \frac{{\rm d}\bar k_\perp}{(2\pi)^\frac{D}{2}} \bar k_\perp^\frac{D-2}{2}
              J_\frac{D-4}{2}\big(\bar k_\perp \|\Delta\vec{\bar x}_\perp\|\big)
               K_0\big(\bar\omega_\perp\sqrt{\!-\!\Delta u_\epsilon\Delta v_\epsilon}\,\big)
\,,\quad\;
\label{propagator integral D: 2 Y}
\end{equation}
with $\bar \omega_\perp = \sqrt{\bar k_\perp^2+m^2}$.
The integral over $\bar k_\perp$  can be evaluated by using (6.596.7)
 in Ref.~\cite{Gradshteyn:2014}, see Eq.~(\ref{propagator integral D: 3}),
\begin{eqnarray}
&&\hskip -1cm\int_0^\infty\! {\rm d}\bar k_\perp \bar k_\perp^\frac{D-2}{2}
          J_\frac{D-4}{2}\big(\bar k_\perp \|\Delta\vec{\bar x}_\perp\|\big)
     K_0\big(\bar \omega_\perp\sqrt{-\Delta u_\epsilon\Delta v_\epsilon}\,\big)
\nonumber\\
&&\hskip 3cm =m^{D-2} \|\Delta\vec{\bar x}_\perp\|^\frac{D-4}{2}
  \frac{ K_{\frac{D-2}{2}}\bigg(m\sqrt{\!-\Delta u_\epsilon\Delta v_\epsilon
                      \!+\!\|\Delta\vec{\bar x}_\perp\|^2}\,\bigg)}
                      {\left(m\sqrt{\!-\Delta u_\epsilon\Delta v_\epsilon\!
                         +\!\|\Delta \vec{\bar x}_\perp\|^2}\,
    \right)^\frac{D-2}{2}}
\,.\qquad
\label{propagator integral D: 3 Y}
\end{eqnarray}
Inserting this into~(\ref{propagator integral D: 2 Y}) yields, 
\begin{eqnarray}
i\Delta^{(+)}(x;x')  \!\!\!&=&\!\! 
\frac{\hbar m^{D-2}}{(2\pi)^{D/2}[\gamma(u)\gamma(u')]^{1/4}\sqrt{\Upsilon(u;u')}}
\frac{ K_{\frac{D-2}{2}}\bigg(m\sqrt{\!-(\Delta u\!-\!i\epsilon)(\Delta \!-\!i\epsilon)
                      \!+\!\|\Delta\vec{\bar x}_\perp\|^2}\,\bigg)}
                      {\left(m\sqrt{\!-(\Delta u\!-\!i\epsilon)(\Delta v\!-\!i\epsilon)\!
                         +\!\|\Delta \vec{\bar x}_\perp\|^2}\,\right)^\frac{D-2}{2}}
\,,\qquad\;\;
\label{propagator integral D: 4 Y}
\end{eqnarray}
where we have restored the original $i\epsilon$ prescription and where,
\begin{equation}
\|\Delta \vec{\bar x}_\perp\|^2 = \frac{\Delta {\tilde x}^2}{A_+}
  +\frac{\Delta {\tilde y}^2}{A_-} + \sum_{i=3}^{D-2}\Delta x_i^2
\,,
\end{equation}
where the sum contributes only when $D>4$.

The integration for the negative frequency integral in Eq.~(\ref{propagator integral D: 1 Y})
proceeds in an analogous fashion, with the difference that the $i\epsilon$ prescription 
gets reversed, resulting in the negative frequency Wightman function,
\begin{eqnarray}
i\Delta^{(-)}(x;x')  \!\!\!&=&\!\! \frac{\hbar m^{D-2}}
               {(2\pi)^\frac{D}{2}[\gamma(u)\gamma(u')]^\frac{1}{4}\sqrt{\Upsilon(u;u')}}
\frac{ K_{\frac{D-2}{2}}\bigg(m\sqrt{\!-(\Delta u\!+\!i\epsilon)(\Delta v\!+\!i\epsilon)
                      \!+\!\|\Delta\vec{\bar x}_\perp\|^2}\,\bigg)}
                      {\left(m\sqrt{\!-(\Delta u\!+\! i\epsilon)(\Delta v\!+\! i\epsilon)\!
                         +\!\|\Delta \vec{\bar x}_\perp\|^2}\,\right)^\frac{D-2}{2}}
\;.\qquad\;\;
\label{propagator integral D: 4 negative Y}
\end{eqnarray}
These results can be also written as,  
\begin{equation}
i\Delta^{(\pm)}(x;x') = \frac{\hbar m^{D-2}}
        {(2\pi)^\frac{D}{2}[\gamma(u)\gamma(u')]^\frac{1}{4}\sqrt{\Upsilon(u;u')}}
   \frac{K_{\frac{D-2}{2}}\big(m\sqrt{\Delta {\bar x}_{(\pm)}^2}\,\big)}
   {\big(m\sqrt{\Delta {\bar x}_{(\pm)}^2}\big)^\frac{D-2}{2}}
\,,
\label{propagator integral D: 5 Y}
\end{equation}
where 
\begin{eqnarray}
\Delta {\bar x}_{(\pm)}^2(x;x') \!\!&=&\!\!
 -(\Delta u\!\mp\!i\epsilon)(\Delta v\!\mp\!i\epsilon)
   + \frac{\Delta \tilde x^2}{A_+(u;u')}
  +\frac{\Delta \tilde y^2}{A_-(u;u')} + \sum_{i=3}^{D-2}\Delta x_i^2
\nonumber\\
  \!\!&=&\!\! -(\Delta t \mp i\epsilon)^2+ \frac{\Delta \tilde x^2}{A_+(u;u')}
  +\frac{\Delta \tilde y^2}{A_-(u;u')} + \sum_{i=3}^{D-1}\Delta x_i^2
\nonumber\\
  \!\!&=&\!\! -(\Delta t \mp i\epsilon)^2+ 
  \left(\Delta x\;\;\Delta y\right)\!\cdot\! {\mathbf\Upsilon}^{-1}
    \!\cdot\!\left(\begin{array}{c}
                         \Delta x\cr 
                        \Delta y\cr
                       \end{array}
                \right)
 + \sum_{i=3}^{D-1}\Delta x_i^2
\,,\qquad
\label{Distance functions: break Lorentz sym}
\end{eqnarray}
are the distance functions which {\it break Lorentz symmetry}.
Note that the $(u;u')$-dependent deformation matrix,
\begin{equation}
\mathbf{\cal G}(u;u')\equiv {\mathbf\Upsilon}^{-1}(u;u') = \frac{1}{\Upsilon(u;u')}
               \left(\begin{array}{cc}
                      \!   \Upsilon_y & - \Upsilon_{xy}\!\cr 
                      \! - \Upsilon_{xy} &  \Upsilon_x\!\cr
                       \end{array}
                \right)
                \quad \Longrightarrow \quad 
                            {\rm det}\big[\mathbf{\cal G}(u;u')\big] = \frac{1}{\Upsilon(u;u')}
\,,\qquad
\label{Appendix B: inverse rotation matrix}
\end{equation}
which rotates elements of the distance
 function in position space~(\ref{Distance functions: break Lorentz sym}),
is the inverse of the corresponding momentum space
 deformation  matrix ${\mathbf\Upsilon}(u;u')$.

In $D=4$ the expressions in~(\ref{propagator integral D: 5 Y}) simplify to, 
\begin{equation}
i\Delta^{(\pm)}(x;x') = \frac{\hbar m^{2}}
         {(2\pi)^{2}[\gamma(u)\gamma(u')]^\frac{1}{4}\sqrt{\Upsilon(u;u')}}
   \frac{K_{1}\big(m\sqrt{\Delta {\bar x}_{(\pm)}^2}\,\big)}
   {m\sqrt{\Delta {\bar x}_{(\pm)}^2}}
\,.
\label{propagator integral 4: 6 Y}
\end{equation}

The most important results obtained in this Appendix can be summarized as follows: 
\begin{enumerate}
\item[(a)] The amplitude of the Wightman functions in a gravitational
wave background gets amplified by a factor 
 $1/\big([\gamma(u)\gamma(u')]^\frac14[\Upsilon(u;u')]^\frac{1}{2}\big)$, 
 which is in general time dependent; 
 
\item[(b)] The distances in the plane in which gravitational waves oscillate contract as, 
$\Delta \tilde x^2\rightarrow\Delta \tilde x^2/A_+(u;u')$ and 
$\Delta \tilde y^2\rightarrow\Delta \tilde y^2/A_-(u;u')$
thus {\it breaking global Lorentz symmetry} of the Minkowski vacuum.
These corrections are generally time dependent, 
even for nonpolarized gravitational waves.
\end{enumerate}

\bigskip
 
 \section*{Appendix C: Integrals}
\label{Appendix C: Integrals}

The following indefinite integrals are used in section~\ref{Scalar propagator}.
For Eq.~(\ref{mode function:  phase 4 + pol}) we need, 
\begin{equation} 
    \int \frac{1}{1 \mp a\cos(\omega_g u)} du 
    = \frac{2}{\omega_g\sqrt{1-a^2}}   \text{Arctan}\bigg(\sqrt{\frac{1 \pm a}{1 \mp a}} \tan\Big(\frac{\omega_g u}{2}\Big)\bigg)
\,,\qquad (|a|<1)
\,.\quad
\label{integral1}
\end{equation}
For equation~(\ref{mode function: general solution phase 4 x}) we need, 
\begin{equation} 
    \int \frac{1}{1 \pm a\cos^2(\omega u)} {\rm d}u  = \frac{1}{\omega\sqrt{1\pm a}}   
     \text{Arctan}\Bigg(\sqrt{\frac{1}{1 \pm a}} \tan(\omega u)\Bigg)
    \,,\qquad (|a|<1)
\,.
\label{integral2}
\end{equation}
and
\begin{equation} 
    \int \frac{\cos(\omega u)}{1 - a\cos^2(\omega u)} {\rm d}u  = \frac{1}{\omega
    \sqrt{a(1 - a)}}   
     \text{Arctan}\Bigg(\sqrt{\frac{a}{1 - a}} \sin(\omega u)\Bigg)
    \,,\qquad (0<a<1)
\,.
\label{integral3}
\end{equation}
The integral that occurs for general gravitational 
waves~(\ref{mode function: general solution phase 4 B}), 
after using the substitution $\tan(\omega_g u - \frac{\psi}{2}) = t$, becomes:
\begin{eqnarray}
\int \frac{\text{d}t}{\sqrt{1+t^2}} \frac{x\sqrt{1+t^2} + y + zt }{a + b t + c t^2} 
\!\!&=&\!\! \frac{1}{\sqrt{4ac - b^2}}
            \Bigg\{ 2x \text{Arctan}\Bigg[ \frac{b+2ct}{\sqrt{4ac-b^2}}\Bigg]
 \nonumber \\
 &&\!\hskip -5cm 
 +\,  i \frac{\big( 2cy - bz +iz\sqrt{4ac-b^2}\big)}
               {\sqrt{2b^2 + 4c(c-a) \!-\! 2ib\sqrt{4ac-b^2} }} 
 \text{Arctanh}\Bigg[ \frac{2c-bt+ i\sqrt{4ac-b^2}t}
                                     {\sqrt{1+t^2}\sqrt{2b^2 + 4c(c-a) \!-\! 2ib\sqrt{4ac-b^2} }}                 
                     \Bigg]
        \!+\! {\rm c.c} \Bigg\}
\,,\qquad 
\nonumber\\ 
\!&&\!  \hskip 3cm (|a|<1, |b|<1,|c|<1, 4ac\!-\!b^2\!>0)
\,,\qquad
\label{integral6}
\end{eqnarray}
with $4ac-b^2 >0$. For the integral we are evaluating, we have the parameters: $a= 1-h_+^2 - h_\times^2 c_{\psi}^2, \quad b= 2h_\times^2 c_{\psi} s_{\psi}, \quad c = 1-h_\times^2 s_{\psi}^2, \quad x = k_x^2 + k_y^2, \quad  y = h_+(k_y^2 - k_x^2) -2 h_\times k_x k_y c_{\psi},$ and $z= 2h_\times k_x k_y s_{\psi}$. 
 
\bigskip


 \section*{Appendix~D: More general gravitational waves}
\label{Appendix D: More general gravitational waves}

  In this Appendix we generalize the results from 
 section~\ref{Scalar propagator} to $D(D-3)/2$ gravitational wave polarizations.
The analysis presented here utilizes 
an expansion in powers of the gravitational wave amplitude. 
For simplicity, we shall still assume that the waves are planar and propagate in the $x^{D-1}-$direction,
they are in general created by $(D-2)(D-3)/2$ rotating systems in $(D-2)(D-3)/2$ 
orthogonal planes,~\footnote{There are $(D-1)(D-2)/2$ orthogonal planes,
but systems in other planes produce gravitational waves pointing in other
directions, $x^\ell\; (\ell\neq D-1)$.} each system producing
two gravitational wave polarizations, $h_+$ and $h_\times$. While all of the 
$(D-2)(D-3)/2$ {\it cross} polarizations are mutually linearly independent, 
the {\it plus} polarizations are 
in general linearly dependent, such that when superimposed one obtains $D-3$ linearly 
independent plus polarizations, totalling $D(D-3)/2$ polarizations, which is the correct number
in $D$ dimensions.

{\bf Exponential representation.} In this appendix we use the more convenient exponential
representation~(\ref{metric exponential representation}), but now with 
\begin{equation}
\tilde h_{ij}(u) = \sum_{\alpha}\epsilon_{ij}^\alpha \tilde h_\alpha c_\alpha(u)
\,,
\label{Appendix D: exponential representation hij}
\end{equation}
where $\epsilon_{ij}^\alpha$ are polarization tensors,
$\tilde h_\alpha$ denote gravitational wave amplitude of polarization $\alpha\in 1,\cdots D(D-3)/2$, and 
\begin{equation}
c_\alpha(u) = \cos(\omega_g u + \psi_\alpha)
\,,
\label{c alpha}
\end{equation}
where $\psi_\alpha$ are phases.
For general gravitational waves
it is still true that $\gamma(u) = {\rm det}[\tilde g_{ij}(u)]=1$,
which follows from the tracelessness of $\tilde h_{ij}$,
$\gamma(u)=\exp\left({\rm Tr}[\log(\tilde h_{ij})]\right)=1$.
The polarization tensors
$\epsilon_{ij}^\alpha $ satisfy, 
$\delta_{ij} \epsilon_{ij}^\alpha = 0\,,\;
  \epsilon_{j\,D-1}^\alpha = 0\;( j=1,\cdots D-1)$. A convenient basis is, 
\begin{eqnarray}
\epsilon_{ij}^{\times (mn)} 
    \!\!&=&\!\! \frac{1}{\sqrt{2}}\big[\delta_{im}\delta_{jn}
                      +\delta_{in}\delta_{jm}\big]
                      \,,\quad (m,n = 1,\cdots, D-2;\; m\neq n,\; m <n)
\label{cross polarizations in D}\\
\epsilon_{ij}^{+ (\ell)} 
   \!\!&=&\!\! \frac{1}{\sqrt{\ell(\ell+1)}}{\rm diag}\big[\underbrace{1,1,\cdots,1}_{\ell}
             ,-\ell,\underbrace{0,0,\cdots,0}_{D-2-\ell}\big]
                      \,,\quad (\ell = 1,\cdots, D-3)
\,,
\label{plus polarizations in D}
\end{eqnarray}
where the plus polarizations are written as a $D-$dimensional generalization of 
the diagonal  Gell-Mann matrices. It is not hard to show that
the polarization matrices~(\ref{cross polarizations in D}--\ref{plus polarizations in D})
form an orthonormal basis,
\begin{eqnarray}
{\rm Tr}\big[\epsilon^{+ (\ell)}\!\cdot\! \epsilon^{+ (\ell')} \big]
       \!\!&=&\!\!\delta^{\ell\ell'}
\,,\quad
{\rm Tr}\big[\epsilon^{+ (\ell)}\!\cdot \epsilon^{\times (mn)} \big] = 0
\,,\quad
{\rm Tr}\big[\epsilon^{\times (mn)}\!\cdot \epsilon^{\times (m'n')} \big] 
 = \frac12[\delta_{mm'}\delta_{nn'}+\delta_{mn'}\delta_{nm'}]
\nonumber\\
&&\hskip -0.7cm
\Longrightarrow \;{\rm Tr}\big[\epsilon^{\alpha}\!\cdot \epsilon^{\alpha'} \big]
= \delta_{\alpha\alpha'}.\quad
\label{orthonormality for basis}
\end{eqnarray}
The scalar operator field equation in lightcone coordinates is then, 
\begin{equation}
   \left[-4\partial_u\partial_v  
    + \left(\sum_{n=0}^\infty \frac{(-1)^n ({\mathbf h}^n)_{ij}(u)}{n!}\right)\partial_i\partial_j  
    - m^2\right] \hat\phi(u,\vec x_\perp,v) = 0
\,.
\label{EOM scalar in D}
\end{equation}
Canonical quantization proceeds as in 
section~\ref{Scalar propagator} with $\Pi = 2\sqrt{-g}\partial_v \phi=\partial_v \phi$. 
Expanding the field operator and its canonical momentum
in mode functions ({\it cf.} 
Eqs.~(\ref{field expansion in 4}--\ref{canonical momentum expansion in 4})),
one obtains the following equations for the mode functions, 
\begin{equation}
   \left\{\partial_u
     \pm\frac{i}{2\Omega_\mp}\bigg[\left(\sum_{n=1}^\infty \frac{(-1)^n ({\mathbf h}^n)_{ij}(u)}{n!}\right)k_ik_j \bigg]
      \pm \frac{i}{2} \Omega_\pm\right\}
       \phi_\pm(u,\vec k) = 0
\,,
\label{EOM scalar in D mode functions}
\end{equation}
The solution of Eq.~(\ref{EOM scalar in D mode functions}) can be formally written as
({\it cf.} Eqs.~(\ref{mode function: general solution 4 B}--\ref{mode function: general solution phase 4 B})), 
\begin{eqnarray}
 \phi_\pm(u,\vec k) \!\!&=&\!\! \frac{1}{\sqrt{2\omega}}
 \exp\Big[\mp \frac{ i}{2} \Big(\Omega_\pm(\vec k\,)u 
              + \frac{1}{\Omega_\mp(\vec k\,)}\Psi(u)\Big)\Big]
 \,,\qquad
 \label{mode function: general solution D}
 \\
 \Psi(u) \!\!&=&\!\!\int^u {\rm d}\bar u\left(\sum_{n=1}^\infty \frac{(-1)^n ({\mathbf h}^n)_{ij}(\bar u)}{n!}\right)k_ik_j
\,.
\label{mode function phase: general solution D}
\end{eqnarray}
From this one can write elements $\Upsilon_{ij}(u;u')$ of the deformation
matrix $\mathbf \Upsilon$ as,
\begin{eqnarray}
\Upsilon_{ij}(u;u')  \!\!&=&\!\!\delta_{ij}+\frac{1}{\Delta u}
  \left[\int^u {\rm d}\bar u\left(\sum_{n=1}^\infty \frac{(-1)^n ({\mathbf h}^n)_{ij}(\bar u)}{n!}\right)
             - (u\rightarrow u')\right]
\,,
\label{elements of Upsilon matrix: ij}
\end{eqnarray}
which can be evaluated order by order in $h_{ij}$, 
\begin{eqnarray}
\Upsilon_{ij}(u;u') \!\!&=&\!\!\delta_{ij}
+\frac14\sum_{\alpha\beta} \epsilon^\alpha_{ik}\epsilon^\beta_{kj} \tilde h_\alpha \tilde h_{\beta} 
    \cos(\psi_\alpha-\psi_\beta)
\nonumber\\
  \!\!&+&\!\!\!\frac{1}{\Delta u \omega_g}\bigg\{\!\! -\!  \sum_\alpha \epsilon^\alpha_{ij} 
     \tilde h_\alpha\sin(\omega_gu\!+\!\psi_\alpha)
    +\frac18\sum_{\alpha\beta} \epsilon^\alpha_{ik}\epsilon^\beta_{kj} \tilde h_\alpha \tilde h_{\beta} 
    \sin(2\omega_gu\!+\!\psi_\alpha\!+\!\psi_\beta)
             - (u\rightarrow u')\bigg\}
\qquad
\nonumber\\
  \!\!&+&\!\! {\cal O}(\tilde h_{ij}^3)
\,.\qquad\qquad
\label{elements of Upsilon matrix: ij 2}
\end{eqnarray}
By making use of the l'Hospital rule, 
one sees that at coincidence the matrix~(\ref{elements of Upsilon matrix: ij}) simplifies,
\begin{eqnarray}
\Upsilon_{ij}(u;u) \!\!&=&\!\!\delta_{ij}
  + \sum_{n=1}^\infty \frac{(-1)^n ({\mathbf h}^n)_{ij}(u)}{n!}
    = \exp(-{\mathbf h})_{ij}(u)
\,.\qquad\qquad
\label{elements of Upsilon matrix: ij 3}
\end{eqnarray}
such that its determinant equals unity,
\begin{eqnarray}
\Upsilon(u;u) = {\rm det}[\Upsilon_{ij}(u;u)] =1
\,.\qquad\qquad
\label{elements of Upsilon matrix: ij 4}
\end{eqnarray}
We have gathered enough evidence to conjecture 
that this is a generic property of any unimodular representation for gravitational waves.
Upon recalling that the spatial distances are deformed by ${\mathbf \Upsilon}^{-1}$, 
the results of this appendix can be inserted into Eqs.~(\ref{Wightman functions: general Lor viol solution})
and~(\ref{Feynman propagator: lightcone coordinates}--\ref{Feynman propagator: Cartesian coordinates}) 
to obtain the Wightman functions and the propagators
up to the desired accuracy, respectively.

\end{document}